%% file: main.tex
\documentclass[a4paper, 11pt]{article}

\usepackage[a4paper,includeheadfoot,lmargin=2cm,rmargin=2cm,tmargin=2cm,bmargin=2cm]{geometry}
\usepackage{amsmath}
\usepackage{mathtools}
\usepackage{amssymb}
\usepackage{xcolor}
\definecolor{myLightBlue}{HTML}{ADD8E6} 
\definecolor{myDarkerBlue}{HTML}{D1F6FF} 
\definecolor{darkgreen}{rgb}{0.0, 0.39, 0.0}
\definecolor{myblue}{rgb}{0, 0.3, 0.5}
\definecolor{myred}{RGB}{130, 0, 0} 
\usepackage[colorlinks=true, linkcolor=myred, citecolor=myblue, urlcolor=myblue, linktoc=page]{hyperref}
\usepackage{graphicx}
\usepackage{subcaption}
\usepackage[textfont=sl,labelfont=bf]{caption}
\usepackage{float}
\usepackage{physics}
\usepackage{feynmp-auto}
\usepackage{fancyhdr}
\usepackage{slashed}
\usepackage{xspace}
\usepackage{appendix}
\usepackage{blindtext}

\usepackage{booktabs}
\usepackage{array}
\usepackage{tabularx}
\usepackage[table]{xcolor}

\usepackage{multirow}
\usepackage{multicol}
\usepackage{ytableau}

\usepackage{fontspec}

\usepackage[style=numeric-comp,sorting=none,maxnames=999,minnames=999,maxcitenames=999,maxbibnames=999]{biblatex}
\usepackage{listings}
\definecolor{codegreen}{rgb}{0,0.6,0}
\definecolor{codegray}{rgb}{0.5,0.5,0.5}
\definecolor{codepurple}{rgb}{0.58,0,0.82}
\definecolor{backcolour}{rgb}{0.95,0.95,0.92}

\lstdefinestyle{mystyle}{
    backgroundcolor=\color{backcolour},   
    commentstyle=\color{codegreen},
    keywordstyle=\color{magenta},
    numberstyle=\tiny\color{codegray},
    stringstyle=\color{codepurple},
    basicstyle=\ttfamily\footnotesize,
    breakatwhitespace=false,         
    breaklines=true,                 
    captionpos=b,                    
    keepspaces=true,                 
    numbers=left,                    
    numbersep=5pt,                  
    showspaces=false,                
    showstringspaces=false,
    showtabs=false,                  
    tabsize=2
}
\numberwithin{equation}{section} 
\usepackage{microtype}

\begin{document}

\begin{flushright}
\end{flushright}

\begin{center}

{\color{myred}{\Large\bf  Completely asymptotically free chiral theories with scalars}}

\vspace{1cm}

\renewcommand{\thefootnote}{\fnsymbol{footnote}}
\setcounter{footnote}{0}

\centerline{Giacomo Cacciapaglia$^{1,2,}$\footnote{{\color{myred}cacciapa@lpthe.jussieu.fr}}, Francesco Sannino$^{2,3,4,5,}$\footnote{{\color{myred}sannino@cp3.sdu.dk}}, Sophie Wagner$^{2,}$\footnote{{\color{myred}sowag21@student.sdu.dk}}}

\vspace{0.5cm}
{\small $^1$ Laboratoire de Physique Theorique et Hautes Energies, UMR 7589, Sorbonne Universit\'e \& CNRS, \\ 4 place Jussieu, 75252 Paris Cedex 05, France}\\
{\small $^2$ $\hbar$QTC \& the Danish Institute for Advanced Study (DIAS),  University of Southern Denmark,\\ Campusvej 55, DK-5230 Odense M, Denmark}\\
{\small $^3$ Dept. of Physics E. Pancini, Universit\`a di Napoli Federico II, via Cintia, 80126 Napoli, Italy}\\
{\small $^4$ INFN sezione di Napoli, via Cintia, 80126 Napoli, Italy}\\
{\small $^5$ Scuola Superiore Meridionale, Largo S. Marcellino, 10, 80138 Napoli, Italy}
\end{center}

\bigskip

\begin{abstract}
We provide the conditions for complete asymptotic freedom for chiral gauge theories including scalars, as motivated by grand unified models. These are generalised Georgi--Glashow and Bars--Yankielowicz theories that feature a scalar field transforming either in the fundamental or in the adjoint of the gauge group. In both scenarios, we consider the addition of multiple chiral fermion families. We systematically analyse the interplay between gauge, Yukawa, and quartic couplings required for all interactions to remain asymptotically free at short distances. We find that for both scalar representations, complete asymptotic free models can be obtained for a specific number of colours and multiplicity of vector-like and chiral families. 
\end{abstract}

\setcounter{footnote}{0}
\renewcommand{\thefootnote}{\arabic{footnote}}

\newpage 
\tableofcontents
\newpage

\input{Sections/Introduction.tex}

\input{Sections/CAF.tex}
\input{Sections/Chiral_theories.tex}
\input{Sections/fundamental_scalar.tex}

\input{Sections/fundamental_scalar_and_families.tex}
\input{Sections/adjoint_scalar.tex}\label{sec:adj}
\input{Sections/adjoint_scalar_and_families.tex}

\input{Sections/Conclusion.tex}
\printbibliography
\newpage
\setcounter{figure}{0}
\renewcommand{\thefigure}{A.\arabic{figure}}
\input{Sections/Appendix.tex}

\end{document}

%% file: Sections/Introduction.tex
\section{Introduction}
The Standard Model (SM) \cite{Glashow:1961tr, Weinberg:1967tq, Higgs:1964ia, Higgs:1966ev, ATLAS:2012yve} is the most successful theory when describing elementary particle interactions. Since it is a chiral gauge-Yukawa field theory, gaining a deeper theoretical understanding of these models is paramount. It is well known that the SM must be an effective field theory since it does feature neither a free nor an interacting ultraviolet fixed point for all its couplings. In fact, the Higgs doublet quartic coupling runs negative~\cite{Degrassi:2012ry,Antipin:2013sga}, indicating a metastable vacuum, while the hypercharge hits a Landau pole at short distances\footnote{The Weyl consistency conditions \cite{Antipin:2013sga} were violated in the analysis of \cite{Degrassi:2012ry} and properly implemented in \cite{Antipin:2013sga}.}. On the phenomenological side, several fundamental questions are left unanswered, such as the origin of dark matter, the flavour mass hierarchy, matter-antimatter asymmetry, and the interplay with quantum gravity. To partially address these limitations, extensive efforts have been devoted to unifying the electromagnetic, weak and strong forces into a single gauge group at high energy scales, known as Grand Unified Theories (GUT)s \cite{Georgi:1974sy,Georgi:1974yf, Fritzsch:1974nn, Pati:1974yy, Langacker:1980js}. In these grand unified schemes, gauge couplings merge into a single coupling that is often asymptotically free\footnote{Instances of asymptotic unification driven by ultraviolet fixed points have also been considered, see~\cite{Bajc:2016efj,Cacciapaglia:2020qky}.}, while typically Yukawa and scalar couplings are untamed at high energy. It is therefore relevant to study the conditions under which Yukawa and scalar couplings can be rendered asymptotically free in chiral gauge theories. This would lead to a classification of chiral gauge-Yukawa theories according to whether they exhibit ultraviolet (UV) fixed points in all couplings. Theories which achieve this are known as \textit{complete asymptotically free}, or CAF, models \cite{Cheng:1973nv, Gross:1973ju, Holdom_2015, Giudice:2014tma, Pica:2016krb, Callaway:1988ya, Hansen:2017pwe}. In a CAF theory, all couplings asymptotically vanish towards the UV at the same or at a higher rate than the gauge coupling. The UV fixed points ensure that the model validity can be extended to arbitrarily high energy scales. For gauge-fermion theories (without scalars), such an analysis was investigated in  \cite{Cacciapaglia:2025nff} where a complete classification of GUT-like theories with massless fermions is offered. Besides requiring gauge asymptotic freedom, extra constraints were introduced as a toolkit to select the UV theories, consisting of an effective counting of the degrees of freedom based on either free energy \cite{Appelquist:1999vs,Appelquist:2000qg} or a-theorem \cite{Cardy:1988cwa,Dondi:2017civ} counting functions. This framework allows us to explore a broad landscape of GUTs and identify promising unification theory candidates.

In this work we focus on two time-honoured classes of chiral theories: the generalised Georgi--Glashow (GG) \cite{Georgi:1974sy,Georgi:1974yf} and the Bars--Yankielowicz (BY) \cite{Bars:1981se} models, whose low energy dynamics in absence of scalar degrees of freedom has been investigated in  \cite{Appelquist:2000qg,Bolognesi:2020mpe,Csaki:2021xhi, Bolognesi:2021yni,   Bolognesi:2021hmg, Bai:2021tgl,Bolognesi:2023xxv,Li:2025tvu,Li:2026ayh}. Here, we are interested in the high-energy behaviour of these models once we include scalar degrees of freedom, thus extending the work in \cite{Molgaard:2016bqf}. We consider either a scalar in the fundamental or in the adjoint representation. Furthermore, we allow for an arbitrary multiplicity of chiral families and add vector-like pairs in the fundamental representation. Both scalar representations help achieve more realistic extensions of the SM, with the fundamental containing the SM Higgs and the adjoint permitting, at low energy, to recover the SM gauge groups \cite{Georgi:1974yf, Li:1973mq}.
We focus primarily on the UV behaviour of these models in the search for the CAF parameter space.  We systematically investigate the required number of colours, chiral families and vector-like fermions to realise a consistent and valid CAF theory. 

This paper is organised as follows. In Section \ref{sec:CAF}, we review the derivations of the conditions arising from the requirement of CAF theories for generic gauge-Yukawa theories. In Section \ref{sec:noscalar}, we consider the GG and BY model without scalars, then in Section \ref{sec:fundscalar} with one scalar in the anti-fundamental representation and lastly in Section \ref{sec:adj} with one scalar in the adjoint representation. In each of these sections, we first describe the models, then perform a CAF analysis using the framework introduced in Section \ref{sec:CAF}. A summary table stating CAF viability of all considered models can be found in Section \ref{sec:conc}, where we offer our conclusions.

%% file: Sections/CAF.tex
\section{Framework of Complete Asymptotic Freedom}\label{sec:CAF}

To constrain the coefficients of the renormalisation group (RG) equations of a gauge-Yukawa theory with scalars, we will start by considering a pure gauge theory and work our way to the full model, following a similar approach to \cite{Giudice:2014tma, Pica:2016krb}.  As we are looking for vanishing values of the couplings, it is justified to retain only one-loop beta functions.

\subsection{Gauge Coupling}\label{sec:CAFgauge}
Consider first the gauge coupling beta function $\beta_g$ to one-loop order  
\begin{equation}
    \beta_g:=\mu\frac{\text{d}\alpha_g}{\text{d}\mu}\implies \beta _g= b_0\alpha_g^2\;,\label{eq:betag1loopcaf}
\end{equation}
where $\mu$ is the RG scale, $\alpha_g=\frac{g^2}{(4\pi)^2}$. The coefficient $b_0$ is determined by group factors and the Dynkin index of the fermion and scalar representations. We will consider $b_0$ a constant in the CAF analysis below. For the pure gauge theory to be asymptotically free \cite{Politzer:1974fr}, the beta function must be negative, i.e.\ with our definition\footnote{Note that the standard convention often defines $\beta_g=-b_0\alpha_g^2$} in Eq.\ \eqref{eq:betag1loopcaf}, $b_0<0$. 
The solution of the differential equation is
\begin{equation}
    \alpha_g=\frac{\alpha_{g_0}}{1-b_0 \alpha_{g_0} \log \left(\frac{\mu}{\mu_0 }\right)}\;,
    \label{eq:solbetag1loop}
\end{equation}
where the initial condition is $\alpha_g(\mu_0)=\alpha_{g_0}$, with $\mu_0$ being the reference RG scale. 

In the case $\mu<\mu_L$, where $\mu_L=\mu_0\exp(1/{b_0\alpha_{g_0}})$, the denominator is negative since $b_0<0$, implying a negative coupling, resulting in a non-physical branch. In the case $\mu=\mu_L$, we have a Landau pole in the infrared (IR). Finally, when $\mu>\mu_L$, the coupling is positive and will vanish as $\mu\to\infty$, signifying asymptotic freedom.

\subsection{Yukawa Coupling }\label{sec:CAFYukawa}
An analogous analysis can be carried out using the RG equations governing the Yukawa couplings, which depend on the gauge coupling at one loop. For simplicity, we consider a single Yukawa coupling $y$, with the following beta function 
\begin{equation}
    \beta_y:=\mu \frac{\text{d}\alpha_y}{\text{d}\mu}=\alpha_y(c_1\alpha_g+c_2\alpha_y)\;,\label{eq:betay1loop}
\end{equation}
where $\alpha_y=\frac{y^2}{(4\pi)^2}$, and $c_1$, $c_2$ are constants.  Generally, one finds $c_1<0$ and $c_2>0$. First, let us consider the case of vanishing gauge coupling, i.e. $\alpha_g=0$: the solution will be similar to that of the pure gauge theory Eq.\ \eqref{eq:solbetag1loop}, yielding
\begin{equation}
    \alpha_y=\frac{\alpha_{y_0}}{1-c_2 \alpha_{y_0} \log \left(\frac{\mu}{\mu_0 }\right)}\;,
\end{equation}
where the difference from the previous solution Eq.\ \eqref{eq:solbetag1loop} is that the constant $c_2$ is positive. Therefore, in the high energy limit $\mu\to\infty$ the denominator becomes negative, resulting in a non-physical branch. For $\mu<\mu_L$ with $\mu_L=\mu_0\exp(1/{c_2\alpha_{y_0}})$ the coupling is positive. Thus, the coupling flows to the Gaussian fixed point in the IR and flows towards a Landau pole in the UV. This establishes that the Yukawa coupling can never be asymptotically free in the absence of a gauge coupling. When the gauge coupling runs to zero in the UV, therefore, the Yukawa coupling must also run to zero as fast as $\alpha_g$, or faster, in order for the contribution of the gauge coupling to the beta function Eq.~\eqref{eq:betay1loop} to remain sizeable.

Let us turn to the case of $\alpha_g\neq0$. We now have a system of two coupled differential equations. A solution can be found utilising the ratio method \cite{Pendleton:1980as} on the Yukawa Eq.\ \eqref{eq:betay1loop} and gauge Eq.\ \eqref{eq:betag1loopcaf} beta functions 
\begin{align}
    \frac{\beta_y}{\beta_g}=\frac{\text{d}\alpha_y}{\text{d}\alpha_g}&= \frac{\alpha_y(c_1 \alpha_g + c_2 \alpha_y)}{b_0 \alpha_g^2} = \frac{\alpha_y}{b_0 \alpha_g} \left( c_1 + c_2 \frac{\alpha_y}{\alpha_g} \right)\;.\label{eq:betayfrac}
\end{align}
Analogously to the case with a single coupling, the RG equation solution will allow us to derive restrictions on the parameters $b_0$, $c_1$ and $c_2$. We refer the reader to Appendix \ref{app:caf} for the full derivation. 

The CAF conditions in a theory consisting of gauge and Yukawa couplings depend on whether the two couplings follow a fixed-flow trajectory \cite{Giudice:2014tma}, or not. Along the fixed flow, the Yukawa coupling vanishes with the same rate as the gauge coupling, i.e. the ratio $\alpha_y/\alpha_g$ flows to a constant. Conversely, in the off-fixed-flow regime, the Yukawa coupling is driven to zero at a rate faster than the gauge coupling. 

In the case of off-fixed-flow behaviour, the CAF conditions are met only when $c_1/b_0 > 1$, C.f. Eq.\ \eqref{eq:appYukawacondoff}. Hence, we find the following solution at high energies
\begin{equation}
    \alpha_y\sim\frac{\alpha_{y_0}}{ \alpha_{g_0}^{\frac{c_1}{b_0}}\left(1-\frac{c_2}{b_0-c_1}\frac{\alpha_{y_0}}{\alpha_{g_0}}\right)}\alpha_g^{\frac{c_1}{b_0}}\;,\quad \mbox{with} \quad c_1<b_0<0\;,\quad\quad \frac{\alpha_{y_0}}{\alpha_{g_0}}<\frac{b_0-c_1}{c_2}\;.\label{eq:ygnotfixed}
\end{equation}
where $\alpha_g$ is in Eq.~\eqref{eq:solbetag1loop}.
Note that whether the Yukawa coupling is asymptotically free also depends on the values of the coupling at the reference RG scale: the last inequality above remains true at any energy. The fixed-flow CAF occurs when the above inequality between the two couplings is saturated, C.f.  Eq.\ \eqref{eq:ygfixed2}. In this case, the Yukawa coupling is locked to run following the gauge one, as follows 
\begin{equation}
    \alpha_y=\frac{b_0-c_1}{ c_2 }\alpha_g\;,\quad \mbox{with} \quad  c_1<b_0<0\;.\label{eq:ygfixed}
\end{equation}
In the fixed-flow regime, the values of the gauge and Yukawa couplings are more tightly constrained than in the off-fixed-flow regime.

\subsection{Quartic Scalar Coupling}\label{sec:CAGquartic}
A similar analysis can now be made for the quartic scalar coupling, with the general beta function 
\begin{equation}
    \beta_\lambda:=\mu \frac{\text{d} \alpha_\lambda}{\text{d}\mu}=\alpha_\lambda\left(d_1\alpha_\lambda + d_2\alpha_g+d_3\alpha_y\right)+d_4\alpha_g^2+d_5\alpha_y^2\;,\label{eq:betalam1loop}
\end{equation}
where $d_1,d_3,d_4\ge0$ and $d_2,d_5\le 0$. The quartic coupling, unlike the gauge beta function in Eq.\ \eqref{eq:betag1loopcaf} and the Yukawa beta function in Eq.\ \eqref{eq:betay1loop}, has terms that are not proportional to the coupling itself. Consequently, setting the quartic coupling to zero does not yield a trivial solution. 

Like the Yukawa and gauge couplings, we can first consider whether the quartic coupling can be asymptotically free on its own. In this case, the solution will be similar to that of the two other couplings
\begin{equation}
    \alpha_\lambda=\frac{\alpha_{\lambda_0}}{1-d_1 \alpha_{\lambda_0} \log \left(\frac{\mu}{\mu_0 }\right)}\;.
\end{equation}
The logarithmic term in the denominator is positive $d_1>0$, thus the analysis is similar to that of the Yukawa coupling in isolation. At the energy scale $\mu_L=\mu_0\exp(1/{d_1\alpha_{\lambda_0}})$ the coupling will approach a UV Landau pole, while at energies $\mu<\mu_L$ the coupling is positive with a trivial Gaussian fixed point in the IR. Finally, above the Landau pole, $\mu>\mu_L$, the coupling is negative and runs to zero in the UV: we consider this an unphysical branch since the potential would be unbounded from below.  Consequently, the quartic coupling cannot exhibit asymptotic freedom without the gauge and Yukawa couplings. 

Let us now consider a theory with both gauge, Yukawa, and a quartic scalar coupling. The beta equation of the quartic coupling Eq.\ \eqref{eq:betalam1loop} can be solved using a similar approach to Eq.\ \eqref{eq:betayfrac} by taking the ratio $\frac{\beta_\lambda}{\beta_g}$ and we find
\begin{equation}
    \frac{\text{d} \alpha_\lambda}{\text{d}\alpha_g}=\frac{\alpha_\lambda}{b_0\alpha_g}\left(d_1\frac{\alpha_\lambda}{\alpha_g} + d_2+d_3\frac{\alpha_y}{\alpha_g}\right)+\frac{d_4}{b_0}+\frac{d_5 \alpha_y^2}{b_0\alpha_g^2}\;.
\end{equation}
To find the CAF conditions in a theory with a gauge, Yukawa, and quartic coupling, it is important to consider whether the Yukawa and gauge coupling are on or off fixed flow. This gives two different differential equations to solve and thus two sets of CAF conditions, see Appendix \ref{app:caf} for more details.

The CAF conditions when the Yukawa and gauge couplings are off fixed flow (derived in Eq.\ \eqref{eq:appCAF1con}) are found to be
\begin{equation}
    b_0<0,\quad\quad b_0-c_1>0,\quad\quad k\geq 0,\quad\quad b_0-d_2+\sqrt{k}>0\;,\label{eq:CAF}
\end{equation}
where $k=(b_0-d_2)^2-4d_1d_4$. Additionally, the constraints on the gauge, Yukawa, and quartic coupling at the reference energy scale (derived in Eqs. \eqref{eq:appke0con} and \eqref{eq:appkg0con}) are given by:
\begin{align}
    &\text{For }k=0: \quad \alpha_{\lambda} = \frac{ \alpha_{\lambda0}  }{ \alpha_{g_0}  } \alpha_g\;,\quad\quad k_0=0\;,\label{eq:CAFalpha}\\&
    \text{For }k>0: \quad 
    \frac{b_0-d_2-\sqrt{k}}{2d_1}\le \frac{\alpha_{\lambda_0}}{\alpha_{g_0}}\le \frac{b_0-d_2+\sqrt{k}}{2d_1}\;, \label{eq:CAFalphafixed}
\end{align}
where $k_0 = (b_0 - d_2)\alpha_{g_0} - 2d_1\alpha_{\lambda0}$. 

Finally, the CAF conditions on fixed flow (see Eq.\ \eqref{eq:appCAF1conon}) are given by
\begin{align}
    &b_0<0,\quad\quad b_0-c_1>0,\quad\quad k'\geq 0,\quad \quad  b_0-d_2'+\sqrt{k'}>0\;,\label{eq:CAFfixed}
\end{align}  
where 
\begin{align}
    d_2'&=d_2+d_3\frac{b_0-c_1}{c_2}\;,\quad
    d_4'=d_4+d_5\left(\frac{b_0-c_1}{c_2}\right)^2\;,\quad
    k'= \left(b_0-d_2'\right)^2-4d_1d_4'\;.
\end{align}
It is important to note that, for the Yukawa on fixed flow, the signs of the primed coefficients are not determined by those of the coefficients of the beta function. This point was missing from the analysis in \cite{Pica:2016krb}. As noted above, the $d$ coefficients satisfy $d_2\le0$ and $d_4\ge0$. However, on fixed flow, this will change since we obtain
\begin{align}
     &d_2'>0 \quad \text{  if}\quad d_3\frac{b_0-c_1}{c_2}>|d_2|\;,
     \\&d_4'<0 \quad \text{  if}\quad |d_5|\frac{(b_0-c_1)^2}{c_2^2}>d_4\;.
\end{align}
A sign change on the primed coefficients can, in some cases, ensure that $k'$ and $b_0-d_2'+\sqrt k$ are positive for all parameter values. Consequently, only the first two conditions may restrict the possibility of the theory exhibiting CAF. We will see the effect of this in later sections.

The analysis has thus far been restricted to simple scenarios involving either a single gauge, Yukawa, or quartic scalar coupling or combinations thereof. We now generalise this approach to the case of multiple couplings, where, as it will turn out, the CAF conditions cannot always be found analytically from the coupled differential equations. 

\subsection{Interplay of Multiple Couplings}\label{sec:CAF2}
To address the complexity due to the presence of multiple couplings of the same type,  we follow the framework of \cite{Giudice:2014tma} to derive the CAF conditions. First, we introduce RG time $t$ so that the beta functions can be rewritten as
\begin{equation}
    \beta_n:=\frac{\text{d}\alpha_n}{\text{d}t}\;,
\end{equation}
where $t=\ln(\mu^2/\mu_0^2)$ and $n=({g_i},{y_a},{\lambda_m})$. Then, we redefine the couplings so that the coupled differential equations can be rewritten as coupled algebraic equations \cite{Zimmermann:1984sx}. This is done by introducing rescaled couplings
\begin{equation}
    \alpha_{g_i} = \frac{\tilde \alpha_{g_i}}{t}, \quad \quad  \alpha_{y_a} = \frac{\tilde \alpha_{y_a}}{t}, \quad \quad \alpha_{\lambda_m} = \frac{\tilde \alpha_{\lambda_m}}{t} \;.
\end{equation}
This lets us define the rescaled coupling vector 
\begin{equation}
    \boldsymbol{x}^T=(\tilde\alpha_{g_1},\dots,\tilde\alpha_{g_{N_g}},\tilde\alpha_{y_1},\dots\,\tilde\alpha_{y_{N_y}},\tilde\alpha_{\lambda_1},\dots,\tilde\alpha_{\lambda_{N_\lambda}})=:(\tilde\alpha_{\boldsymbol{g}},\tilde\alpha_{\boldsymbol{y}},\tilde\alpha_{\boldsymbol{\lambda}})\;,
\end{equation}
which obeys the differential equation
\begin{equation}
    \frac{\text{d}\boldsymbol{x}}{\text{d}\ln t}=\boldsymbol{V}(\boldsymbol{x})\;,\quad
    \boldsymbol{V}(\boldsymbol{x})=\begin{pmatrix}
        \tilde \alpha_{\boldsymbol{g}}+\beta_{\boldsymbol{g}}(\tilde\alpha_{g})\\\tilde \alpha_{\boldsymbol{y}}+\beta_{\boldsymbol{y}}(\tilde\alpha_{g},\tilde\alpha_{y})\\\tilde \alpha_{\boldsymbol{\lambda}}+\beta_{\boldsymbol{\lambda}}(\tilde\alpha_{g},\tilde\alpha_{y},\tilde\alpha_{\lambda})
    \end{pmatrix}\;,\label{eq:CAF2V}
\end{equation}
The fixed-flow terminology from the first CAF analysis can be extended to this case to help in solving the coupled differential equations. That is, if they all go as $1/t$, then the differential in Eq.\ \eqref{eq:CAF2V} will be zero. Then the fixed flows are determined by the solutions to $\boldsymbol{V}(\boldsymbol{ x}_\infty)=0$ yielding 
\begin{equation}
     \boldsymbol{x}_\infty=\begin{pmatrix}
        -\beta_{{\boldsymbol{g}}}(\tilde\alpha_{g\infty})\\-\beta_{{\boldsymbol{y}}}(\tilde\alpha_{g\infty},\tilde\alpha_{y\infty})\\-\beta_{\boldsymbol{\lambda}}(\tilde\alpha_{g\infty},\tilde\alpha_{y\infty},\tilde\alpha_{\lambda\infty})
    \end{pmatrix}\label{eq:solCAF2}\;.
\end{equation}
As in previous CAF analyses, the equations couple in such a way that we can first solve the first equations for the gauge couplings, then the ones for the Yukawas, knowing the solutions for $\alpha_{gi}$, and finally for the quartic couplings.
 
In the case of multiple gauge couplings, we can start by finding the possible solutions to Eq.\ \eqref{eq:solCAF2}. Using the gauge beta function in Eq.\ \eqref{eq:betag1loopcaf}, the possible solutions are
\begin{equation}
   (\tilde\alpha_{g\infty})_i= \left\{0,-\frac 1{(b_0)_i}\right\}
        \;,
\end{equation}
It should be noted that the non-zero solution $-\frac{1}{(b_0)_i}$ exists only for $(b_0)_i<0$. Thus, it is only under this condition that the theory admits a non-trivial solution. Additionally, and unsurprisingly, the CAF condition coincides with the first CAF analysis (see Section \ref{sec:CAFgauge}).

Let us continue with the case of one gauge coupling and multiple Yukawa couplings. For this, we use the Yukawa beta function from Eq.\ \eqref{eq:betay1loop}. The fixed-flow trajectories of the Yukawa couplings are given by
\begin{equation}
   \tilde\alpha_{\boldsymbol{y}\infty}= - \beta_{\boldsymbol y}(-b_0^{-1}, \tilde\alpha_{y\infty})\;. \label{eq:CAF2y}
\end{equation}
We can find a solution to Eq.\  \eqref{eq:CAF2y} for the case of one Yukawa. Only for $c_1-b_0<0$ will there be non-trivial solutions. As expected, the CAF condition matches that obtained in the first CAF analysis (see Eq.\ \eqref{eq:ygfixed}):
\begin{align} 
    (\tilde\alpha_{y\infty})_a=\begin{cases}
       0\\\frac {(c_1)_{a}-b_0}{b_0(c_2)_a}
    \end{cases}.
\end{align}
Finally, the case of one gauge, one Yukawa and multiple quartic scalar couplings can be considered. The beta equation of the quartic couplings can be written as
\begin{equation}
    \beta_{{\lambda_m}} = d^{\lambda}_{mnp} \alpha_{\lambda_n} \alpha_{\lambda_p} + \alpha_{\lambda_m} \left( d^{\lambda_y}_{ma} \alpha_{y_a} + d^{\lambda_g}_{mi} \alpha_{g_i} \right) + d^{y}_{mab} \alpha_{y_a} \alpha_{y_b}+ d^{g}_{mij} \alpha_{g_i} \alpha_{g_j}\;,\label{eq:multquartic}
\end{equation}
where $m$ indicates the $m$th quartic coupling. This can be rewritten by setting Eq.\ \eqref{eq:CAF2V} equal to zero ensuring that all the couplings are on fixed flow, which yields
\begin{equation}
    \tilde\alpha_{\lambda _{m\infty}} + d^{\lambda}_{mnp}\tilde\alpha_{\lambda _{m\infty}}\tilde\alpha_{\lambda _{p\infty}}+ \tilde\alpha_{\lambda _{m\infty}}(d^{\lambda y}_m \tilde\alpha_{y _{\infty}} +d^{\lambda_g}_m\tilde\alpha_{y _{\infty}}) + d^{y}_m\tilde\alpha_{y\infty}^2+ d^{g}_m \tilde\alpha_{g\infty}^2 = 0    \;.\label{eq:CAF2quartic}
\end{equation}
Next, we will use this approach on a specific case of one gauge, one Yukawa and two quartic couplings.

\subsubsection{Two quartic scalar couplings}
We consider the case of $m=1,2$ corresponding to two quartic couplings. In this case, the algebraic Eq. \eqref{eq:CAF2quartic} can be written as two coupled quadratic equations
\begin{equation}
    a  {x}^2 + b  {z}^2 + c  {x}{z} + d  {x} + e = 0\;,
\end{equation}
\begin{equation}
    f  {z}^2 + g  {x}^2 + h  {x}{z} + i  {z} + j = 0\;,\label{eq:CAF2poly2}
\end{equation}
where $x=\tilde\alpha_{\lambda_{1\infty}}$ and $z=\tilde\alpha_{\lambda_{2\infty}}$. The coefficients are given by
\begin{align}
    a=d^{\lambda}_{111},\quad b=d^{\lambda}_{122},\quad c=d^{\lambda}_{112},\quad d=1+d^{\lambda y}_1 \tilde\alpha_{y _{\infty}} +d^{\lambda_g}_1\tilde\alpha_{g _{\infty}},\quad e=d^{y}_1\tilde\alpha_{y\infty}^2+ d^{g}_1 \tilde\alpha_{g\infty}^2\;,\label{eq:coef2quartic}
\end{align}
with $f,\ g,\ h,\ i$ and $j$ given by analogous equations with $1\Leftrightarrow 2$.
The two coupled polynomial equations can be rewritten as a single fourth-degree polynomial:
\begin{equation}
    \alpha x^4+ \beta x^3 +\gamma x^2 +\delta x+\epsilon=0\;,\label{eq:fourthdegpoly}
\end{equation}
where $x=\tilde\alpha_{\lambda \infty}$. The definitions of the coefficients can be found in Appendix \ref{sec:appCafadj}. 

To extract the CAF region we need real solutions to the polynomial equation. Therefore we consider the nature of the roots by using Rees' approach in \cite{Rees1922}. Furthermore, we can exploit Descartes' rule of signs \cite{Wang2004} to identify positive real-valued quartic scalar couplings that satisfy CAF conditions. This CAF analysis will prove useful in the case of a chiral model with one adjoint scalar (see Section \ref{sec:adj}). 

We have now considered all the couplings on fixed flow. Similar to the CAF analyses in Sections \ref{sec:CAFgauge}, \ref{sec:CAFYukawa} and \ref{sec:CAGquartic}, we can also consider the Yukawa couplings to be off fixed flow. 

\paragraph{Yukawas off fixed flow.}
We consider the Yukawa coupling off fixed flow. This implies that the Yukawa coupling runs to zero faster than the others, hence $\tilde\alpha_{y\infty} = 0$. The equations determining the CAF conditions on the quartic couplings remain unchanged, with the only difference that the coefficients $d,\ e,\ i,\ j$ lose their dependence on $\tilde\alpha_{y\infty}$. 

%% file: Sections/Chiral_theories.tex
\section{Introduction to the Chiral Models} \label{sec:noscalar}

We will now turn our attention to specific chiral gauge models, namely a generalised Georgi-Glashow (GG) \cite{Georgi:1974sy,Georgi:1974yf} and Bars-Yankielowicz (BY) \cite{Bars:1981se}. They are defined in terms of a $\mathrm{SU}(N)$ gauge theory with an anomaly-free combination of fermions in the anti-fundamental and anti-symmetric/symmetric representations, respectively. We will also allow for the presence of an arbitrary number of vector-like families consisting of pairs in the fundamental and anti-fundamental.

\subsection{Generalised Georgi--Glashow model}
We start by considering a generalised GG model \cite{Georgi:1974sy,Georgi:1974yf}, with fermion content summarised in Table \ref{tab:GG}. It therefore includes three representations: the anti-fundamental fermions $\tilde\psi$, the fundamental fermions $\psi$, and the two-index antisymmetric fermion $\Psi$, under the gauge symmetry $\mathrm{SU}(N)$, with $N\ge5$. The anti-fundamental fermions also transform under a global symmetry group $\mathrm{SU}(N-4+p)$ given by its multiplicity, while the fundamental fermions under $\mathrm{SU}(p)$. All fields are also charged under two Abelian global symmetries, $\mathrm{U}_1(1)$ and $\mathrm{U}_2(1)$, which have no mixed anomaly with the gauge symmetry. In terms of these fields, the Lagrangian is given by
\begin{equation}
    \mathcal{L}= -\frac{1}{4} F_{\mu\nu}^a F^{a\mu\nu}+ i \bar{\Psi}^{ij} \slashed{D} \Psi_{ij} 
    + i \bar{\psi}^i_\alpha \slashed{D} \psi_i^\alpha 
    + i \bar{\tilde{\psi}}_i^\beta \slashed{D} \tilde{\psi}^i_\beta\;,\label{eq:lGGnoscalar}
\end{equation}
where we use the following index conventions: The Lorentz indices are denoted by $\mu,\nu$, the adjoint gauge index by $a=1,\dots,N^2-1$, the fundamental gauge indices from $\mathrm{SU}(N)$ by $i,j=1,\dots,N$, and the flavour indices are given by $\alpha=1,\dots, p$ and $\beta=1,\dots,N-4+p$. 

This theory has one gauge coupling, with beta function \cite{Gross:1973ju}
\begin{equation}
    \beta_g=-{2\alpha_g^2}\left[\frac{11}{3}C_2(G)-\frac{2}{3}\sum_f T(R_f)-\frac{1}{3}\sum_s T(R_s)\right]\;, \label{eq:betagaugeformula}
\end{equation}
where the sums are taken over fermions $f$ and scalars $s$.
Since there are no scalars in the theory, the one-loop coefficient is given by:
\begin{equation}
b_0 = -2 \left[ \frac{11}{3} C_2(G) - \frac{2}{3} \sum_f T(R_f)\right] \;.
\end{equation}
Recall that the model contains the fermion $\Psi$, which is in the anti-symmetric representation of $\mathrm{SU}(N)$ ($1$ copy) with dimension $\frac{N(N-1)}{2}$; the fermion $\tilde{\psi}$ which is in the anti-fundamental representation ($N-4+p$ copies) with dimension $\bar{N}$; lastly, $\psi$ which is in the fundamental representation ($p$ copies) with dimensions $N$.
Using the standard group theory factors for $\mathrm{SU}(N)$: $C_2(G) = N$, $T(\text{Fund}) = T(\text{Anti-Fund}) = 1/2$, and $T(\text{Anti-Symm}) = \frac{N-2}{2}$ and substituting in the field content, we obtain
\begin{align}
\sum T(R_f) &= 1 \cdot T(A) + (N-4+p) \cdot T(\bar{F}) + p \cdot T(F)  = N - 3 + p\;.
\end{align}
Therefore, 
we find
\begin{equation}
    b_0=-\frac{2}{3}\left({9}N+6-2p\right)\;. \label{eq:betagauge}
\end{equation}
The model contains only a single gauge coupling. Consequently, the theory is constrained solely by the condition on the gauge beta function coefficient, $b_0<0$. In this particular model, this translates into a constraint on the parameters $N$ and $p$, which can be rewritten as a function $p(N)$. This function describes the upper limit of vector-like fermions needed to be added in an $\mathrm{SU}(N)$ theory to ensure CAF, and is given by
\begin{equation}
    p_{\text{max}}(N)=\frac{9N+6}{2} \;.
    \label{eq:pNnoscalarGG}
\end{equation} 
\begin{table}[t!]
    \centering
    \renewcommand{\arraystretch}{1.5} 
    \begin{tabular}{|c|c|c|c|c|c|}
    \hline
    Fields & $\mathrm{SU}(N)$ & $\mathrm{SU}(N - 4 + p)$ & $\mathrm{SU}(p)$ & $\mathrm{U}_1(1)$ & $\mathrm{U}_2(1)$ \\
    \hline
    $\Psi$ & $\ytableausetup{boxsize=0.25cm, centertableaux}\begin{ytableau}{}\\{}\end{ytableau}$ & $1$ & $1$ & $N - 4$ & $2p$ \\

    $\tilde\psi$ & $\overline{\begin{ytableau}{}\end{ytableau}\rule{0pt}{7.5pt}}$ & $\overline{\begin{ytableau}{}\end{ytableau}\rule{0pt}{7.5pt}}$ & $1$ & $-(N - 2)$ & $-p$ \\

    $\psi$ & ${\begin{ytableau}{}\end{ytableau}}$ & $1$ & ${\begin{ytableau}{}\end{ytableau}}$ & $N - 2$ & $-(N - p)$ \\
    \hline
    \end{tabular}
    \caption{Field content of the GG model, where spinors are left-handed Weyl. The table shows how the fields transform under the gauge and global symmetry groups, and their charges under the anomaly-free $\mathrm{U}_1(1)$ and $\mathrm{U}_2(1)$ symmetries.}
    \label{tab:GG} 
\end{table}
The gauge coupling thus creates an upper limit for the CAF region of the model and the GG model without scalars can therefore be CAF for all $N$.


\subsection{Generalised Bars--Yankielowicz model}
We now focus on the BY model \cite{Bars:1981se}. As shown in Table \ref{tab:BY}, the difference from the GG model is that the chiral fermion is in the two-index symmetric representation, i.e.\ $\chi_{ij}=\chi_{ji}$ with dimensions $\frac{N(N+1)}{2}$, defined for $N\ge 3$, and the multiplicity of the anti-fundamental is different. Thus, the Lagrangian is similar to that of the GG model and can be expressed as
\begin{align}\mathcal{L}= -\frac{1}{4} F_{\mu\nu}^a F^{a\mu\nu}+ i \bar{\chi}^{ij} \slashed{D} \chi_{ij} 
    + i \bar{\psi}^i_\alpha \slashed{D} \psi_i^\alpha 
    + i \bar{\tilde{\psi}}_i^\beta \slashed{D} \tilde{\psi}^i_\beta
    \;,
\end{align}
with similar index definitions as in Eq.\ \eqref{eq:lGGnoscalar} except for the flavour index of the anti-fundamental fermion now given by $\beta=1,\dots, N+4+p$.
\begin{table}[b!]
    \centering
    \renewcommand{\arraystretch}{1.5} 
    \begin{tabular}{|c|c|c|c|c|c|}
    \hline
    Fields & $\mathrm{SU}(N)$ & $\mathrm{SU}(N + 4 + p)$ & $\mathrm{SU}(p)$ & $\mathrm{U}_1(1)$ & $\mathrm{U}_2(1)$ \\
    \hline
    $\chi$ & $\ytableausetup{boxsize=0.25cm}\begin{ytableau}{}&{}\end{ytableau}$ & $1$ & $1$ & $N + 4$ & $2p$ \\

    $\tilde\psi$ & $\overline{\begin{ytableau}{}\end{ytableau}\rule{0pt}{7.5pt}}$ & $\overline{\begin{ytableau}{}\end{ytableau}\rule{0pt}{7.5pt}}$ & $1$ & $-(N + 2)$ & $-p$ \\

    $\psi$ & ${\begin{ytableau}{}\end{ytableau}}$ & $1$ & ${\begin{ytableau}{}\end{ytableau}}$ & $N +2$ & $-(N - p)$ \\
    \hline
    \end{tabular}
    \caption{Field content of the BY model. The table shows how the fields transform under the gauge and global symmetry groups, and their charges under the anomaly-free $\mathrm{U}_1(1)$ and $\mathrm{U}_2(1)$ symmetries.}
    \label{tab:BY} 
\end{table}
The gauge beta function can be found similarly to the GG model using Eq.\ \eqref{eq:betagaugeformula}. The computation only differs in the number of anti-fundamental fermions: $N+4+p$, and the Dynkin index of the chiral fermion $T(\text{Symm})=\frac{N+2}{2}$, which yields
\begin{align}
\sum T(R_f) &= \frac{N+2}{2} + \frac{N+4+p}{2} + \frac{p}{2} = N + 3 + p\;.
\end{align}
Thus, we obtain 
\begin{equation}
    b_0=-\frac{2}{3}\left({9}N-6-2p\right) \;. \label{eq:betagaugeBY}
\end{equation}
This expression mirrors the $b_0$ of the GG model Eq.\ \eqref{eq:betagauge}, differing only by the sign of the second term. Therefore, the CAF region for the BY model is similar to that of the GG model. The conditions on the gauge beta function create a comparable upper boundary, which is given by
\begin{equation}
    p_{\text{max}}(N)=\frac{9N-6}{2} \;.\label{eq:pNnoscalarBY}
\end{equation} 
Hence, a CAF region exists for any $N$. 

%% file: Sections/fundamental_scalar.tex
\section{Adding a Single Fundamental Scalar} \label{sec:fundscalar}
We first extend the GG and BY models with one scalar field in the fundamental representation. Thus, by introducing the scalar field $\phi$ in the Lagrangian, we can write a Yukawa coupling with the fermions in the two-index and anti-fundamental representations, plus one quartic coupling: 
\begin{multline}
    \mathcal{L}= -\frac{1}{4} F_{\mu\nu}^a F^{a\mu\nu}+ i \bar{\Lambda}^{ij} \slashed{D} \Lambda_{ij} 
    + i \bar{\psi}^i_\alpha \slashed{D} \psi_i^\alpha 
    + i \bar{\tilde{\psi}}_i^\beta \slashed{D} \tilde{\psi}^i_\beta +(D_\mu \phi_i)^\dagger (D^\mu\phi_i)+ \\(y^\beta\Lambda_{ij} \tilde\psi ^{i}_{\beta} (\phi_j)^\dagger + h.c.)+\lambda ((\phi_i)^\dagger \phi^i)^2 
    \;,
\end{multline}
where $\Lambda_{ij}$ refers to either $\Psi_{ij}$ for the case of the antisymmetric two-index tensor of the GG model or $\chi_{ij}$ in the case of the two-index symmetric tensor of the BY model. The field content is summarised in Table~\ref{tab:fund}, where we see that both global Abelian charges are preserved by the Yukawa couplings.
\begin{table}[ht]
    \centering
    \renewcommand{\arraystretch}{1.5} 
    \begin{tabular}{|c|c|c|c|c|c|}
    \hline
    Fields & $\mathrm{SU}(N)$ & $\mathrm{SU}(N \mp 4 + p)$ & $\mathrm{SU}(p)$ & $\mathrm{U}_1(1)$ & $\mathrm{U}_2(1)$ \\
    \hline
    $\Lambda$ & $\ytableausetup{boxsize=0.25cm, centertableaux}\begin{ytableau}{}\\{}\end{ytableau} \;\, / \;\, \begin{ytableau}{}&{}\end{ytableau}$ & 1 & 1 & $N \mp 4$ & $2p$ \\

    $\tilde\psi$ & $\overline{\begin{ytableau}{}\end{ytableau}\rule{0pt}{7.5pt}}$ & $\overline{\begin{ytableau}{}\end{ytableau}\rule{0pt}{7.5pt}}$ & 1 & $-(N \mp 2)$ & $-p$ \\

    $\psi$ & ${\begin{ytableau}{}\end{ytableau}}$ & 1 & ${\begin{ytableau}{}\end{ytableau}}$ & $N \mp 2$ & $-(N - p)$ \\
    \hline
    $\phi$ & $\begin{ytableau}{}\end{ytableau}\rule{0pt}{7.5pt}$ & 1 & 1 & $\mp 2$ & $p$ \\
    \hline
    \end{tabular}
    \caption{Field content of the GG (upper signs, antisymmetric $\Lambda$) and BY (lower signs, symmetric $\Lambda$) models with a scalar $\phi$ in the fundamental representation.  The table shows how the fields transform under the gauge and global symmetry groups, and their charges under the anomaly-free $\mathrm{U}_1(1)$ and $\mathrm{U}_2(1)$ symmetries. }
    \label{tab:fund} 
\end{table}

The model has both Yukawas and a single quartic coupling, as well as a gauge coupling. It should be noted that the Yukawa term explicitly breaks the global symmetry $\mathrm{SU}(N\mp 4+p)$, where the upper (lower) sign refers to the GG (BY) model. 
The Yukawa couplings is a vector $y^\beta$ in the flavour space and, without loss of generality, it can be rotated using a $\mathrm{SU}(N\mp4+p)$ transformation so that only its first component is non-zero\footnote{The flavour symmetry group is therefore broken to $\mathrm{SU}(N\mp4+p) \to \mathrm{SU}(N\mp 4+p-1)$.}. Hence, physically there is a single coupling constant, which we define as $y$, which greatly simplifies the computation of the beta functions since $y_\beta y^\beta=y^2$.

For the gauge RG equations, the only change from the computation in the previous section is the inclusion of the complex scalar in the fundamental, which contributes with $T(S) = 1/2$. Using \eqref{eq:betagauge} we obtain 
\begin{equation}
\Delta b_0^{\text{scalar}} =  \frac{2}{3} T(S) =  \frac{1}{3}\;.\label{eq:scaleff}
\end{equation}
Thus, the beta function of the gauge coupling is given by
\begin{equation}
    b_0=-\frac{1}{3}\left(18N-4p\pm{12}-1\right)\;,\label{eq:betagwithp}
\end{equation}where the $(-1)$ inside the bracket corresponds to the scalar contribution.

The remaining beta functions are computed using the ``RGBeta'' library in Mathematica \cite{Thomsen:2021ncy} and are given by
\begin{align}
    \beta_y&= \frac{\alpha_y}{2}\left[(3 N+2\mp3)\alpha_y+6 \alpha_g\left(-3 N\pm2+\frac 5N\right) \right]\;,\label{eq:betaywithp}
   \\
   \beta_\lambda&=\alpha_\lambda\left[2(N\mp1)\alpha_y-\frac{6 \alpha_g 
   \left(N^2-1\right)}{N}+4 \alpha_\lambda (N+4)\right]\nonumber\\&\;\;\;\;\;\;-\frac{1}{2} (N+1\mp 2 )
  \alpha_y^2+\frac{3 \alpha_g^2 \left(N^3+N^2-4 N+2\right)}{4 N^2}\;. \label{eq:betalamwithp}
\end{align}
\begin{figure}[t]
    \centering
    \begin{subfigure}[b]{0.49\textwidth}
        \centering
        \includegraphics[width=\textwidth]{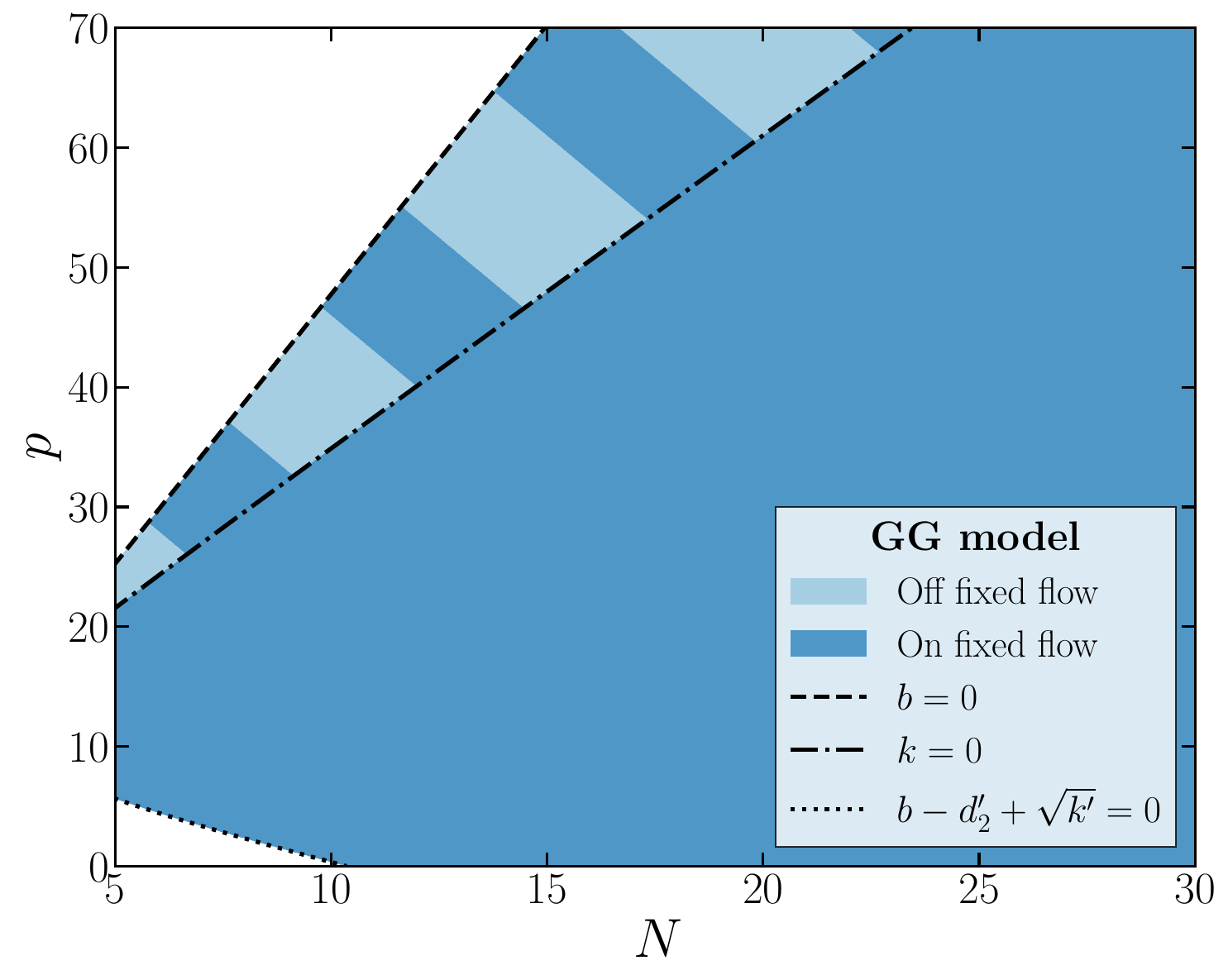}
        \caption{GG model} 
        \label{fig:cafgg_fund}
    \end{subfigure}
    \hspace{0.1cm}
    \begin{subfigure}[b]{0.49\textwidth}
        \centering
        \includegraphics[width=\textwidth]{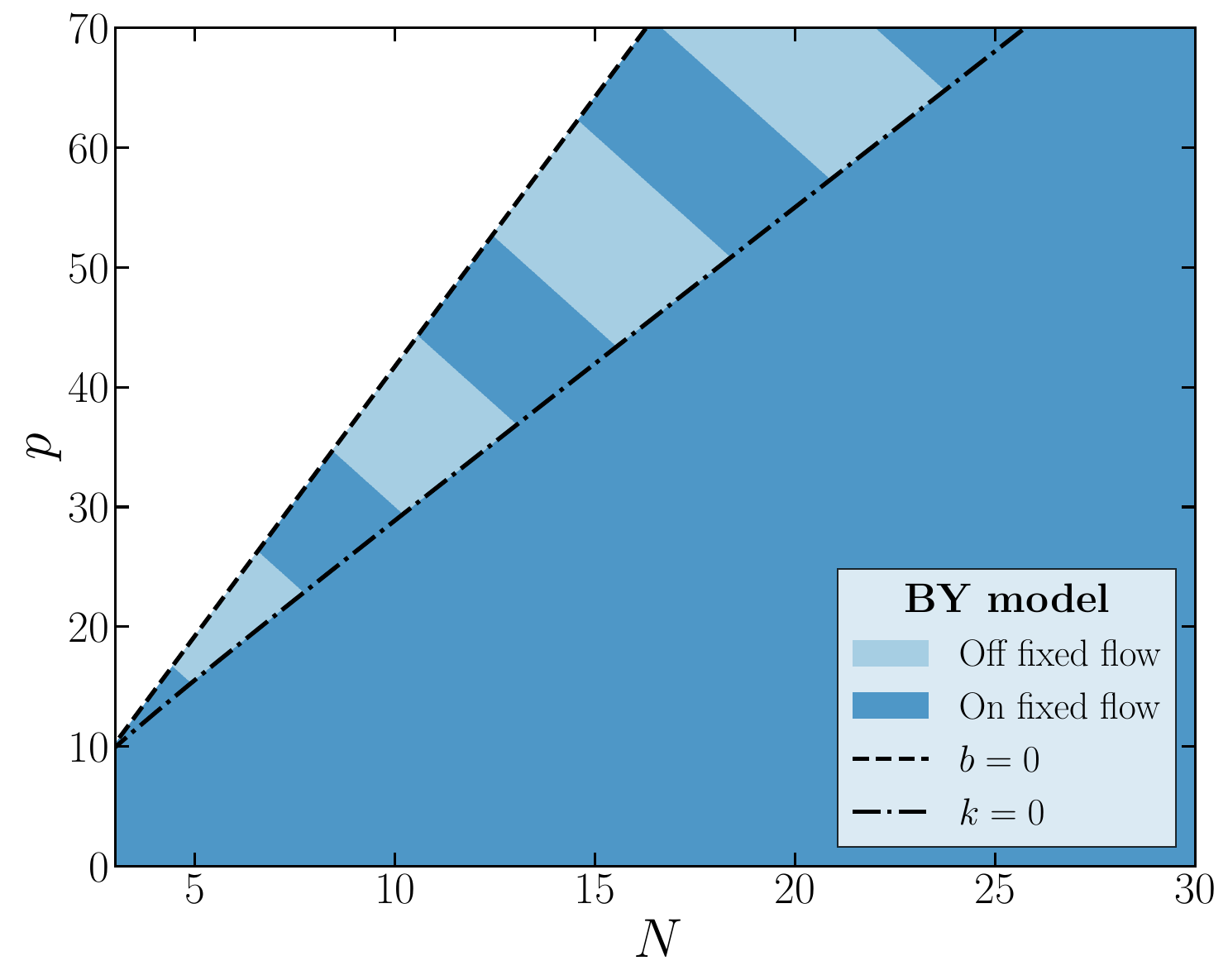}
        \caption{BY model} 
        \label{fig:cafby_fund}
    \end{subfigure}
    \caption{CAF regions in the parameter space $(N,p)$ for (a) the GG model and (b) the BY model, with one scalar in the fundamental representation. The dark blue area represents the fixed-flow CAF region, bounded above by the $b_0=0$ condition (dashed black line) and below by the $b_0-d_2'+\sqrt{k'}=0$ condition (dotted black line) for the GG model. The light blue region highlights the CAF region off fixed flow, bounded above by the $b_0=0$ condition and below by the $k=0$ condition (dash-dotted black line). The striped region indicates that the model exhibits CAF both on and off fixed flow.  The boundaries are given by the $p_{\max}$, $p^{\text{off}}_{\min}$, and $p^{\text{on}}_{\min}$ functions Eqs. \eqref{eq:pmaxfundoff}, \eqref{eq:pminfundoff}, and  \eqref{eq:pminfundon}.}
    \label{fig:cafby}
\end{figure}
The limits of the CAF regions can be found by considering the boundary of the three conditions $b_0=0$, $k=0$, and $b_0-d_2'+\sqrt{k'}=0$ (see Section~\ref{sec:CAF}). In the off-fixed-flow regime, for every $N$, a CAF region is found when $p$ is between the two values
\begin{align}
    p_{\text{max}}(N)&=\frac{18N\pm12-1}{4}\label{eq:pmaxfundoff}\;,\\
   p^{\text {off}}_{\text{min}}(N) &= -\frac{1}{4}  \pm 3  + \frac{9}{2N}  + \frac{3}{2N} \sqrt{ 3 (N+4)(N^3 + N^2 - 4N + 2) }\;.\label{eq:pminfundoff}
\end{align}
There is also a second branch of $p^{\text {off}}_{\text{min}}(N)$ with a minus in front of the square-root term; this region is unphysical as it always results in a negative $p \,\forall\, N$. 

The boundary functions on fixed flow have the same upper boundary as in the off-fixed-flow regime (see Eq.\ \eqref{eq:pmaxfundoff}). The lower boundary found from the $b_0-d_2'+\sqrt{k'}=0$ condition is given by
\begin{align}
    p^{\text{on}}_{\text{min}}(N)&= \frac{-720 + 4N(19 + 168N + 2N^2 - 9N^3)}{16N(N - 1)(N + 4)}\nonumber\\&\quad\;+\frac{ 3\sqrt{6}\sqrt{(1 - 3N)^2(N - 1)^2(N + 4)^2(N^2 + 2N - 2)}}{16N(N - 1)(N + 4)}\label{eq:pminfundon}\;,
\end{align}
where the boundary is only relevant for the GG model with $N < 10$, as can be seen from Figure \ref{fig:cafgg_fund}.

We have now obtained the necessary beta functions and can continue the analysis of the models by performing a CAF analysis using the conditions mentioned in Eq.\ \eqref{eq:CAF} and Eq.\ \eqref{eq:CAFfixed}. Deploying these conditions on the beta functions Eqs. \eqref{eq:betagwithp}, \eqref{eq:betaywithp}, and \eqref{eq:betalamwithp}. It should be noted that the CAF regions found in this analysis \textit{may} correspond to fully asymptotically free theories. Whether these solutions correspond to CAF regions also depends on initial conditions on the couplings at a given RG scale. These conditions are provided in Eqs. \eqref{eq:ygnotfixed}, \eqref{eq:ygfixed}, \eqref{eq:CAFalpha} and \eqref{eq:CAFalphafixed}. Only when these are satisfied does one realise a genuine CAF regime. 

In Figure \ref{fig:cafby}, the results of the CAF analysis are shown in the parameter space $(N,p)$. The light blue region in both figures represents the region where the coefficients of the beta functions satisfy the CAF conditions, when the Yukawa and gauge couplings are not on fixed flow. In other words, terms proportional to the Yukawa coupling have only a subleading effect in the beta function of the quartic couplings in this region.

Another feature of Figure \ref{fig:cafby} is the dark blue region, which extends up to the dashed black line. This is the region where the CAF conditions are satisfied for the Yukawa and gauge coupling on fixed flow. Thus, in this region, neither of the couplings is subleading. As mentioned in Section \ref{sec:CAF}, on fixed flow, the coefficients $d_2'$ and $d_4'$ can change sign. This is exactly what we observe in these models, effectively rendering the CAF conditions coming from the quartic coupling trivially upheld. Thus, the fixed-flow region closely resembles that of the models without a scalar. 

To better understand these conditions and these regions, flow diagrams have been made both within and outside of each region. These can be found in Appendix \ref{app:flow}. The analysis gives the same result as the CAF analysis and clearly illustrates how the values of the couplings at the RG scale are restricted. These conditions are Eq.\ \eqref{eq:CAFalpha} and Eq.\ \eqref{eq:CAFalphafixed}. 

Next, we extend the model to include multiple chiral families motivated by the SM, which has three chiral families.

%% file: Sections/fundamental_scalar_and_families.tex
\subsection{Inclusion of Multiple Chiral Families }\label{sec:fundscalarNg}
To further explore the GG  and BY models with a scalar in the fundamental representation, we now extend the analysis to include multiple chiral families. The field content of the models is shown in Table \ref{tab:fund_Ng}. The flavour symmetry $\mathrm{SU}(N\mp4+p)$ is now extended to $\mathrm{SU}(N_g(N\mp4)+p)$ ensuring anomaly cancellation. Furthermore, another group, $\mathrm{SU}(N_g)$, appears, under which the chiral fermionic field transforms as an indicator of the chiral flavour symmetry. This will also change the terms involving the (anti-)symmetric two-index fermionic field in the Lagrangian, by introducing a new flavour index $\gamma=1,\dots,N_g$, where $N_g$ is the number of chiral families, putting this together we find the following Lagrangian
\begin{multline}
    \mathcal{L} =  -\frac{1}{4} F_{\mu\nu}^a F^{a\mu\nu}+ i \bar \Lambda^{ij\gamma}\slashed{D}\Lambda_{ij\gamma}
    + i \bar{\psi}^i_\alpha \slashed{D} \psi_i^\alpha 
    + i \bar{\tilde{\psi}}_i^\beta \slashed{D} \tilde{\psi}^i_\beta +(D_\mu \phi_i)^\dagger (D^\mu\phi_i)+ \\(y^{\beta\gamma}\Lambda_{ij\gamma} \tilde\psi ^{i}_{\beta} \phi^{\dagger j} + h.c.)+\lambda (\phi^{\dagger i} \phi_i)^2 
    \;,
\end{multline}
where the Yukawa coupling is a $N_g\times (N_g(N\mp4)+p)$ matrix in flavour space.

\noindent We begin the analysis by considering the gauge beta function, which is given by
\begin{equation}
    b_0=-\frac{22N-1\pm 12 N_g-4N_gN -4p}{3}\;,
\end{equation}
where the beta function now depends on the number of chiral families $N_g$ together with $p$ and $N$. 
Hence, we can derive an upper limit on $p$ as a function of $N$ and $N_g$:
\begin{equation} \label{eq:pmaxfundNg}
    p_{\max} (N,N_g) = \frac{22N-1\pm12N_g-4N_g N}{4}\,.
\end{equation}
When $p_{\max} < 0$, no CAF region can exist as the gauge coupling is not asymptotically free, giving a well-known upper bound on the number of chiral generations:
\begin{equation}
    N_{g} \leq \frac{22N-1}{4(N\mp 3)}\,. 
\end{equation}
\begin{table}[b]
    \centering
    \renewcommand{\arraystretch}{1.5} 
    \begin{tabular}{|c|c|c|c|c|c|c|}
    \hline
    Fields & $\mathrm{SU}(N)$ & $\mathrm{SU}(N_g(N \mp 4) + p)$ & $\mathrm{SU}(p)$ & $\mathrm{SU}(N_g)$ & $\mathrm{U}_1(1)$ & $\mathrm{U}_2(1)$ \\
    \hline
    $\Lambda$ & $\ytableausetup{boxsize=0.25cm, centertableaux}\begin{ytableau}{}\\{}\end{ytableau} \;\, / \;\, \begin{ytableau}{}&{}\end{ytableau}$ & $1$ & $1$ &$\overline{\begin{ytableau}{}\end{ytableau}\rule{0pt}{7.5pt}}$ & $N \mp 4$ & $2p$ \\

    $\tilde\psi$ & $\overline{\begin{ytableau}{}\end{ytableau}\rule{0pt}{7.5pt}}$ & $\overline{\begin{ytableau}{}\end{ytableau}\rule{0pt}{7.5pt}}$ & 1&1 & $-(N \mp 2)$ & $-p$ \\

    $\psi$ & ${\begin{ytableau}{}\end{ytableau}}$ & 1 & ${\begin{ytableau}{}\end{ytableau}}$ & 1&$N \mp 2$ & $-(NN_g - p)$ \\
    \hline
    $\phi$ & $\begin{ytableau}{}\end{ytableau}\rule{0pt}{7.5pt}$ & 1 & 1 &1& $\mp 2$ & $p$ \\
    \hline
    \end{tabular}
    \caption{Field content of the GG (upper signs, antisymmetric $\Lambda$) and BY (lower signs, symmetric $\Lambda$) models with $N_g$ chiral families and one fundamental scalar $\phi$.  The table shows how the fields transform under the gauge and global symmetry groups, and their charges under the $\mathrm{U}_1(1)$ and $\mathrm{U}_2(1)$ symmetries. }
    \label{tab:fund_Ng} 
\end{table}
\begin{figure}[h!]
    \centering
    \begin{subfigure}[b]{0.49\textwidth}
        \centering
        \includegraphics[width=\textwidth]{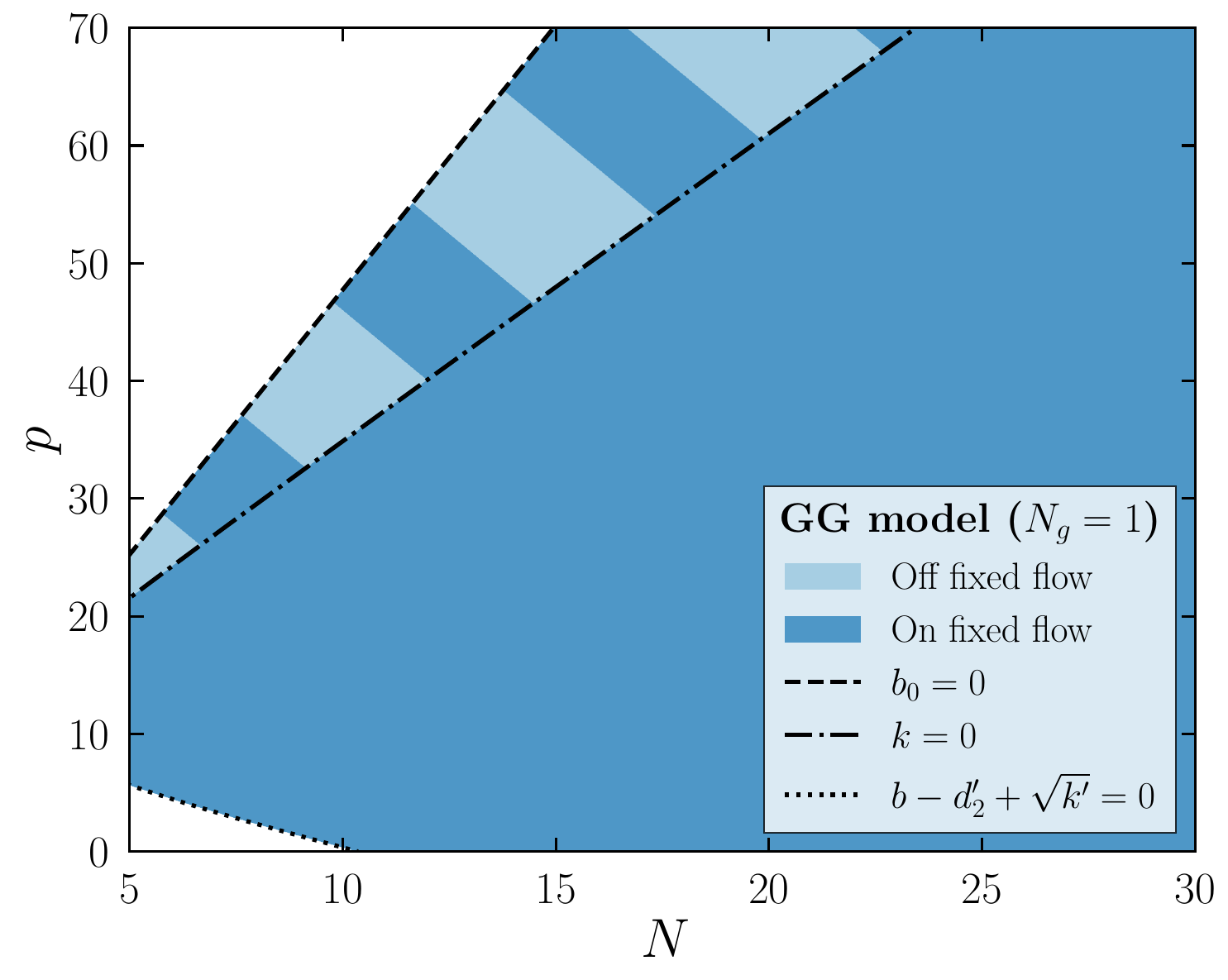}
        \caption{$N_g=1$} 
        \label{fig:gg_ng1}
    \end{subfigure}%
    \hspace{0.1cm}
    \begin{subfigure}[b]{0.49\textwidth}
        \centering
        \includegraphics[width=\textwidth]{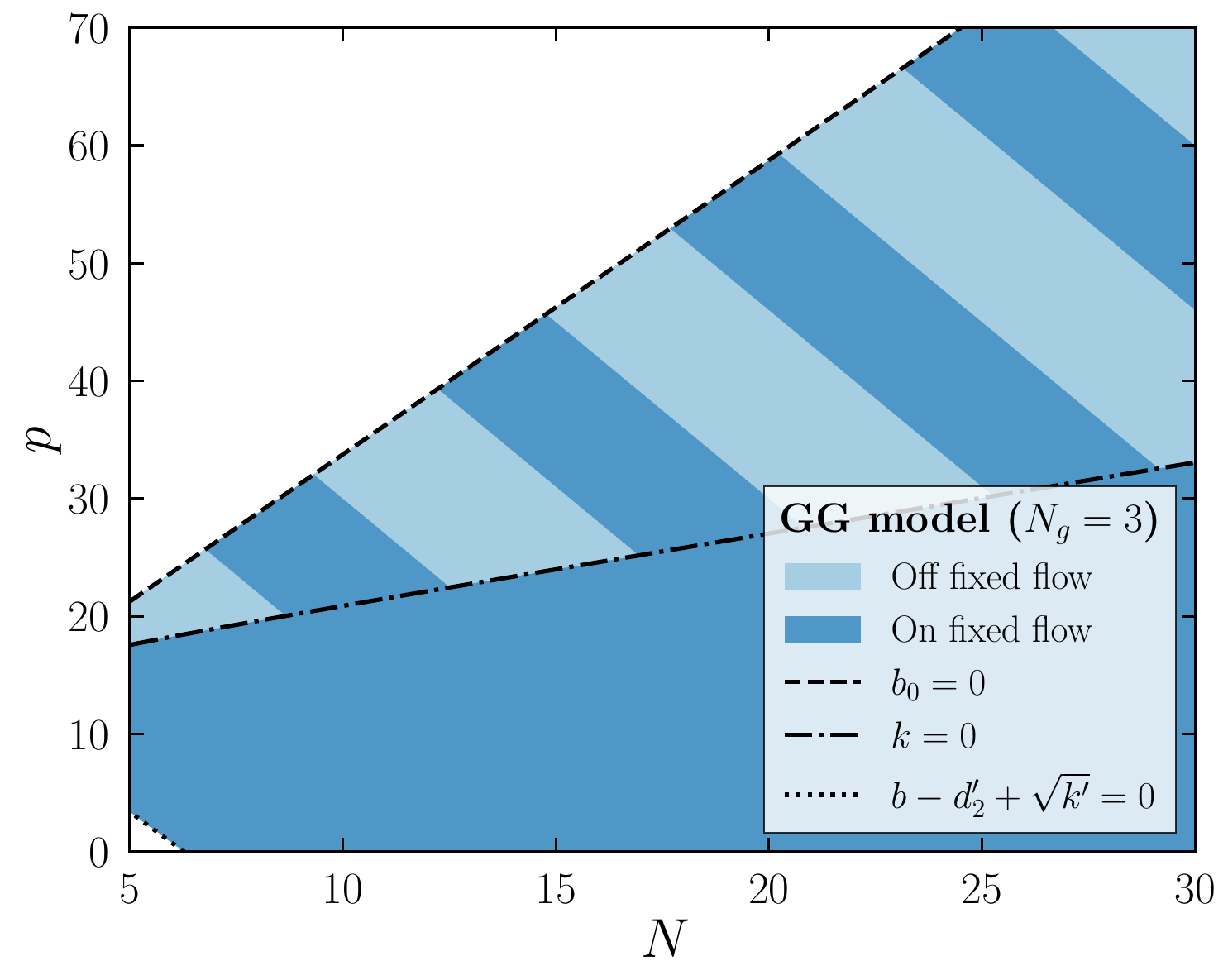}
        \caption{$N_g=3$}
        \label{fig:gg_ng3}
    \end{subfigure}
    
    \vspace{0.5cm} 
    \begin{subfigure}[b]{0.49\textwidth}
        \centering
        \includegraphics[width=\textwidth]{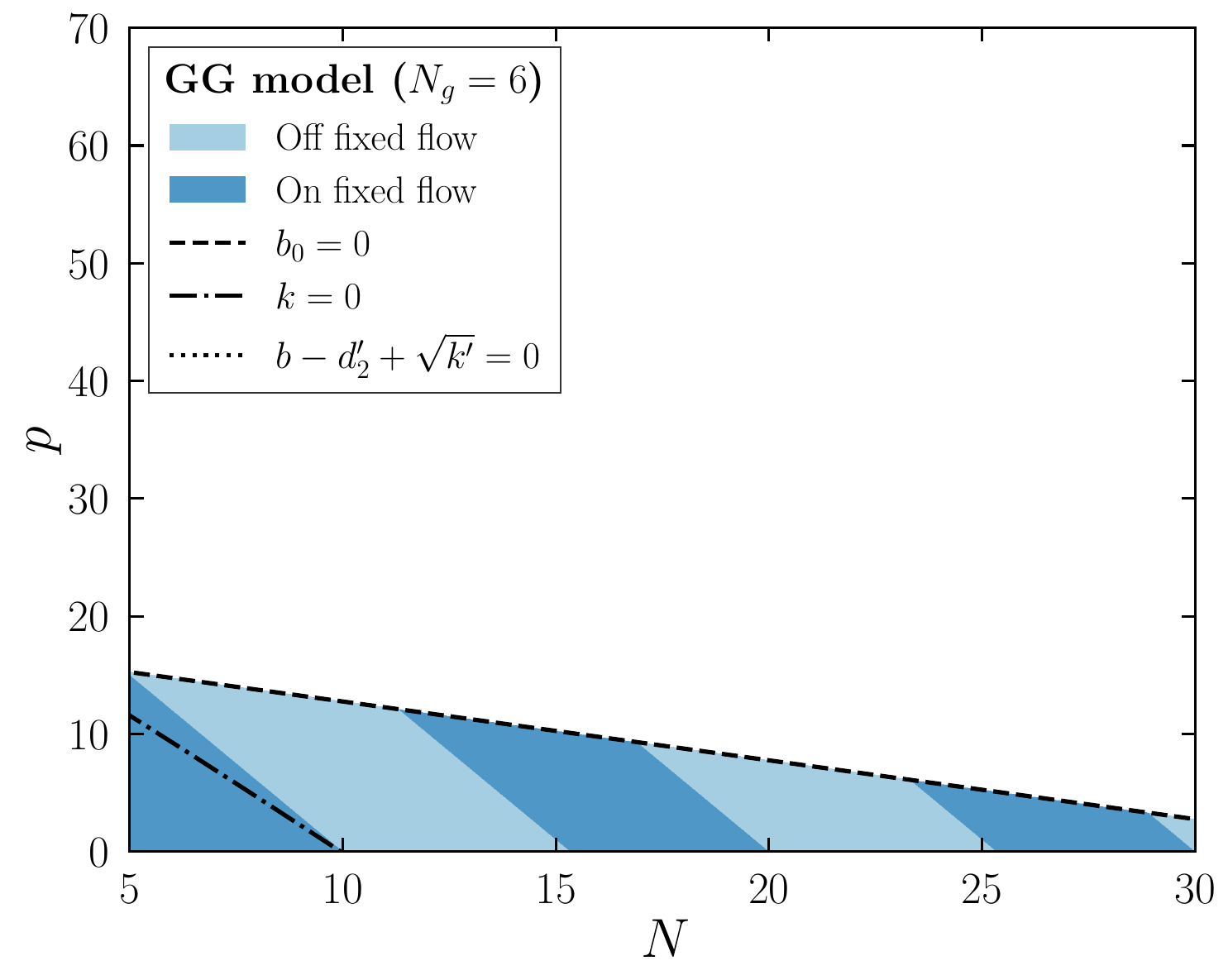}
        \caption{$N_g=5$} 
        \label{fig:gg_ng6}
    \end{subfigure}%
    \hspace{0.1cm}
    \begin{subfigure}[b]{0.49\textwidth}
        \centering
        \includegraphics[width=\textwidth]{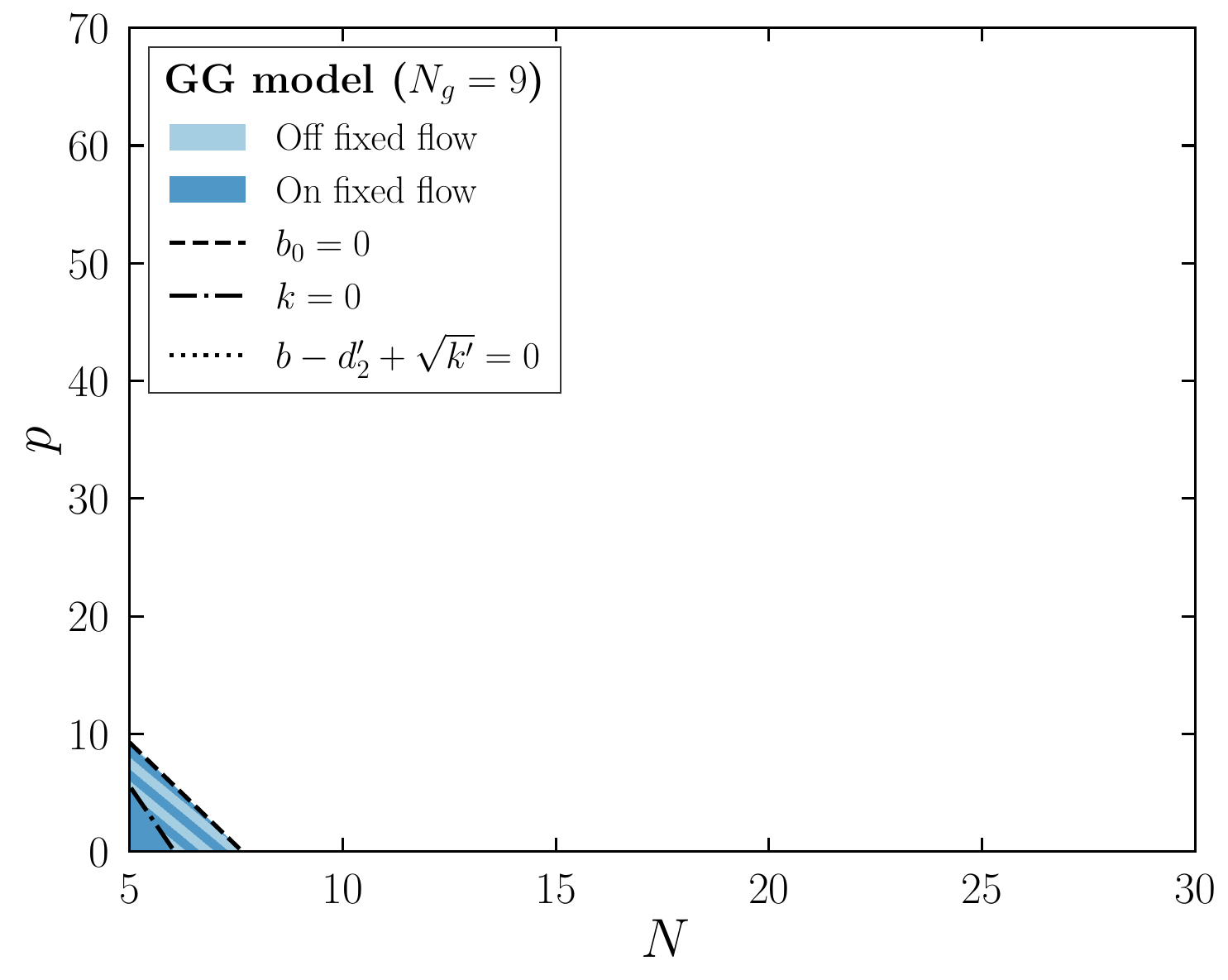}
        \caption{$N_g=6$}
        \label{fig:gg_ng9}
    \end{subfigure}
    
    \caption{CAF regions in the ($N$, $p$) parameter space of the GG model with one fundamental scalar for $N_g = \{1, 3, 5, 6\}$. The dark blue area indicates the CAF region on fixed flow, which is bounded above by the $b_0=0$ condition (dashed black line) and below by the $b_0-c_1>0$ condition (dotted black line). The light blue area highlights the CAF region off the fixed flow, bounded strictly between $b_0=0$ (above) and $k=0$ (dash-dotted black line, below). The striped region indicates the existence of CAF both on and off fixed flow. The theoretical boundary curves correspond to the $p_{\max}$, $p^{\text{off}}_{\min}$, and $p^{\text{on}}_{\min}$ functions detailed in Eqs. \eqref{eq:pmaxfundNg}, \eqref{eq:pminfundNg}, and \eqref{eq:pminfundNgon}.}
\label{fig:CAFGGNG9}
\end{figure}
Next, we consider the Yukawa beta function. 
In the current models with multiple chiral generations, the Yukawa coupling is an $N_g\times (N_g(N\mp 4)+p)$ matrix. We can again use the flavour symmetries to decompose the matrix\footnote{We can do this using singular value decomposition (SVD). Any complex matrix $A$ can be written as $A=U\Sigma V^T$, where $U$ and $V$ are unitary matrices and $\Sigma$ is a rectangular diagonal matrix with non-negative real entries. } so that, without loss of generality, the Yukawa matrix consists on $N_g$ independent and decoupled couplings. 
To further simplify the beta function, we assume flavour universality, i.e.\ all the Yukawa couplings have the same values for all generations.  This effectively reduces the RG evolution to a single beta function of the form 
\begin{equation}
    \beta_{y}={\alpha_y}
    \left(3\alpha_g\left(\frac{5}{N}\pm2-3N\right)
    +\alpha_y\left(\mp N_g+N_gN+\frac{2\mp1+N}{2}\right)\right)\;,
\end{equation} 
where the factor $N_g$ comes from the trace over flavours.
Within the same assumption, the beta function of the quartic coupling is given by
\begin{align}
    \beta_{\lambda}=&\alpha_\lambda\left[2N_g(N\mp1)\alpha_y+\left(\frac{6}{N}-6N\right)\alpha_g+4 \alpha_\lambda (N+4)\right]-\frac{N_g}{2} (N+1\mp2)
  \alpha_y^2\nonumber\\&\quad+\frac{3 \alpha_g^2 \left(N^3+N^2-4 N+2\right)}{4 N^2}\;.
\end{align}
These beta functions now include a dependency on the number of chiral families, effectively introducing an $N_g$-dependence in the CAF regions.

\begin{figure}[b!]
    \centering
    \begin{subfigure}[b]{0.49\textwidth}
        \centering
        \includegraphics[width=\textwidth]{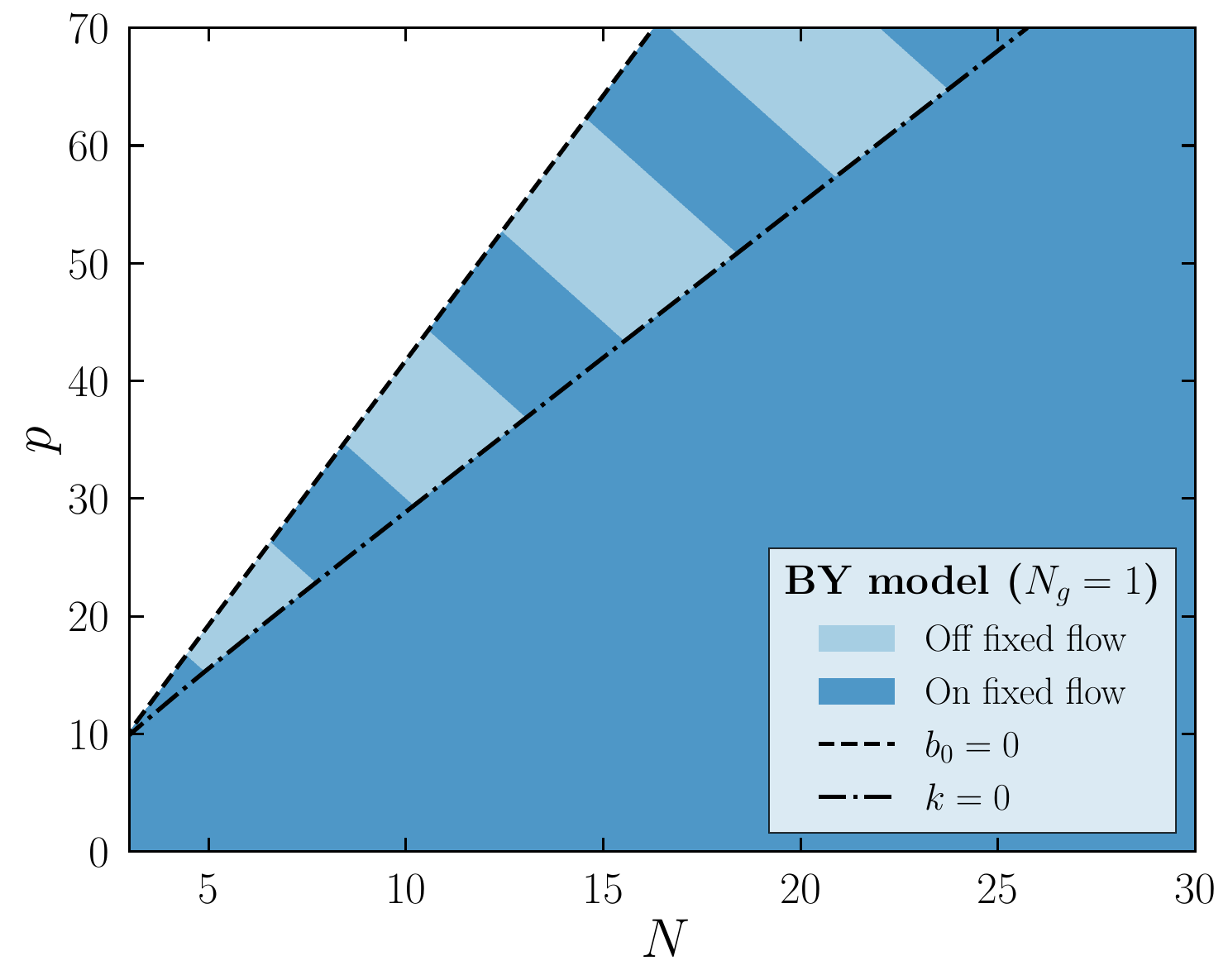}
        \caption{$N_g=1$} 
        \label{fig:by_ng1}
    \end{subfigure}
    \hspace{0.1cm}
    \begin{subfigure}[b]{0.49\textwidth}
        \centering
        \includegraphics[width=\textwidth]{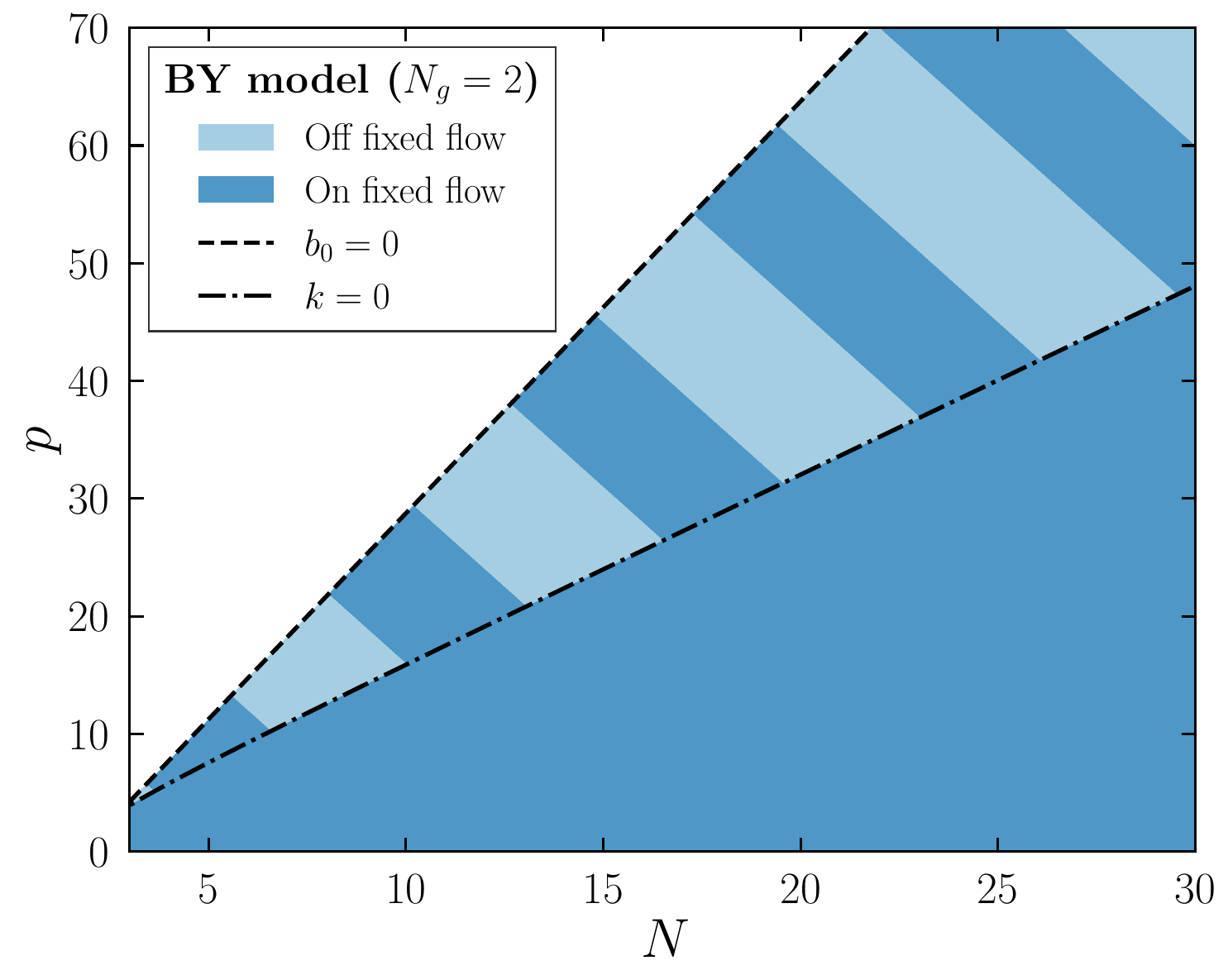}
        \caption{$N_g=2$} 
        \label{fig:by_ng2}
    \end{subfigure}
    
    \vspace{0.5cm} 
    
    \begin{subfigure}[b]{0.49\textwidth}
        \centering
        \includegraphics[width=\textwidth]{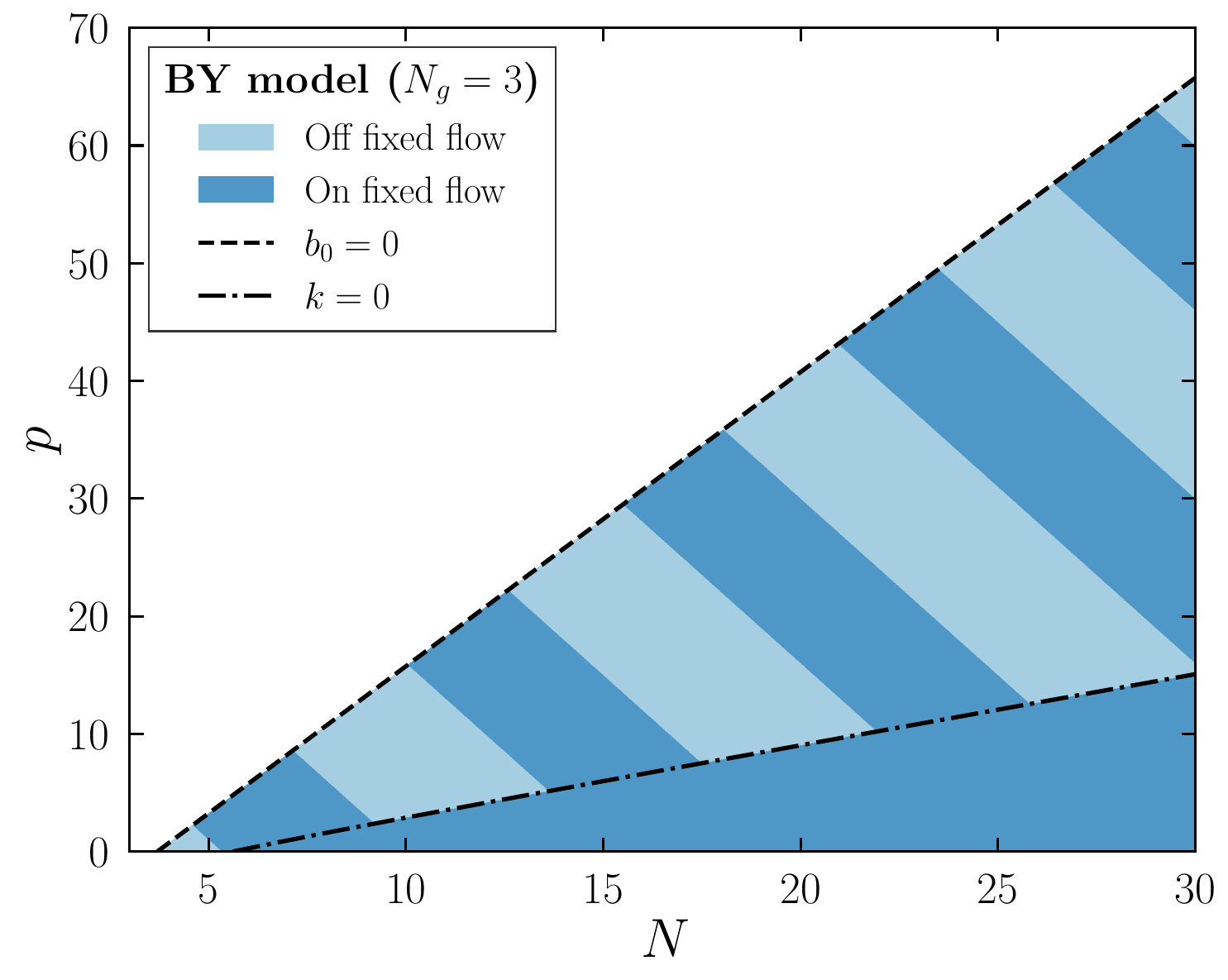}
        \caption{$N_g=3$} 
        \label{fig:by_ng3}
    \end{subfigure}
    \hspace{0.1cm}
    \begin{subfigure}[b]{0.49\textwidth}
        \centering
        \includegraphics[width=\textwidth]{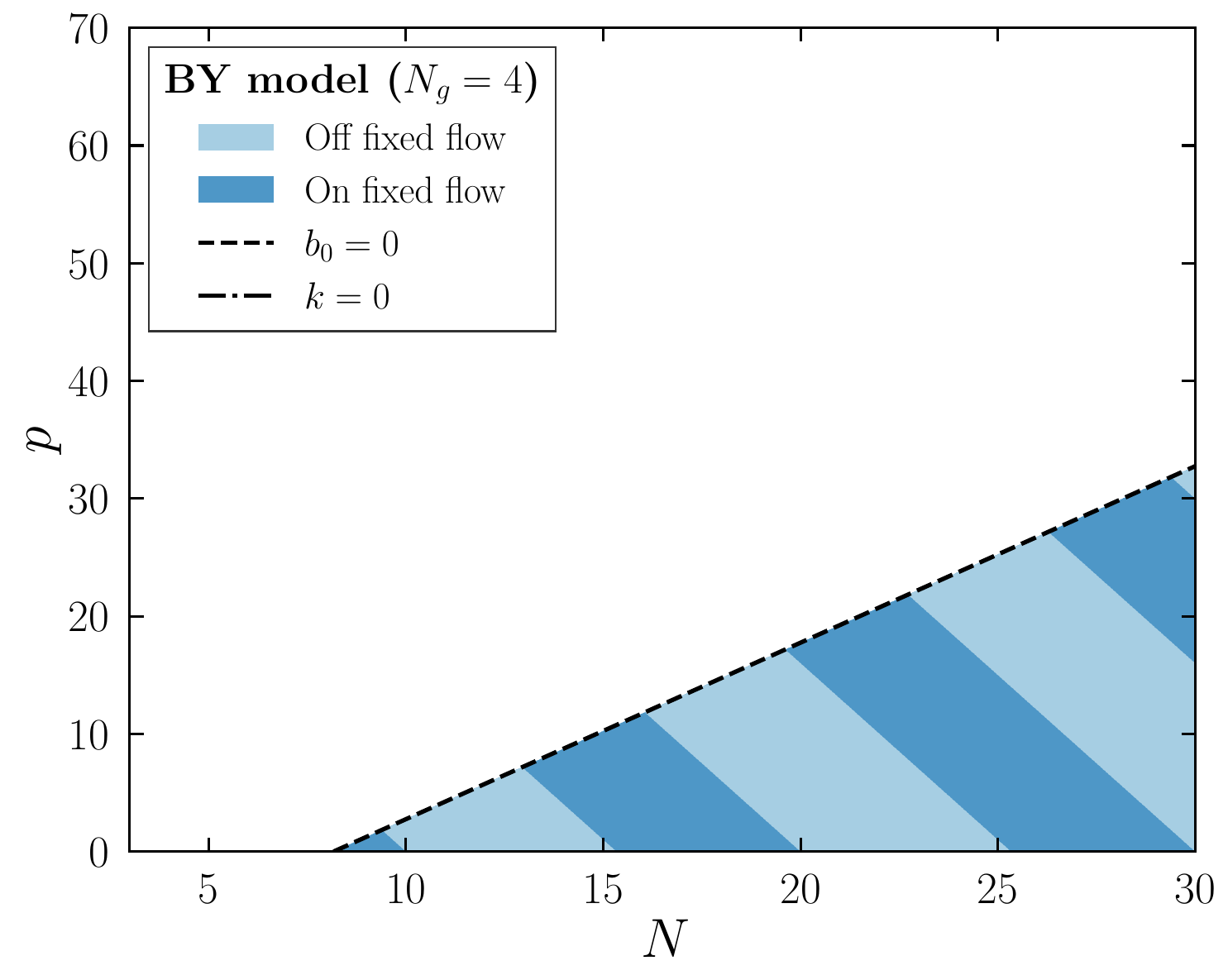}
        \caption{$N_g=4$} 
        \label{fig:by_ng4}
    \end{subfigure}
    
    \caption{CAF regions in the ($N$, $p$) parameter space of the BY model with one fundamental scalar for $N_g = \{1, 2, 3, 4\}$. The dark blue area indicates the CAF region on fixed flow, which is bounded above by the $b_0=0$ condition (dashed black line). The light blue area highlights the CAF region off fixed flow, bounded strictly between $b_0=0$ (above) and $k=0$ (dash-dotted black line, below). The striped region indicates the existence of CAF both on and off fixed flow. The theoretical boundary curves correspond to the $p_{\max}$ and $p^\text{on}_{\min}$ functions detailed in Eqs. \eqref{eq:pmaxfundNg} and \eqref{eq:pminfundNg}.}
    \label{fig:CAFNG9}
\end{figure}

We can again describe the boundaries of the on- and off-fixed-flow CAF regions using maximum and minimum $p$ functions, where $p_{\max}$ is given by the asymptotic freedom of the gauge coupling, C.f. Eq~\eqref{eq:pmaxfundNg}. 
For the off-fixed-flow regimes, the minimum $p(N,N_g)$ is given by the condition $k=0$, yielding
\begin{align}
    p^{\text{off}}_{\min}(N,N_g)&= -\frac{1}{4} + N + \frac{3\left(3 + \sqrt{3}\sqrt{8 - 14N + 5N^3 + N^4}\right)}{2N} \pm 3N_g - N N_g\;.\label{eq:pminfundNg}
\end{align}
In the case on fixed flow, instead, the CAF region extends all the way to $p=0$ for the BY models, while for the GG models a lower boundary stems from the condition $b_0-d_2'+\sqrt{k'}=0$, yielding
\begin{align}
    p^{\text{on}}_{\min}(N,N_g)& = \frac{1}{16} \left( 68 + \frac{180}{N} - 20N + 48N_g - 16NN_g + \frac{3\sqrt{6}\sqrt{\frac{-2 + N(2+N)}{N_g}}(1 + N + 2(-1+N)N_g)}{N} \right)\,.\label{eq:pminfundNgon}
\end{align}
The CAF regions in the parameter space $(N,p)$ are shown in Fig.~\ref{fig:CAFGGNG9} and~\ref{fig:CAFNG9} for GG and BY models, respectively, and for a sample $N_g$ values.

As shown in Figure \ref{fig:CAFGGNG9} for the GG model, the CAF region becomes smaller as $N_g$ becomes larger, due to the positive contribution of fermions to the gauge beta function. The panels show the parameter space of the GG model for $N_g=(1,3,5,6)$, noting that CAF regimes can be found as long as $N_g\leq{13}$. 
It is of interest to consider the case of $\mathrm{SU}(5)$ with three chiral families, which corresponds to a minimal GUT with the Higgs multiplet scalar \cite{Georgi:1974yf}. In this setup, the fundamental scalar $\phi$ can break the electroweak symmetry. Our analysis shows that the GG model features CAF regions both on and off fixed flow, see Fig.~\ref{fig:gg_ng3}. Specifically, when the Yukawa coupling is on fixed flow, the smallest amount of vector-like fermions needed to realise CAF is $p=4$. However, when the Yukawa coupling is off fixed flow the requirement increases to $p=18$. This highlights the viability of the $\mathrm{SU}(5)$ GUT when supplied with a sufficient vector-like sector. 

The BY model can be seen in Figure \ref{fig:CAFNG9}. The figure displays four values of $N_g=(1,2,3,4)$ while CAF regions disappear for $N_g\ge 5$. In all the cases, the theory exhibits CAF both on and off fixed flow. 

%% file: Sections/adjoint_scalar.tex
\section{Adding a Single Adjoint scalar} 
Consider now the case of a scalar field in the adjoint representation of $\mathrm{SU}(N)$. This scalar field will be represented by an $N\times N$ matrix in gauge space, with $N^2-1$ independent components written as $\Phi=\phi^bT^b$.  
The adjoint representation is real; consequently, it cannot carry any charges under the Abelian groups. 

The adjoint scalar allows us to write a Yukawa coupling between the vector-like fermions $\psi$ and $\tilde{\psi}$, hence the Lagrangian of the models reads
\begin{align}
    \mathcal{L}= &-\frac{1}{4} F_{\mu\nu}^a F^{a\mu\nu}+ i \bar{\Lambda}^{ij} \slashed{D} \Lambda_{ij} 
    + i \bar{\psi}^i_\alpha \slashed{D} \psi_i^\alpha 
    + i \bar{\tilde{\psi}}_i^\beta \slashed{D} \tilde{\psi}^i_\beta\nonumber\\&+\tr((D^\mu\Phi)(D_\mu\Phi))+ (y^\beta_\alpha \tilde\psi_\beta \Phi\psi^{\alpha}+h.c.)+\lambda_1\tr(\Phi^2)^2+\lambda_2\tr(\Phi^4)
    \;.\label{eq:Ladj}
\end{align}
Furthermore, two independent scalar quartic couplings can be written\footnote{For $N=3$, completeness identities that contract the indices imply that the two operators are equivalent, hence only one quartic coupling exists. The correct beta function can be found by summing $\beta_{\lambda_1} + \beta_{\lambda_2} \equiv \beta_\lambda$ and identifying $\lambda = \lambda_1+\lambda_2$. We checked explicitly that doing the analysis with one or two quartic couplings yields the same result. }. The field content of the models is summarised in Table \ref{tab:adj}: note that the $\mathrm{U}_2(1)$ symmetry, originally present in the scalar-less models, is missing as it is explicitly broken by the Yukawa coupling. 

From Eq.\ \eqref{eq:Ladj} we see that the Yukawa term involves only the fermions in the fundamental and anti-fundamental representations of $\mathrm{SU}(N)$, hence the Yukawa couplings appear as a $p\times (N\mp 4+p)$ matrix. Analogously to the cases in Section~\ref{sec:fundscalarNg}, we can decompose the matrix in terms of $p$ independent real couplings. We will again apply the simplification that all couplings are equal.
With this, we find the beta functions of the models to be 
\begin{align}
    \beta_g&=-\frac{1}{3}\,\alpha_g^2\left(17N \pm 12 -4p\right)\;,\\
    \beta_{y}&=\alpha_y\left[\left(\frac 6N-6N\right) \alpha_g+ 2p \alpha_{y}+ \left(N-\frac 3N\right )\alpha_y\right]\;,\\
    \beta_{\lambda_1} &=4\frac{N^2-9}{N}\alpha_{\lambda_1}^2+12\alpha_{\lambda_1}\alpha_{\lambda_2}+4(p\alpha_{y}-3N\alpha_{g})\alpha_{\lambda_1}+3N\alpha_g^2-2p\alpha_{y}^2\;,
    \\
    \beta_{{\lambda_2}}&=(7+N^2)\alpha_{\lambda_2}^2+\frac{12(3+N^2)}{N^2}\alpha_{\lambda_1}^2+\frac{4(2N^2-3)}{N}\alpha_{\lambda_1}\alpha_{\lambda_2}+4(p\alpha_{y}-3N\alpha_{g})\alpha_{\lambda_2}+18\alpha_{g}^2\;.
\end{align}
 \begin{table}[ht]
    \centering
    \renewcommand{\arraystretch}{1.5} 
    \begin{tabular}{|c|c|c|c|c|}
    \hline
    Fields & $\mathrm{SU}(N)$ & $\mathrm{SU}(N \mp 4 + p)$ & $\mathrm{SU}(p)$ & $\mathrm{U}_1(1)$  \\
    \hline
    $\Lambda$ & $\ytableausetup{boxsize=0.25cm, centertableaux}\begin{ytableau}{}\\{}\end{ytableau} \;\, / \;\, \begin{ytableau}{}&{}\end{ytableau}$ & 1 & 1 & $N \mp 4$  \\

    $\tilde\psi$ & $\overline{\begin{ytableau}{}\end{ytableau}\rule{0pt}{7.5pt}}$ & $\overline{\begin{ytableau}{}\end{ytableau}\rule{0pt}{7.5pt}}$ & 1 & $-(N \mp 2)$ \\

    $\psi$ & ${\begin{ytableau}{}\end{ytableau}}$ & 1 & ${\begin{ytableau}{}\end{ytableau}}$ & $N \mp 2$  \\
    \hline
     $\Phi$ & \textbf{adj} & $1$ & $1$ & $0$  \\
    \hline
    \end{tabular}
    \caption{Field content of the GG (upper signs, antisymmetric $\Lambda$) and BY (lower signs, symmetric $\Lambda$) models with an adjoint scalar $\phi$.  The table shows how the fields transform under the gauge and global symmetry groups, and their charges under the $\mathrm{U}_1(1)$ symmetry. Note that the $\mathrm{U}_2(1)$ is explicitly broken by the Yukawa term in the Lagrangian and, therefore, not listed.}
    \label{tab:adj} 
\end{table}

\begin{figure}[ht]
    \centering
    \begin{subfigure}[b]{0.49\textwidth}
        \centering
        \includegraphics[width=\textwidth]{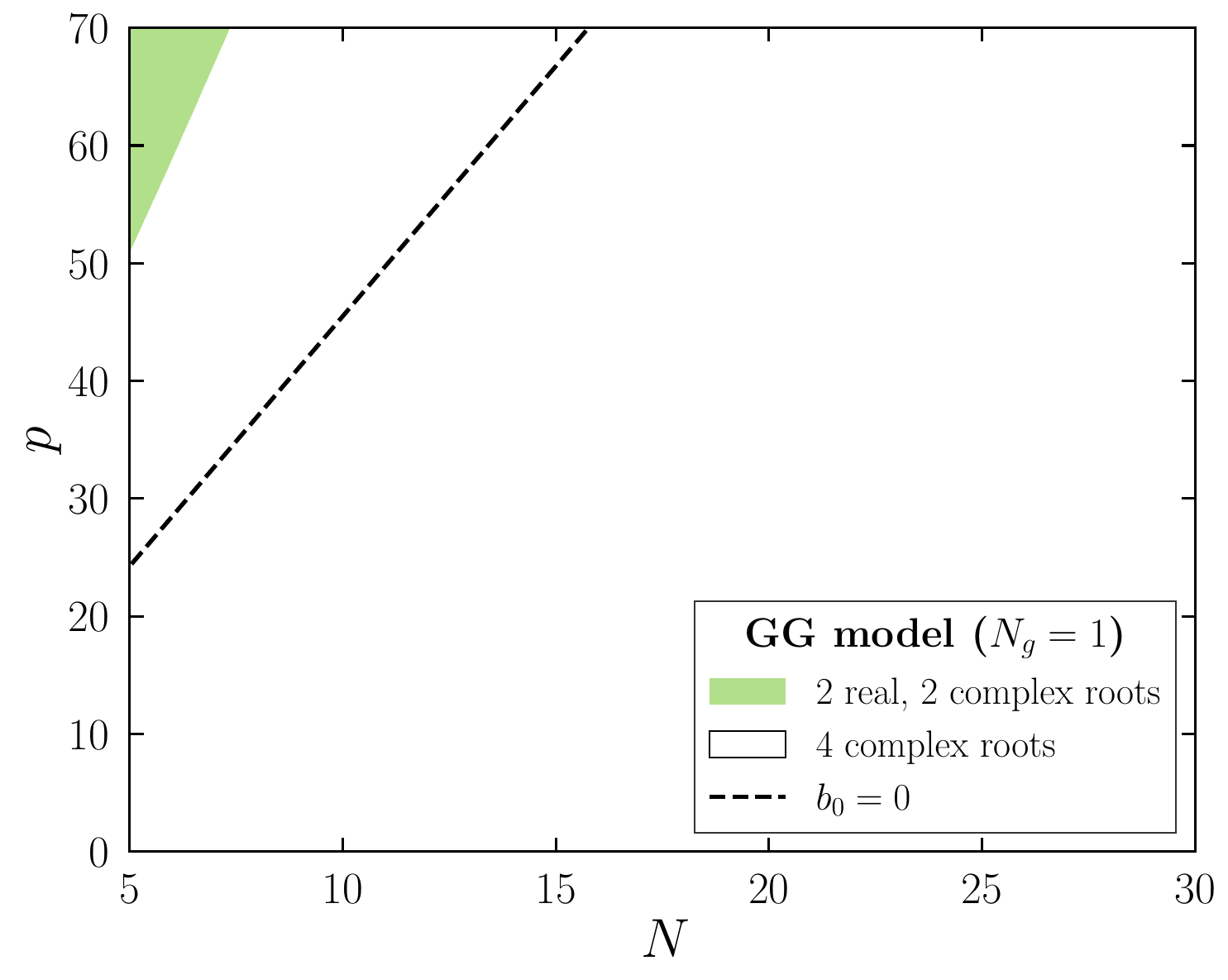}
        \caption{All couplings on fixed flow} 
        \label{fig:roots_fixed}
    \end{subfigure}
    \hspace{0.1cm}
    \begin{subfigure}[b]{0.49\textwidth}
        \centering
        \includegraphics[width=\textwidth]{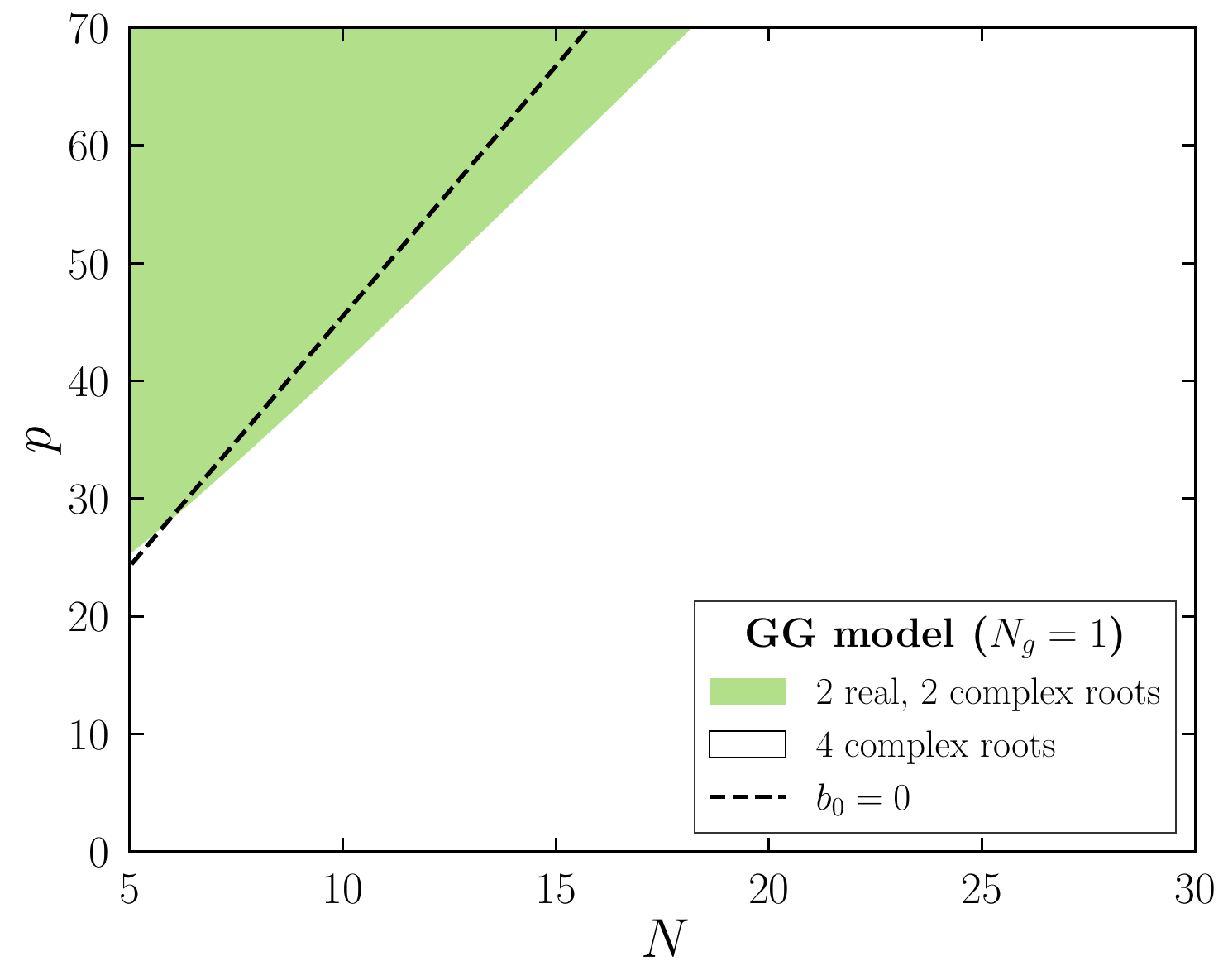}
        \caption{$\alpha_y$ off fixed flow} 
        \label{fig:roots_noy}
    \end{subfigure}
    
    \caption{ Root structure of the fourth-degree polynomial for the GG model, when (a) all couplings are on fixed flow and (b) the Yukawa coupling is off fixed flow. The shaded regions denote the presence of four complex roots (white) and two real and two complex roots (light green). The dashed black line marks the $b_0=0$ condition, underneath which the gauge coupling is asymptotically free.} 
    \label{fig:Deltaadj}
\end{figure}

Having two quartic couplings makes the CAF analysis different from the previous models and requires root analysis of fourth-degree polynomials. We use the approach described in Section \ref{sec:CAF2} and Appendix \ref{sec:appCafadj}. This framework converts the task of solving coupled differential equations to an exercise of finding real positive solutions to a fourth-degree polynomial. Viable CAF regions must feature at least one real solution. For GG models, in Figure~\ref{fig:Deltaadj} we show the result of the analysis both on and off fixed flow, where the viable region in green is delimitated by the condition $\Delta=0$ (see Eqs. \eqref{eq:Delta0noY} and \eqref{eq:Delta0}), where $\Delta$ is the discriminant of the fourth degree polynomial in Eq.\ \eqref{eq:fourthdegpoly}. This region should be compared with the asymptotic freedom condition for the gauge coupling running, which gives an upper bound on $p$ from the condition $b_0=0$
\begin{align}
    p_{\max}(N)&=\frac{17N\pm12}{4}\;, \label{eq:pmaxadj}
\end{align}
shown as a dashed line in the plots. Hence, there is no viable region on fixed flow, while solutions exist off fixed flow.



\begin{figure}[t!]
    \centering
    \includegraphics[width=0.5\textwidth]{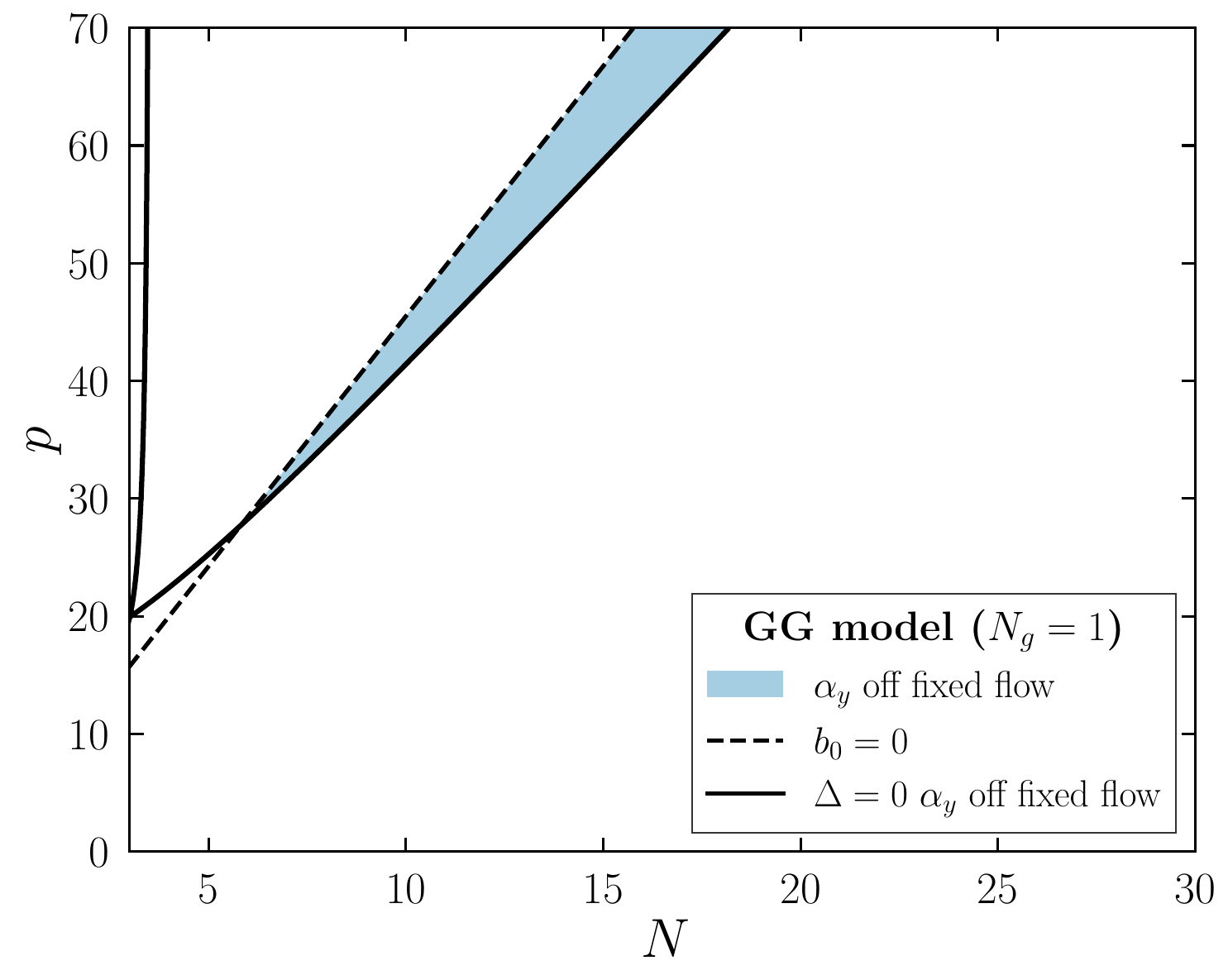}
    \caption{CAF region for the GG model with adjoint scalar when the Yukawa coupling $\alpha_y$ is off fixed flow (light blue). The dashed black line illustrates the $b_0=0$ condition, Eq.\ \eqref{eq:pmaxadj}, while the solid black line the $\Delta=0$ condition for $\alpha_y$ off fixed flow, Eq.~\eqref{eq:Delta0noY}.}
    \label{fig:BYposroots}
\end{figure}

We can now perform the complete CAF analysis of the beta function system. The result for the GG models is shown in Figure~\ref{fig:BYposroots}, where the blue sliver highlights the only available CAF region off fixed flow. Interestingly, CAF models only exist for $N \geq 7$ and $p \geq 32$, where no solution exists for the $\mathrm{SU}(5)$ GUT model with one generation.

The analysis for the BY models provides similar results, with the CAF region shown in Figure~\ref{fig:gg_combined_ng1} (for more details, see Appendix~\ref{sec:appGGplotsadj}). As before, CAF regions only exist off fixed flow for $N\geq 7$ and $p \geq 26$.

We will now proceed to consider multiple generations of chiral families in the next section.


%
%

%% file: Sections/adjoint_scalar_and_families.tex
\subsection{Inclusion of Multiple Chiral Families}\label{sec:adjNg}
 Let us now upgrade the model with an adjoint scalar to include multiple chiral generations. As done in Section \ref{sec:fundscalarNg}, the global symmetry group now depends on the number of families via the piece $\mathrm{SU}(N_g)\times\mathrm{SU}(N_g(N\mp4)+p)$. The updated list of fields and charges is shown in Table \ref{tab:adj_Ng} for both GG an BY models. The Lagrangian is the same as in Eq.~\eqref{eq:Ladj}, except for a new flavour index on the two-index chiral fermion $\Lambda^\gamma$, and the Yukawa matrix being a $p\times (N_g(N\mp4)+p)$, which can again be decomposed in terms of $p$ independent real coupling, which we will consider identical. 

\begin{table}[ht]
    \centering
    \renewcommand{\arraystretch}{1.5} 
    \begin{tabular}{|c|c|c|c|c|c|}
    \hline
    Fields & $\mathrm{SU}(N)$ & $\mathrm{SU}(N_g(N \mp 4) + p)$ & $\mathrm{SU}(p)$ & $\mathrm{SU}(N_g)$ & $\mathrm{U}_1(1)$  \\
    \hline
    $\Lambda$ & $\ytableausetup{boxsize=0.25cm, centertableaux}\begin{ytableau}{}\\{}\end{ytableau} \;\, / \;\, \begin{ytableau}{}&{}\end{ytableau}$ & $1$ & $1$ &$\overline{\begin{ytableau}{}\end{ytableau}\rule{0pt}{7.5pt}}$ & $N \mp 4$ \\

    $\tilde\psi$ & $\overline{\begin{ytableau}{}\end{ytableau}\rule{0pt}{7.5pt}}$ & $\overline{\begin{ytableau}{}\end{ytableau}\rule{0pt}{7.5pt}}$ & 1&1 & $-(N \mp 2)$  \\

    $\psi$ & ${\begin{ytableau}{}\end{ytableau}}$ & 1 & ${\begin{ytableau}{}\end{ytableau}}$ & 1&$N \mp 2$ \\
    \hline
    $\Phi$ & \textbf{adj} & $1$ & $1$ & $1$ & $0$  \\
    \hline
    \end{tabular}
    \caption{Field content of the GG (upper signs, antisymmetric $\Lambda$) and BY (lower signs, symmetric $\Lambda$) models with $N_g$ chiral families and adjoint scalar $\phi$.  The table shows how the fields transform under the gauge and global symmetry groups, and their charges under the $\mathrm{U}_1(1)$ symmetry. }
    \label{tab:adj_Ng} 
\end{table}

Since, the Yukawa interaction does not involve the chiral fermions, only the gauge beta function will be affected by the number of generations, as follows:
\begin{equation}
     \beta_g=-\frac{1}{3}\left(21N \pm 12 N_g-4p-4N_gN)\right)\alpha_g^2\;,\\
\end{equation} 
while the beta functions of the Yukawa and scalar quartic couplings remain the same as for the one-family models.
\begin{figure}[t!]
    \centering
    \begin{subfigure}[b]{0.49\textwidth}
        \centering
        \includegraphics[width=\textwidth]{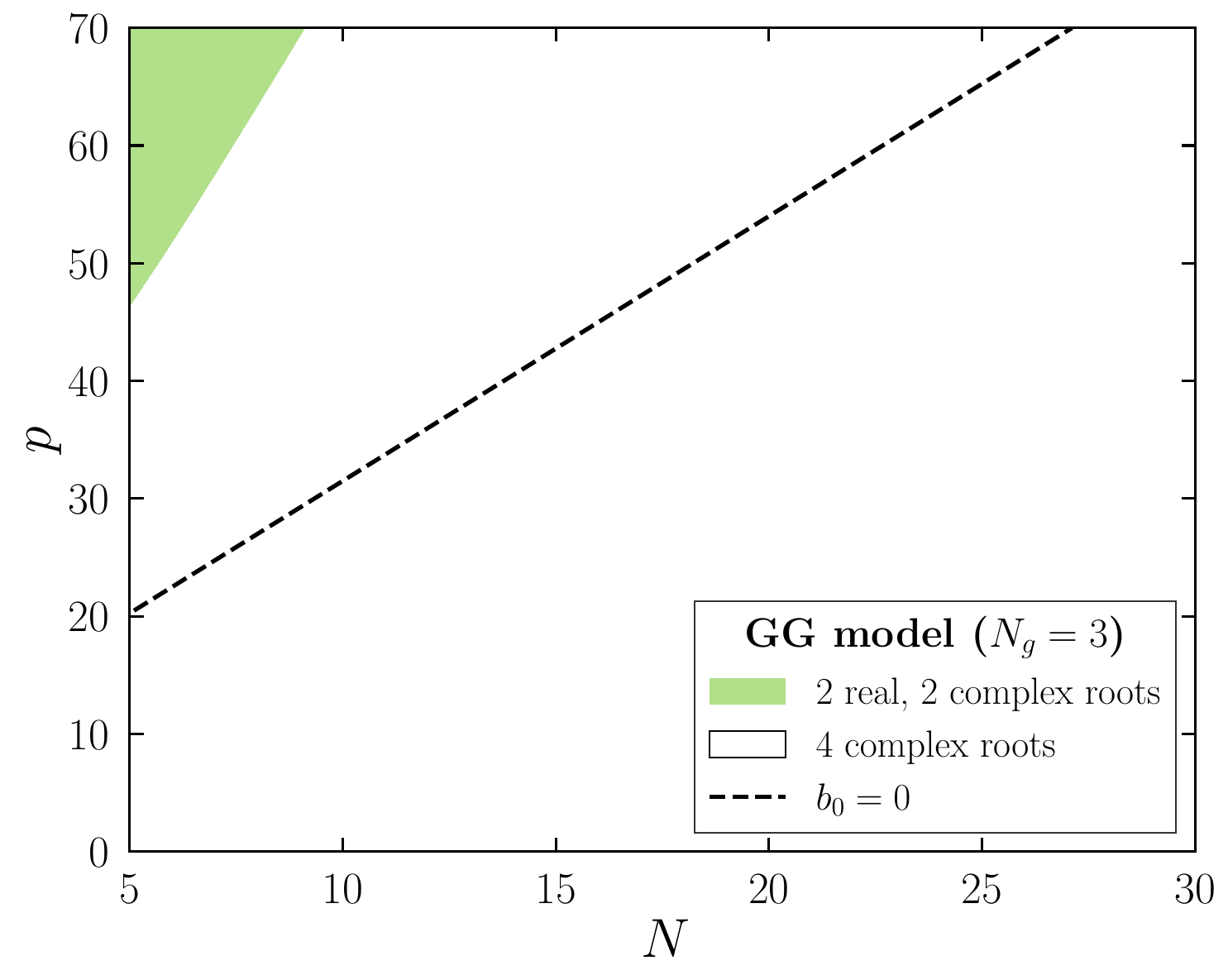}
        \caption{Real roots on fixed flow} 
        \label{fig:roots_fixed_ng2}
    \end{subfigure}
    \hspace{0.1cm}
    \begin{subfigure}[b]{0.49\textwidth}
        \centering
        \includegraphics[width=\textwidth]{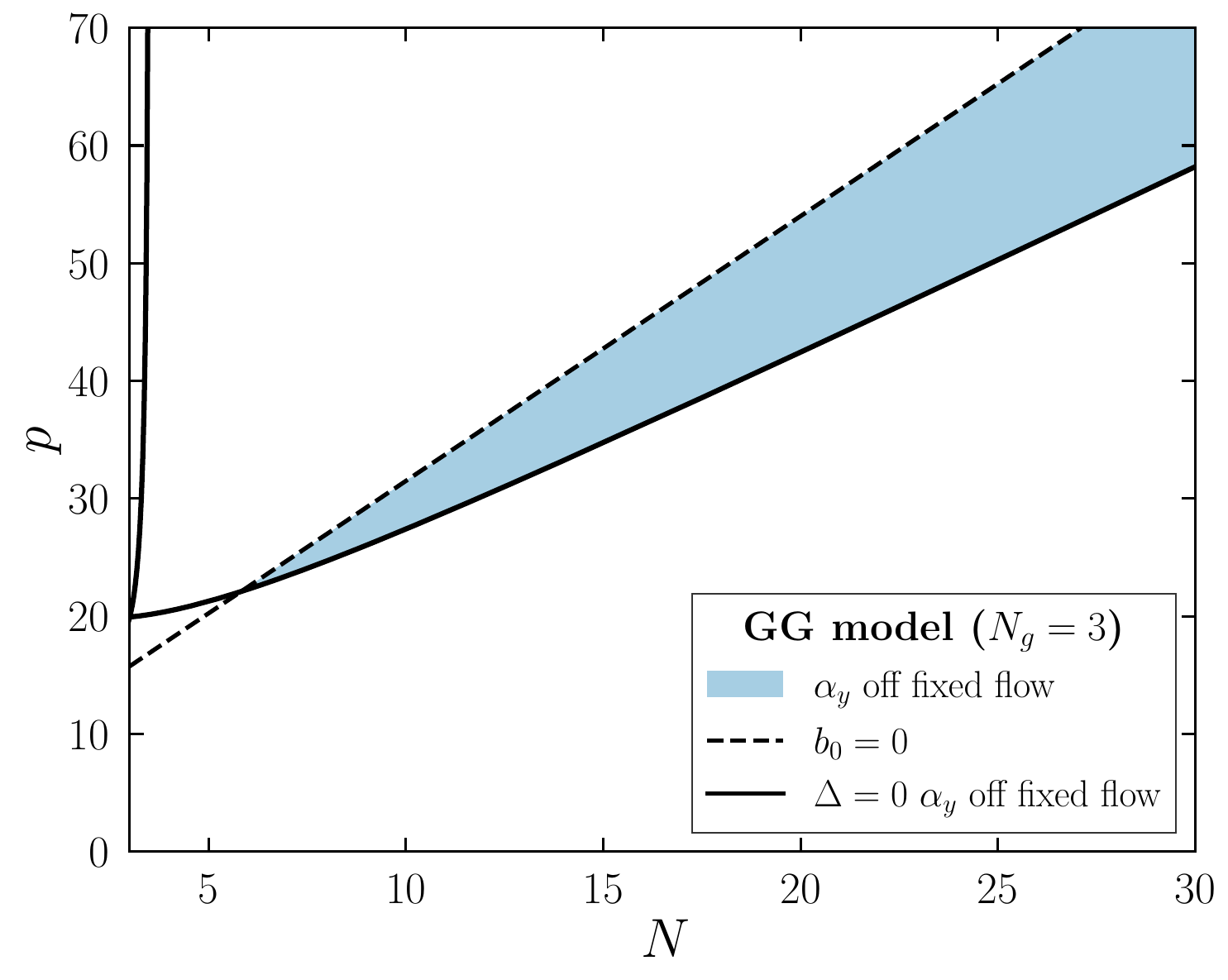}
        \caption{CAF region off fixed flow} 
        \label{fig:final_combined_ng2}
    \end{subfigure}
    
    \caption{ Properties of the GG model with $N_g=3$ chiral families. Panel (a) shows the root structure of the fourth-degree polynomial with all couplings on fixed flow. The shaded regions denote four complex roots (white) or two real and two complex roots (light green). Panel (b) depicts the CAF region with Yukawa coupling $\alpha_y$ off fixed flow (light blue). The dashed black line illustrates the $b_0=0$ condition, Eq.\ \eqref{eq:pmaxadj}, while the solid black line the $\Delta=0$ condition when $\alpha_y$ is off fixed flow, Eq.~\eqref{eq:Delta0noY}.} 
    \label{fig:Ng2}
\end{figure}
\begin{figure}[h!]
    \centering
    \begin{subfigure}[b]{0.49\textwidth}
        \centering
        \includegraphics[width=\textwidth]{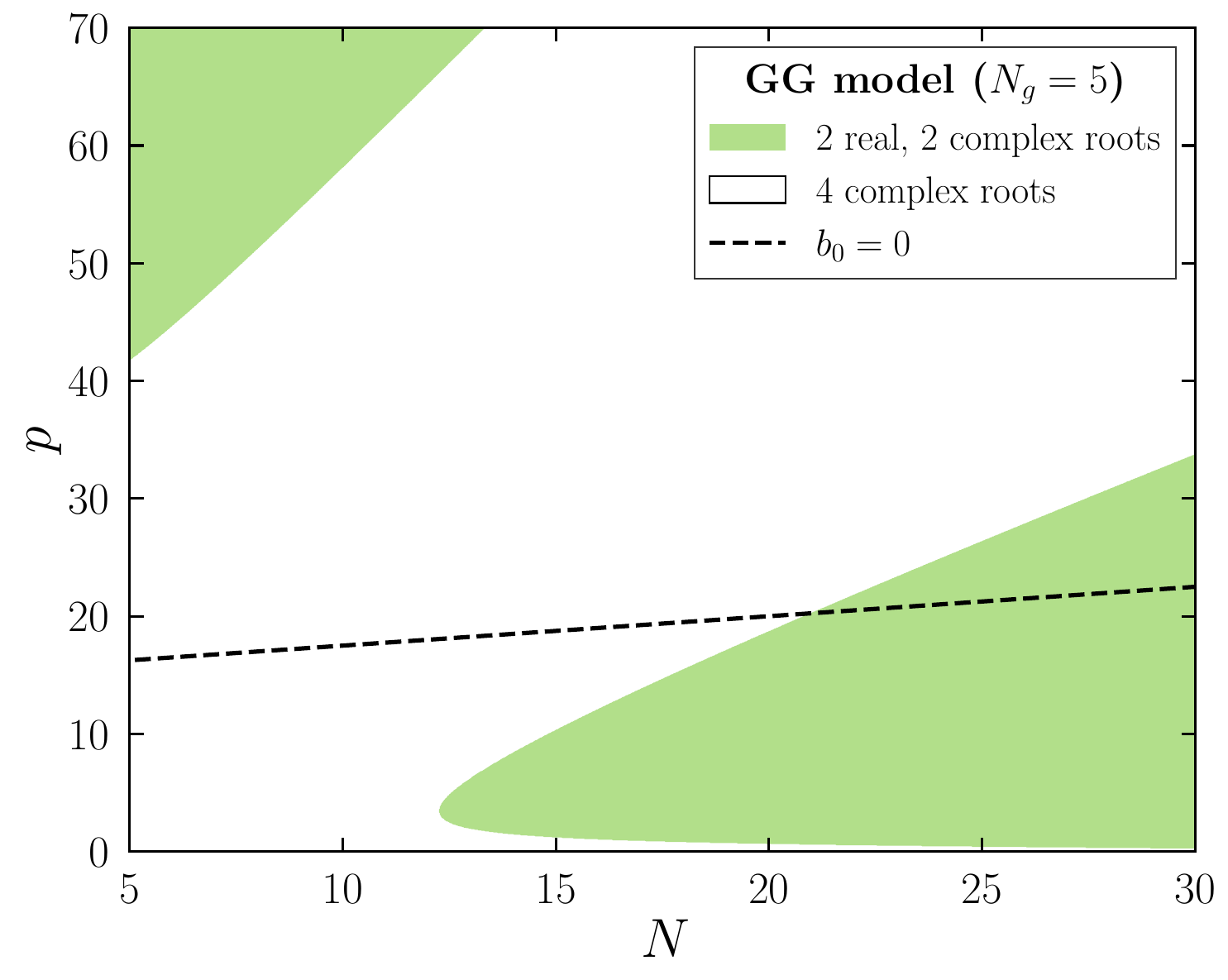}
        \caption{Real roots on fixed flow with $N_g=5$} 
        \label{fig:roots_fixed_ng3}
    \end{subfigure}
    \hspace{0.1cm}
    \begin{subfigure}[b]{0.49\textwidth}
        \centering
        \includegraphics[width=\textwidth]{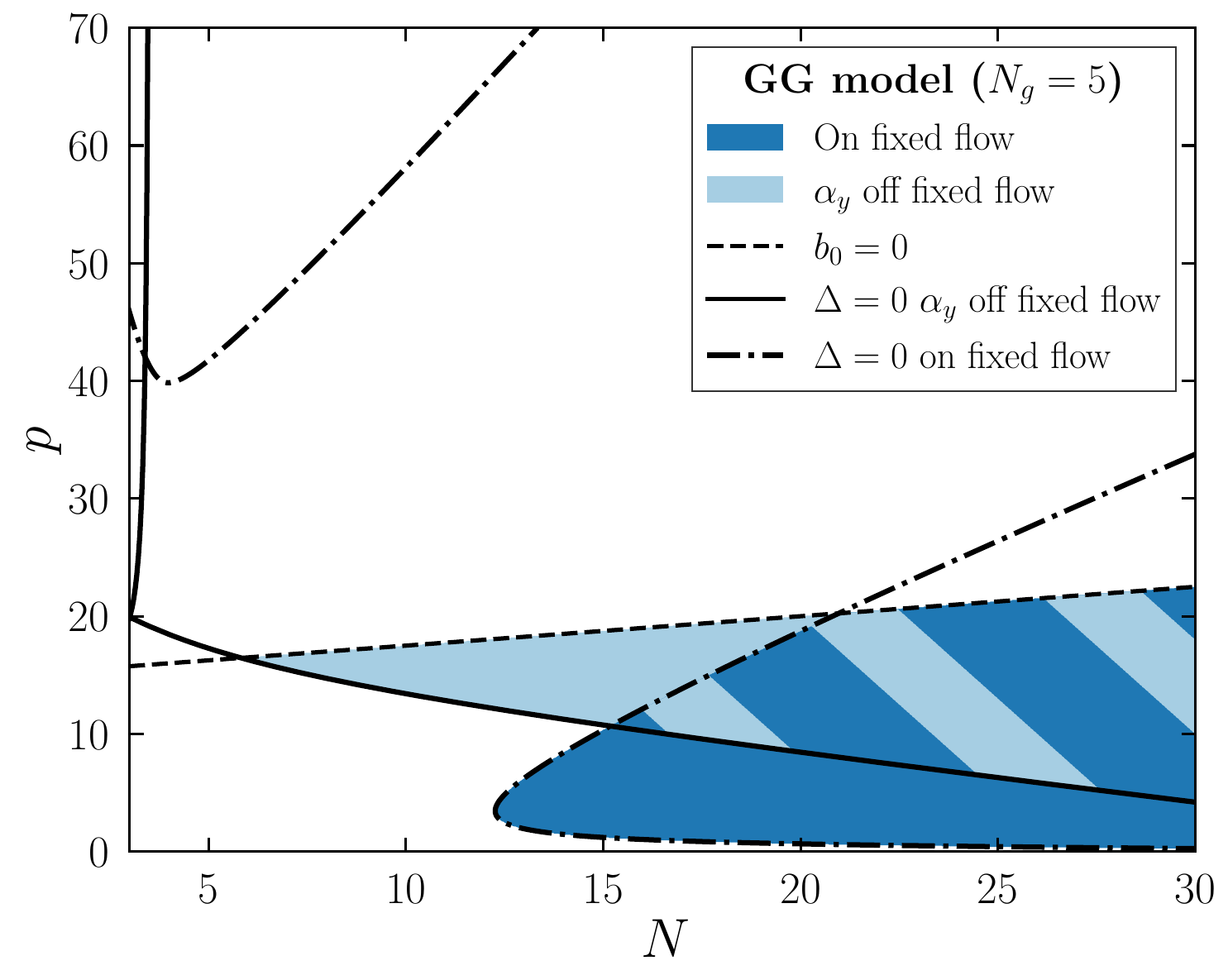}
        \caption{CAF region with $N_g=5$} 
        \label{fig:final_combined_ng3}
    \end{subfigure}

     \vspace{0.5cm} 
    \begin{subfigure}[b]{0.49\textwidth}
        \centering
        \includegraphics[width=\textwidth]{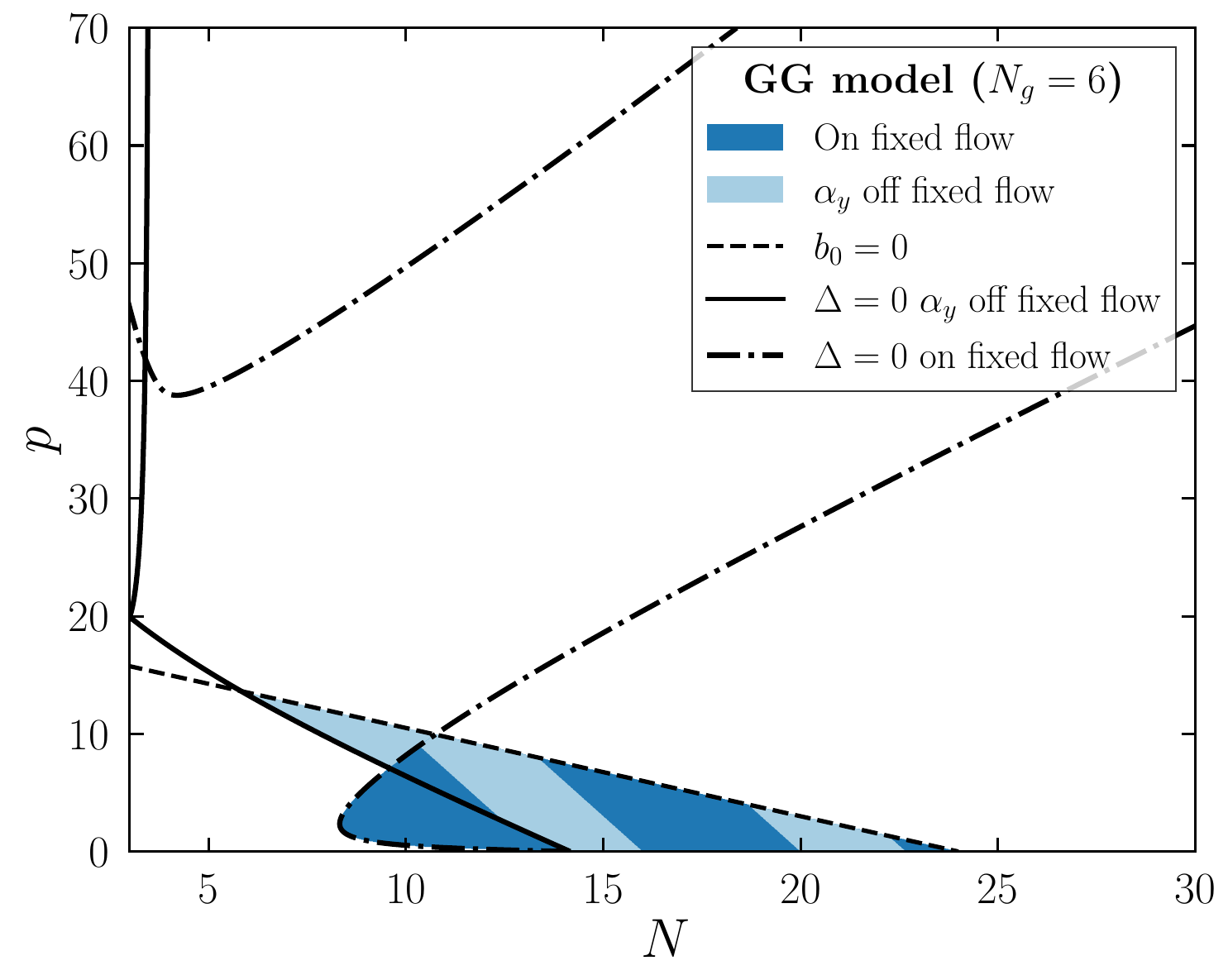}
        \caption{CAF region with $N_g=6$} 
        \label{fig:gg_combined_ng5}
    \end{subfigure}
    \hspace{0.1cm}
    \begin{subfigure}[b]{0.49\textwidth}
        \centering
        \includegraphics[width=\textwidth]{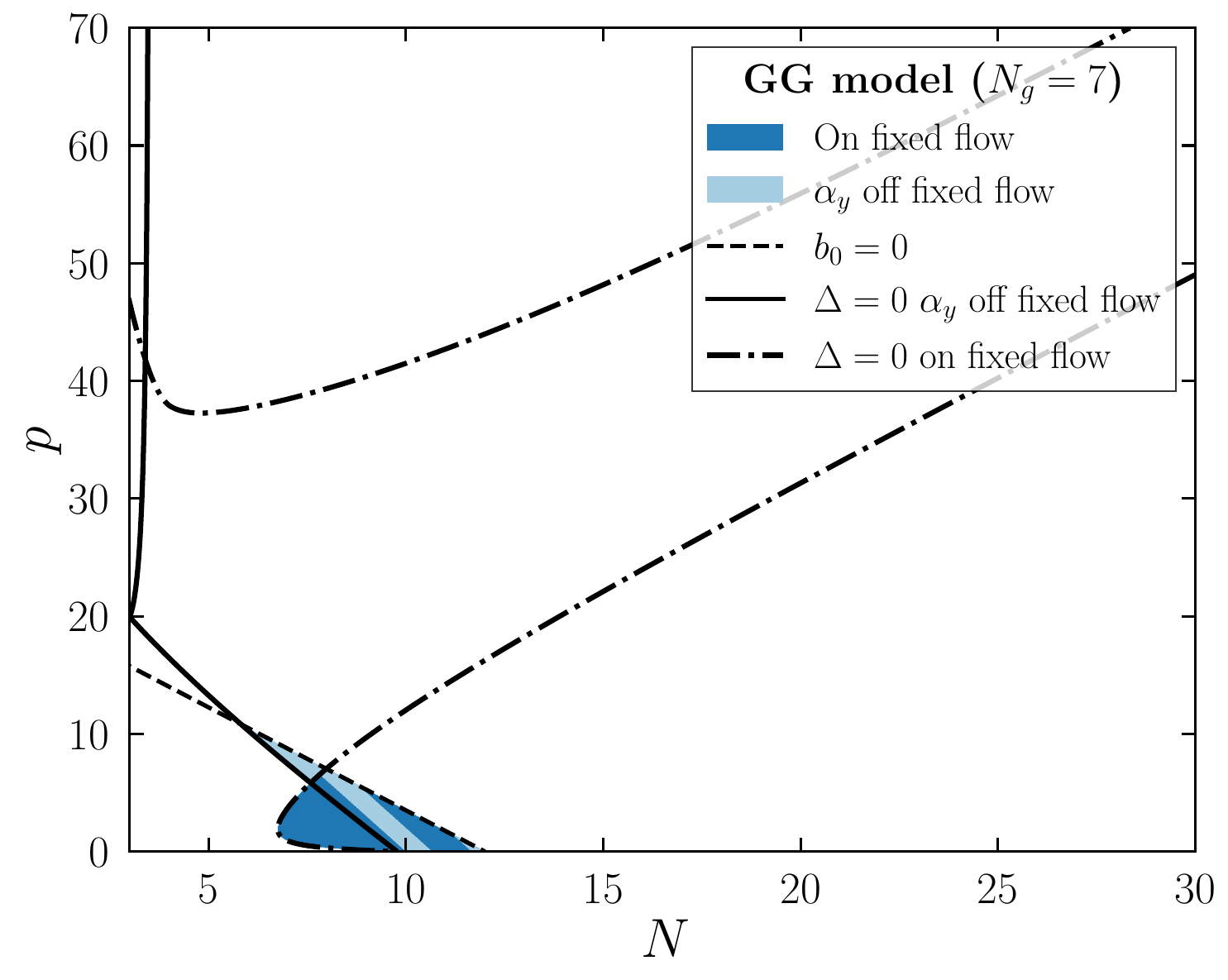}
        \caption{CAF region with $N_g=7$}
        \label{fig:gg_combined_ng6}
    \end{subfigure}
    
    \caption{Properties of the GG model with adjoint scalar when $N_g=\{5,6,7\}$. Panel (a) shows the root structure of the fourth-degree polynomial with all couplings on fixed flow. The shaded regions denote four complex roots (white) or two real and two complex roots (light green). The remaining panels (b,c,d) depict the CAF regions with all couplings on fixed flow (dark blue) and for the Yukawa coupling $\alpha_y$ off fixed flow (light blue). The dashed black line illustrates the $b_0=0$ condition, Eq.\ \eqref{eq:pmaxadj}, the solid black line the $\Delta=0$ condition for $\alpha_y$ off fixed flow, Eq.~\eqref{eq:Delta0noY}, and the dash-dotted line $\Delta=0$ condition on fixed flow, Eq.~\eqref{eq:Delta0}. } 
    \label{fig:Ng3}
\end{figure}
It is only possible to express the upper boundary in a compact analytic form, analogous to the case of a single chiral family, from $b_0=0$, yielding
\begin{align}
    p_{\max}(N,N_g)=\frac{21N\pm12 N_g-4NN_g}{4}\,.\label{eq:padjscalar}
\end{align}
where the other boundaries are determined by $\Delta=0$ (see Eqs. \eqref{eq:Delta0noY} and \eqref{eq:Delta0}).

Now we can perform the complete CAF analysis. First, we consider the GG model with $N_g=3$, which is relevant for GUTs. Figure \ref{fig:Ng2} showcases a similar result as we obtained for one family: CAF solutions only exist off fixed flow. More in detail, Figure \ref{fig:roots_fixed_ng2} illustrates that on fixed flow, all roots are complex within the region where the gauge coupling is asymptotic, i.e. below the dashed line. In Figure \ref{fig:final_combined_ng2}, the light blue region represents the CAF region for the Yukawa coupling off fixed flow. Solutions only exist for $N\geq 7$ and $p\geq 24$, once again excluding the $\mathrm{SU}(5)$ GUT model.

\begin{figure}[t!]
    \centering
    \begin{subfigure}[b]{0.49\textwidth}
        \centering
        \includegraphics[width=\textwidth]{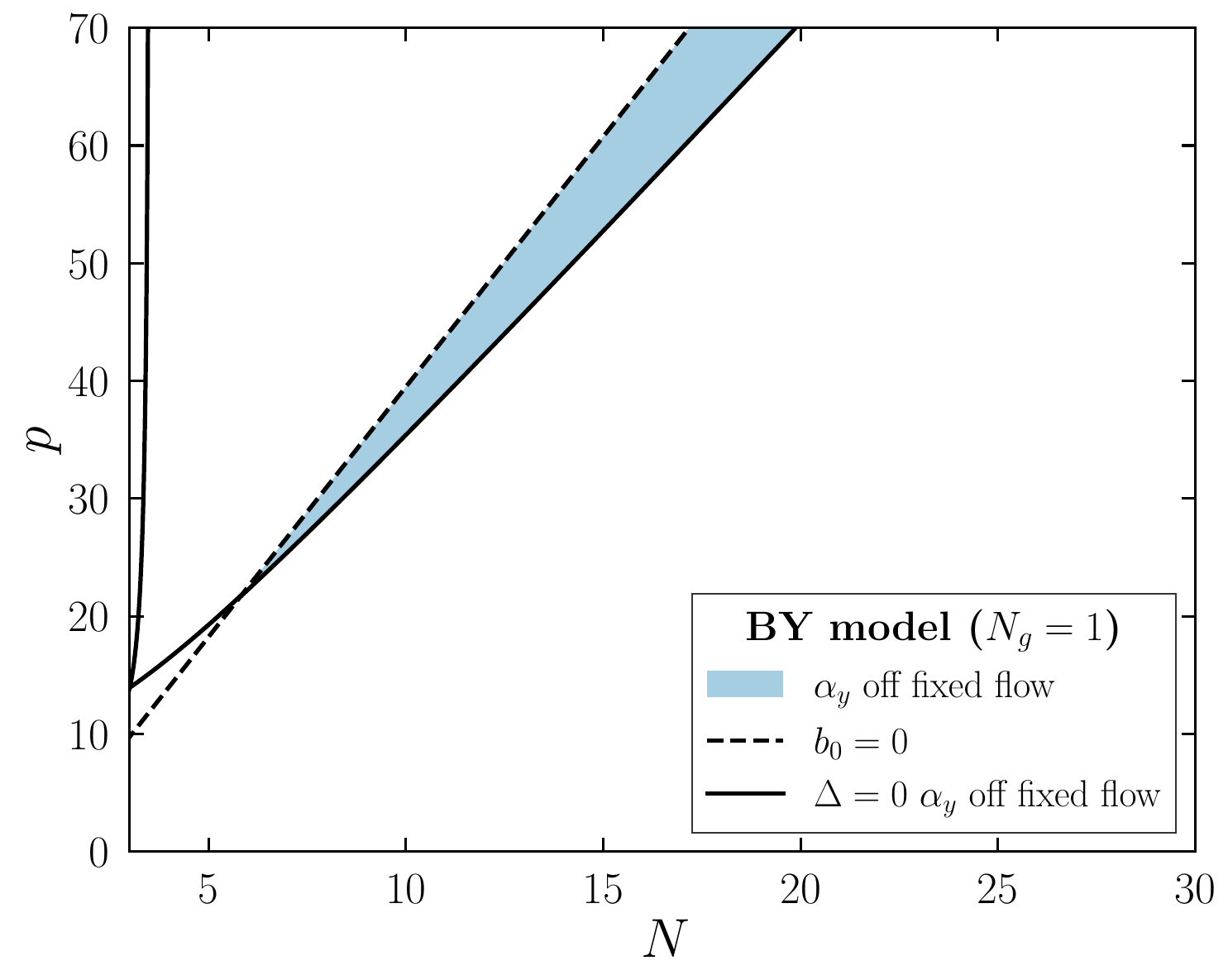}
        \caption{$N_g=1$} 
        \label{fig:gg_combined_ng1}
    \end{subfigure}
    \hspace{0.1cm}
    \begin{subfigure}[b]{0.49\textwidth}
        \centering
        \includegraphics[width=\textwidth]{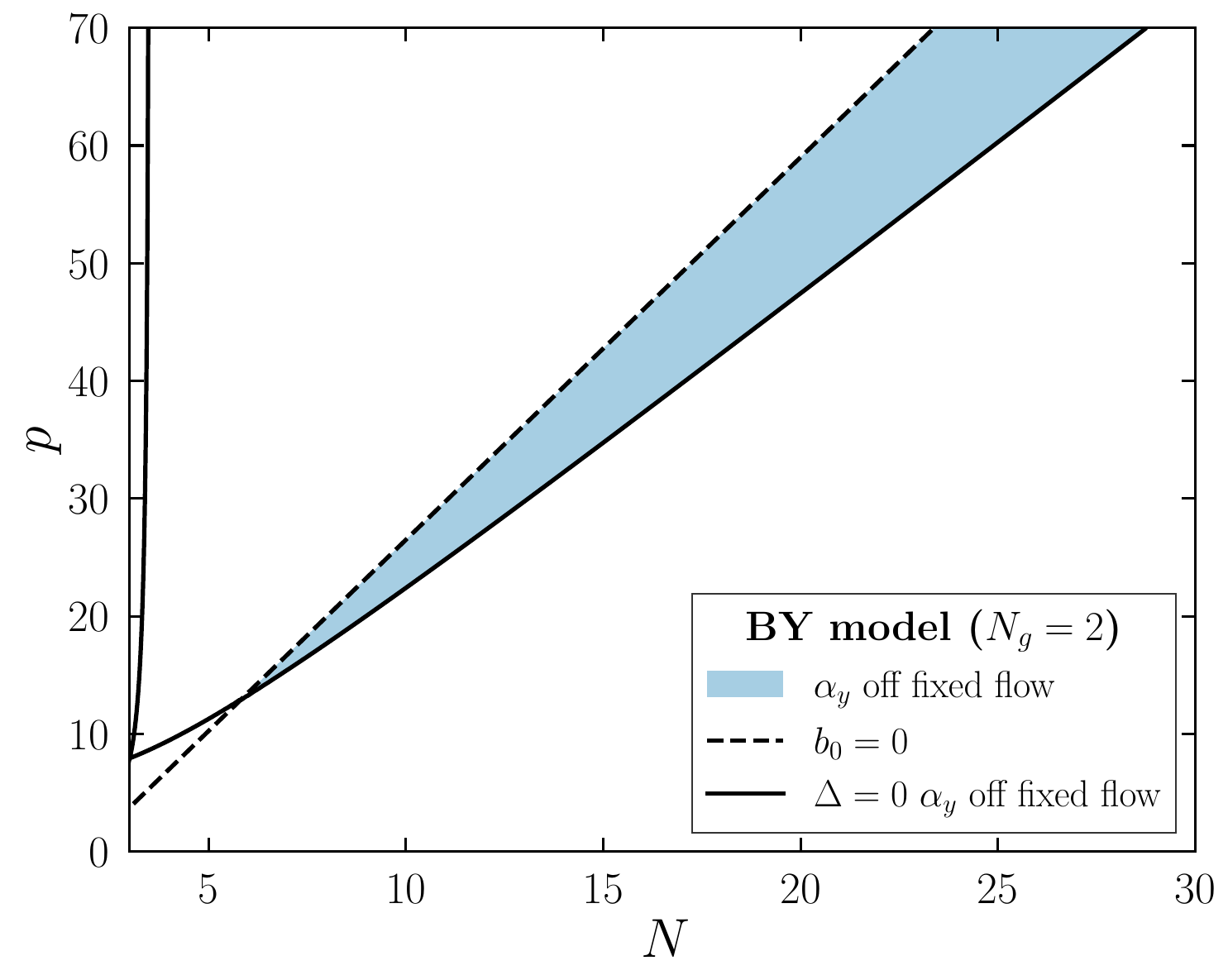}
        \caption{$N_g=2$} 
        \label{fig:gg_combined_ng3}
    \end{subfigure}
    
    \vspace{0.5cm} 
    \begin{subfigure}[b]{0.49\textwidth}
        \centering
        \includegraphics[width=\textwidth]{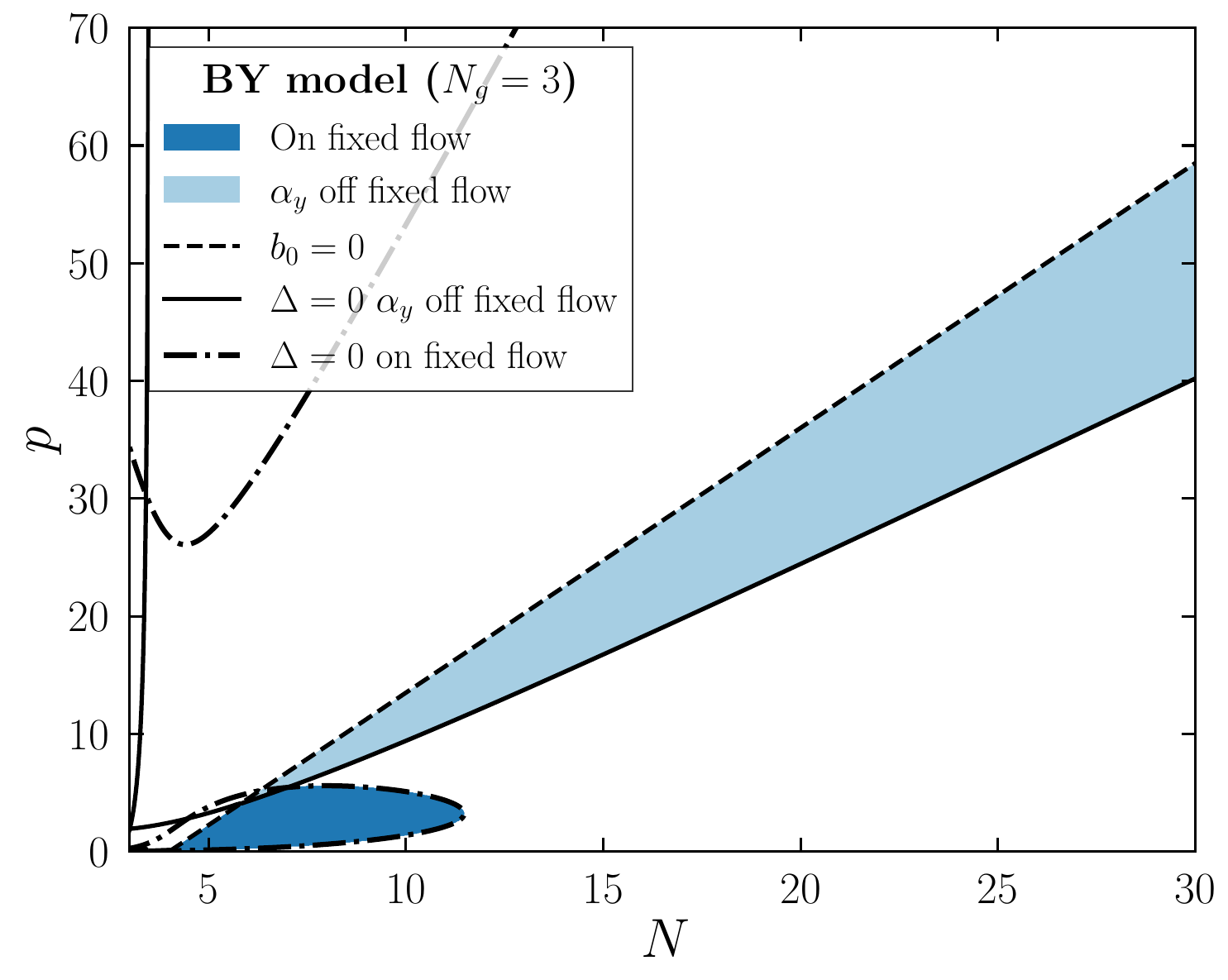}
        \caption{$N_g=3$} 
        \label{fig:by_combined_ng5}
    \end{subfigure}
    \hspace{0.1cm}
    \begin{subfigure}[b]{0.49\textwidth}
        \centering
        \includegraphics[width=\textwidth]{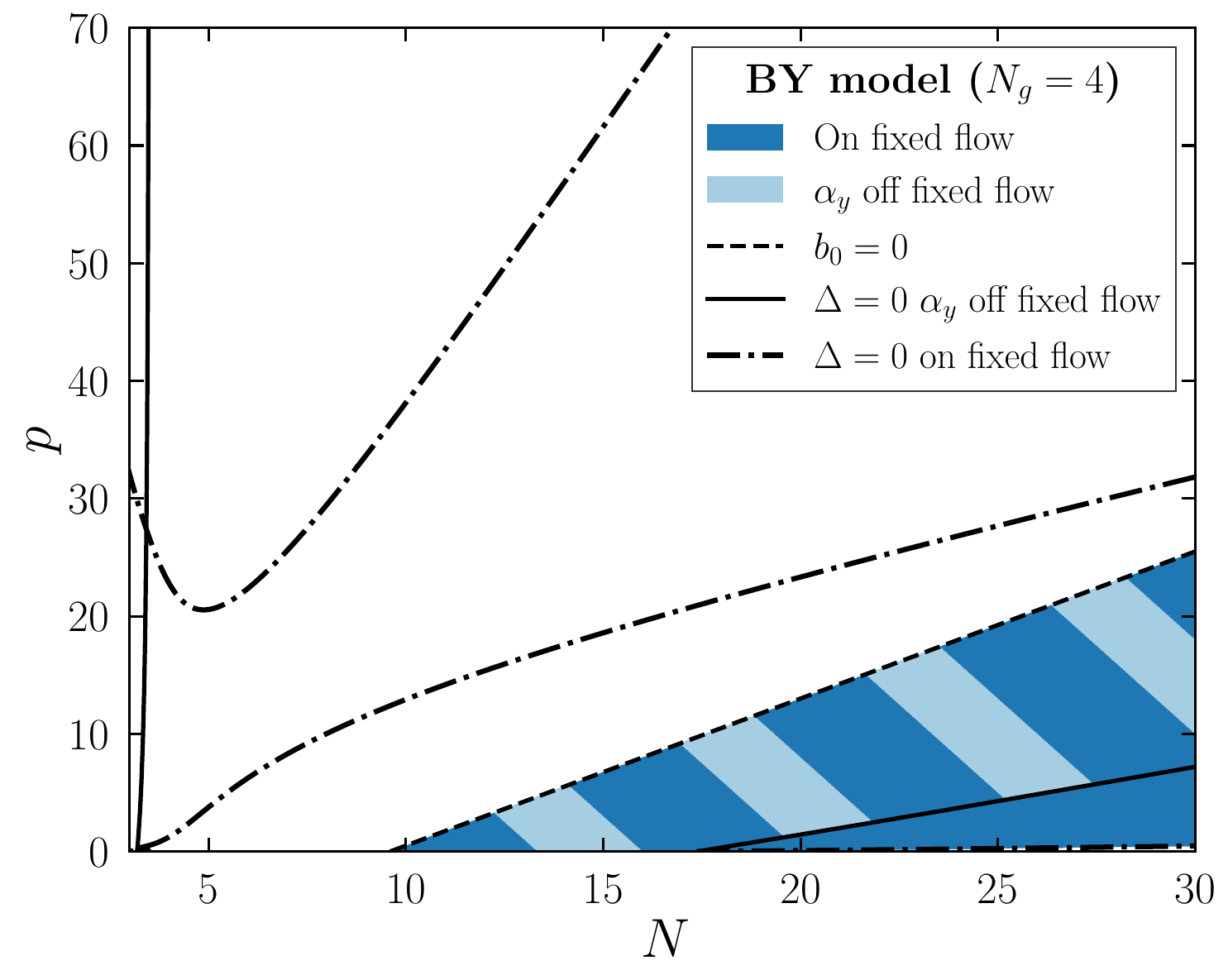}
        \caption{$N_g=4$}
        \label{fig:by_combined_ng6}
    \end{subfigure}
    
    \caption{CAF regions in the ($N$, $p$) parameter space for the BY model with adjoint scalar for $N_g = \{1, 2, 3, 4\}$. The dark blue area indicates the CAF region on the fixed flow, which is bounded above by the $b_0=0$ condition (dashed black line) in Eq.~\eqref{eq:padjscalar} and below by the $\Delta=0$ condition (black dash dotted line) with $\alpha_y$ on fixed flow in  Eq.~\eqref{eq:Delta0}. The light blue area highlights the region of CAF with the Yukawa coupling $\alpha_y$ off the fixed flow, bounded above by the $b_0=0$ condition and below by the $\Delta=0$ condition (solid black line) for $\alpha_y$ off fixed flow in Eq.~\eqref{eq:Delta0noY}. } 
    \label{fig:GGadjNG37}
\end{figure}
Matters change when the number of families increases. For instance, the case $N_g=5$ is illustrated in Figure \ref{fig:roots_fixed_ng3}, where the nature of the roots is shown. In this instance, we identify two regions with two real and two complex roots, one at large $p$ above the gauge asymptotic free region (hence excluded), and one within such region. Further, the polynomial has one sign change within this area, thus it has one positive real root as a consequence of Descartes' rule of signs. As shown in Figure \ref{fig:final_combined_ng3}, the GG model with $N_g=5$ has a CAF region on fixed flow highlighted in dark blue and starting at $N \geq 13$.
Off fixed flow, the situation is similar to the cases with lower $N_g$: a viable region exists below the dashed line, as shown by the light blue region in Figure \ref{fig:final_combined_ng3}. Similar solutions can be found for larger $N_g$, as illustrated in Figure \ref{fig:gg_combined_ng5} and \ref{fig:gg_combined_ng6} for $N_g=6,7$, respectively. Finally, for $N_g>13$, the gauge coupling loses asymptotic freedom.

For the BY models, the gauge coupling looses asymptotic freedom for $N \geq 5$, as was the case in Section \ref{sec:fundscalarNg}. In Figure \ref{fig:GGadjNG37}, the CAF region can be seen for the case of $N_g\in\{1,2,3,4\}$.

%% file: Sections/Conclusion.tex
\section{Conclusion}\label{sec:conc}
\begin{table}[ht]
\centering
\renewcommand{\arraystretch}{1.4}
\renewcommand{\tabularxcolumn}[1]{m{#1}} 

\begin{tabularx}{\textwidth}{@{}ccc >{\centering\arraybackslash}X >{\centering\arraybackslash}m{1.8cm} @{}}
\toprule
\multicolumn{1}{c}{\textbf{Model}} & \multicolumn{1}{c}{\textbf{CAF}} & \multicolumn{1}{c}{\textbf{$p(n)$}} & \multicolumn{1}{c}{\textbf{$(N,p)_{N_g}$}} & \multicolumn{1}{c}{\textbf{Figures}} \\
\midrule

\multicolumn{5}{c}{\textit{without scalars}} \\
\midrule
\rowcolor{myDarkerBlue} GG & CAF & \eqref{eq:pNnoscalarGG} & $(5,0)_1$ &--- \\

\rowcolor{myLightBlue} BY & CAF & \eqref{eq:pNnoscalarBY} & $(3,0)_1$ &--- \\

\midrule
\multicolumn{5}{c}{\textit{with one scalar in the fundamental representation - all on fixed flow}} \\
\midrule
\rowcolor{myDarkerBlue} & Not CAF: $N_g > 13$ & --- & --- & --- \\
\rowcolor{myDarkerBlue} \multirow{-2}{*}{GG} & CAF: $N_g\in\{1,\dots,13\}$ &\eqref{eq:pmaxfundNg} \& \eqref{eq:pminfundNgon} & $(5,6)_1$ $(5,5)_2$ $(5,4)_3$ $(5,3)_4$ $(5,2)_5$ $(5,0)_6$ $(5,0)_7$ $(5,0)_8$ $(5,0)_9$ $(5,0)_{10}$ $(5,0)_{11}$ $(5,0)_{12}$ $(5,0)_{13}$ &\ref{fig:cafgg_fund} \& \ref{fig:CAFGGNG9}  \\

\rowcolor{myLightBlue} & Not CAF: $N_g > 4$ & --- & --- & --- \\
\rowcolor{myLightBlue} \multirow{-2}{*}{BY} & CAF: $N_g\in\{1,\dots,4\}$  & \eqref{eq:pmaxfundNg}  & $(3,0)_1$ $(3,0)_2$ $(4,0)_3$ $(9,0)_4$ &\ref{fig:cafby_fund} \& \ref{fig:CAFNG9} \\

\midrule
\multicolumn{5}{c}{\textit{with one scalar in the fundamental representation - Yukawa off fixed flow}} \\
\midrule
\rowcolor{myDarkerBlue} & Not CAF: $N_g > 13$ & --- & --- & --- \\
\rowcolor{myDarkerBlue} \multirow{-2}{*}{GG} &CAF: $N_g\in\{1,\dots,13\}$  & \eqref{eq:pmaxfundNg} \& \eqref{eq:pminfundNg}& $(5,22)_1$ $(5,20)_2$ $(5,18)_3$ $(5,16)_4$ $(5,14)_5$ $(5,12)_6$ $(5,10)_7$ $(5,8)_8$ $(5,6)_9$ $(5,4)_{10}$ $(5,2)_{11}$ $(5,0)_{12}$ $(5,0)_{13}$ &\ref{fig:cafgg_fund} \& \ref{fig:CAFGGNG9} \\

\rowcolor{myLightBlue} & Not CAF: $N_g > 4$ & --- & --- & --- \\
\rowcolor{myLightBlue} \multirow{-2}{*}{BY} & CAF: $N_g\in\{1,\dots,4\}$  &\eqref{eq:pmaxfundNg} \& \eqref{eq:pminfundNg} & $(3,10)_1$ $(3,4)_2$ $(4,0)_3$ $(9,0)_4$   &\ref{fig:cafby_fund} \& \ref{fig:CAFNG9} \\

\bottomrule
\end{tabularx}
\caption{Summary table of CAF solutions for the GG and BY models with varying multiplicity of chiral families ($N_g$) and vector-like families ($p$). The table shows the CAF status, the equations containing the boundary functions $p(N,N_g)$ that define the CAF regions, the minimal $N$ for each $N_g$ and the required minimal $p$ displayed as $(N,p)_{N_g}$, and links to the relevant figures.\label{tab:conc_part1}}
\end{table}

We studied the short-distance behaviour of chiral gauge-Yukawa theories in the quest for complete asymptotic freedom in all couplings. Motivated by simplicity and grand unified theories, we focused on the generalised Georgi--Glashow and Bars--Yankielowicz models. Besides varying multiplicity of chiral families, we also included vector-like copies in the fundamental representation, as well as one single scalar either in the fundamental or adjoint gauge representation.
We employed general CAF conditions to identify the parameter space where all couplings run to zero at high energies, distinguishing cases where all couplings vanish at the same speed as the gauge one (on fixed flow), or when the Yukawa coupling vanishes faster than the gauge one (off fixed flow).

We discovered that all considered classes of models support complete asymptotically free dynamics, within specific parameter regions in the number of colours, as well as multiplicity of chiral and vector-like families.  
We summarised our results in Table~\ref{tab:conc_part1} for models with a scalar in the fundamental and in Table~\ref{tab:conc_part2} for a scalar in the adjoint. In particular, the fourth column contains the minimal models in terms of the number of colours and vector-like families, for all the allowed numbers of chiral families.   
 
This classification has potential applications in grand unification theories when seeking a simple UV completion of the models. Of particular interest is the Georgi--Glashow model with 3 generations: we find that a CAF dynamics can be found with a fundamental scalar when at least $4$ vector-like families are added on fixed flow, while this number increases to $18$ for off-fixed-flow dynamics. Instead, no such solutions were found for models with the adjoint scalar, where the minimal model with three chiral families would be based on $SU(7)$. We recall that the fundamental scalar contains the usual Higgs doublet of the Standard Model, hence it is employed to trigger the electroweak symmetry breaking, while the adjoint is commonly used to break $SU(5)$ to the standard model gauge group. A realistic model would require both scalars, with a complex coupling structure, hence our results should be considered as a first attempt to find completely asymptotically free grand unified theories.

The case with a single adjoint, moreover, could be relevant for UV completions of the standard model based on non-supersymmetric gauge dualities, see Refs~\cite{Mojaza:2011rw,Sannino:2011mr,Cacciapaglia:2024mfy}. In such models, Yukawa couplings are generated at low energy by the duality, hence dual grand unified theories only require a single scalar for breaking the gauge theory. 

Chiral gauge-Yukawa theories constitute a cornerstone of our understanding of particle interactions, starting from the Standard Model itself. Their dynamics remain mysterious under various aspects. Our results offer a first glimpse of their short-distance behaviour, where complete asymptotic freedom of all couplings renders them truly fundamental (i.e. valid at arbitrarily high energies). The low-energy dynamics of purely fermionic chiral gauge theories remain only partly understood, even for paradigmatic examples such as the Georgi–-Glashow and Bars–-Yankielowicz theories. A first comprehensive investigation of these systems, including an attempt to delineate their conformal window, was carried out in Ref.~\cite{Appelquist:2000qg}, further complementing studies of the conformal window in vector-like gauge theories \cite{Sannino:2004qp,Dietrich:2006cm}.

Our results point towards new classes of fundamental gauge-Yukawa theories, with light scalars, whose low-energy dynamics remain utterly unexplored.

\begin{table}[ht]
\centering
\renewcommand{\arraystretch}{1.4}
\renewcommand{\tabularxcolumn}[1]{m{#1}} 

\begin{tabularx}{\textwidth}{@{}ccc >{\centering\arraybackslash}X >{\centering\arraybackslash}m{1.8cm} @{}}
\toprule
\multicolumn{1}{c}{\textbf{Model}} & \multicolumn{1}{c}{\textbf{CAF}} & \multicolumn{1}{c}{\textbf{$p(n)$}} & \multicolumn{1}{c}{\textbf{$(N,p)_{N_g}$}} & \multicolumn{1}{c}{\textbf{Figures}} \\
\midrule
\multicolumn{5}{c}{\textit{with one scalar in the adjoint representation - all on fixed flow}} \\
\midrule
\rowcolor{myDarkerBlue} & Not CAF: $N_g \in \mathbb{N} \setminus \{5,\dots,12\}$ &--- &--- &--- \\
\rowcolor{myDarkerBlue} \multirow{-2}{*}{GG} & CAF: $N_g\in\{5,\dots,12\}$ &\eqref{eq:padjscalar} \& \eqref{eq:Delta0} & $(13,2)_5$ $(9,1)_6$ $(7,1)_7$ $(6,1)_8$ $(6,1)_9$ $(6,1)_{10}$ $(5,1)_{11}$ $(5,1)_{12}$ & \ref{fig:final_combined_ng3}, \ref{fig:gg_combined_ng5} \& \ref{fig:gg_combined_ng6}\\

\rowcolor{myLightBlue} &Not CAF: $N_g \in \mathbb{N} \setminus \{3,4\}$ &--- &--- &--- \\
\rowcolor{myLightBlue} \multirow{-2}{*}{BY} & CAF: $N_g\in\{3,4\}$ & \eqref{eq:padjscalar} \& \eqref{eq:Delta0} &$(5,1)_3$ $(10,0)_4$ &\ref{fig:by_combined_ng5} \& \ref{fig:by_combined_ng6} \\

\midrule
\multicolumn{5}{c}{\textit{with one scalar in the adjoint representation - Yukawa coupling off fixed flow}} \\
\midrule
\rowcolor{myDarkerBlue} & Not CAF: $N_g > 9$ & --- & --- & --- \\
\rowcolor{myDarkerBlue} \multirow{-2}{*}{GG} & CAF: $N_g \in \{1,\dots,9\}$ & \eqref{eq:padjscalar} \& \eqref{eq:Delta0noY} & $(7,32)_1$ $(7,28)_2$ $(7,24)_3$ $(7,20)_4$ $(7,16)_5$ $(7,12)_6$ $(7,8)_7$ $(7,4)_8$ $(7,0)_9$ & \ref{fig:BYposroots}, \ref{fig:final_combined_ng2}, \ref{fig:final_combined_ng3}, \ref{fig:gg_combined_ng5} \& \ref{fig:gg_combined_ng6}\\

\rowcolor{myLightBlue} & Not CAF: $N_g > 4$ & --- & --- & --- \\
\rowcolor{myLightBlue} \multirow{-2}{*}{BY} & CAF: $N_g \in \{1,\dots,4\}$ & \eqref{eq:padjscalar} \& \eqref{eq:Delta0noY} & $(7,26)_1$ $(7,16)_2$ $(7,6)_3$ $(10,0)_4$ &  \ref{fig:GGadjNG37} \\



\bottomrule
\end{tabularx}
\caption{Same as Table~\ref{tab:conc_part1} for GG and BY models with adjoint scalar.
\label{tab:conc_part2}}
\end{table}
\subsubsection*{Acknowledgments}
The work of F.S. and S.W. is partially supported by the Carlsberg Foundation, Semper Ardens grant CF22-0922.

%% file: Sections/Appendix.tex
\appendix
\section{Detailed CAF analysis}
\subsection{CAF Analysis from Solving ODE's} \label{app:caf}
This approach considers coupled ODE's governing the beta functions, which admit analytic solutions. 
\subsubsection{Yukawa Coupling}
The RG equation, which governs the Yukawa coupling, is given by
\begin{equation}
    \mu \frac{\text{d}\alpha_y}{\text{d}\mu}=\alpha_y(c_1\alpha_g+c_2\alpha_y)\;, \label{eq:betay1loopapp}
\end{equation}
where $\alpha_y=\frac{y^2}{(4\pi)^2}$, and $c_1$, $c_2$ are constants, where in general $c_1<0$ and $c_2>0$.  

We have a system of two coupled differential equations, one for the gauge coupling (see Eq.\ \eqref{eq:betag1loopcaf}), the other for the Yukawa coupling (see Eq.\ \eqref{eq:betay1loopapp}). The ratio of the Yukawa Eq.\ \eqref{eq:betay1loopapp} and gauge Eq.\ \eqref{eq:betag1loopcaf} beta functions, are given by
\begin{align}
   \frac{\text{d}\alpha_y}{\text{d}\alpha_g}&= \frac{\alpha_y}{b_0 \alpha_g} \left( c_1 + c_2 \frac{\alpha_y}{\alpha_g} \right)\;.
\end{align}
The solutions to the differential equation depend on $b_0$ and $c_1$. 

\paragraph{Case 1 $b_0=c_1$:} In this case the solution is given by 
\begin{equation}
    \alpha_y=\frac{\alpha_{y_0}}{\alpha_{g_0}\left(1+\frac{c_2}{c_1}\frac{\alpha_{y_0}}{\alpha_{g_0}}\log\alpha_{g_0}\right)-\frac{c_2}{c_1}\alpha_{y_0}\log \alpha_g }\alpha_g\;.\label{eq:ysolb0c1}
\end{equation}
For $\alpha_g\to0$ the denominator will be dominated by the second term  $(-\frac{c_2}{c_1}\alpha_{y_0}\log \alpha_g) $, as the logarithm of the gauge coupling will go towards negative infinity. Since the constant in front will always be negative $\frac{c_2}{c_1}<0$, the $\alpha_y$ coupling will be negative.

\paragraph{Case 2 $b_0\neq c_1$:} The solution changes to
\begin{equation}
    \alpha_y=\frac{\alpha_{y_0}}{ \alpha_g^{1-\frac{c_1}{b_0}}\alpha_{g_0}^{\frac{c_1}{b_0}}\left(1-\frac{c_2}{b_0-c_1}\frac{\alpha_{y_0}}{\alpha_{g_0}}\right)+\frac{c_2}{b_0-c_1}\alpha_{y_0} }\alpha_g\;,
\end{equation}
with the initial condition $\alpha_y(\alpha_{g_0})=\alpha_{y_{0}}$. The equation can be divided into two parts, one which will be dependent on $\alpha^{\frac{c_1}{b_0}}_g$ and one $\alpha_g$. If $\frac{c_1}{b_0}>1$, the first term will dominate, and for $\frac{c_1}{b_0}<1$ the second term will dominate. The case of $\frac{c_1}{b_0}=1$ has already been ruled out from the analysis of Eq.\ \eqref{eq:ysolb0c1}. In the case of $\frac{c_1}{b_0}<1$, the coupling will go as \begin{equation}
    \alpha_y\sim\frac{b_0-c_1}{c_2}\alpha_g\;,\quad\quad {c_1}>|b_0|>0\;,
\end{equation}
which will always be negative, since $b_0-c_1<0$. This case will thus not have an asymptotically free Yukawa coupling. The last possible case is the case of $\frac{c_1}{b_0}>1$ or $b_0-c_1>0$. In this case, the coupling will go as
\begin{equation}
    \alpha_y\sim\frac{\alpha_{y_0}}{ \alpha_{g_0}^{\frac{c_1}{b_0}}\left(1-\frac{c_2}{b_0-c_1}\frac{\alpha_{y_0}}{\alpha_{g_0}}\right)}\alpha_g^{\frac{c_1}{b_0}}\;,\quad\quad \frac{\alpha_{g_0}}{\alpha_{y_0}}>\frac{c_2}{b_0-c_1}\;,\quad\quad c_1<b_0<0\;.\label{eq:appYukawacondoff}
\end{equation}
To ensure that the coupling will be positive, we need to constrain the values of the couplings at the arbitrary energy level $\frac{c_2}{b_0-c_1}\frac{\alpha_{y_0}}{\alpha_{g_0}}<1$. In the UV, where $\alpha_g\to 0$, the Yukawa coupling will also be asymptotically free, and because $\frac{c_1}{b_0}>1$, the Yukawa coupling will go faster to zero than the gauge coupling at high energies. Another possibility is that the values of the couplings at the arbitrary energy level are fine-tuned $\frac{\alpha_{g_0}}{\alpha_{y_0}}=\frac{c_2}{b_0-c_1}$. In this case, the first term in the denominator vanishes, and the previously dominated term is now the only one left:
\begin{equation}
    \alpha_y=\frac{b_0-c_1}{ c_2 }\alpha_g\;,\quad\quad \frac{\alpha_{g_0}}{\alpha_{y_0}}=\frac{c_2}{b_0-c_1}\;,\quad\quad b_0-c_1>0\;.\label{eq:ygfixed2}
\end{equation}
This case is known as fixed flow \cite{Giudice:2014tma}. On fixed flow, the Yukawa coupling vanishes at the same order as the gauge coupling. Now that we have the CAF conditions for the case of one Yukawa and one gauge coupling, let us turn to the case which includes the scalar.

\subsubsection{Quartic Scalar Coupling}
The remaining coupling can now be added to the analysis, the self-coupling
\begin{equation}
    \mu \frac{\text{d} \alpha_\lambda}{\text{d}\mu}=\alpha_\lambda\left(d_1\alpha_\lambda + d_2\alpha_g+d_3\alpha_y\right)+d_4\alpha_g^2+d_5\alpha_y^2\;.
\end{equation}
This coupling is dependent on both the other two couplings, with the coefficients $d_1,d_3,d_4\ge0$ and $d_2,d_5\le 0$.

To solve the coupled differential equations we consider the ratio $\frac{\beta_\lambda}{\beta_g}$:
\begin{equation}
    \frac{\text{d} \alpha_\lambda}{\text{d}\alpha_g}=\frac{\alpha_\lambda}{b_0\alpha_g}\left(d_1\frac{\alpha_\lambda}{\alpha_g} + d_2+d_3\frac{\alpha_y}{\alpha_g}\right)+\frac{d_4}{b_0}+\frac{d_5\alpha_y^2}{b_0\alpha_g^2}\;.\label{eq:appodelam}
\end{equation}
Recall that the Yukawa coupling can be asymptotically free in two cases: on or off fixed flow with the gauge coupling. First, we consider the case of the Yukawa coupling off fixed flow. Then the differential equation simplifies to
\begin{equation}
    \frac{\text{d} \alpha_\lambda}{\text{d}\alpha_g}=\frac{\alpha_\lambda}{b_0\alpha_g}\left(d_1\frac{\alpha_\lambda}{\alpha_g} + d_2\right)+\frac{d_4}{b_0}\;.
    \label{eq:betalamfracapp}
\end{equation}
To find solutions, we will first define $k = (b_0 - d_2)^2 - 4d_1 d_4$, and $k_0 = (b_0 - d_2)\alpha_{g_0} - 2d_1\alpha_{\lambda0}$.
This will be useful for finding $\alpha_\lambda$ as a function of $\alpha_g$ for cases $k<0$, $k>0$, and $k=0$.
\paragraph{Case 1 $k < 0$:}
We start with the general solution to the differential equation of $\alpha_\lambda(\alpha_g)$ (Equation \ref{eq:betalamfracapp}):
\begin{align}
    \alpha_{\lambda} (\alpha_{g}) = \frac{\alpha_{g}}{2\alpha_{1}}\Bigg[ b_0& - d_2 + 
    \sqrt{-k} \tan \bigg( \frac{1}{2} \bigg( \frac{\sqrt{-k} \log (\alpha_{g})}{b_0} 
    \\-&\sqrt{-k} \bigg( 2b_0 \sqrt{-k} \tan^{-1} \bigg( \frac{(b_0 - d_2) \alpha_{g} \sqrt{-k} - 2d_1 \sqrt{-k} \alpha_{\lambda_0}}{k \alpha_{g_0}} \bigg) + k \log (\alpha_{g_0}) 
    \bigg)\bigg) \frac{1}{b_0 k} \bigg) \Bigg]\;,\nonumber
\end{align}
where the definition of $k_0$ can be used to simplify the expression, yielding
\begin{equation}
    \boxed{\alpha_{\lambda} = \frac{\alpha_{g}}{2d_{1}} \left[ b_0 - d_2 + \sqrt{-k} \tan\left( \frac{\sqrt{-k}}{2b_0} \log\frac{\alpha_g}{\alpha_{g_0}} - \tan^{-1}\frac{k_0}{\sqrt{-k\alpha_{g_0}}} \right) \right] , \quad k<0 }
    \label{eq:alphalamkl0}
\end{equation}
This solution only holds for \( k < 0 \), because we want a real result. In this case, the coupling will reach a Landau pole, because of the nature of the tangent function as one varies the gauge coupling. Thus, no possible CAF regions are found when $k<0$.

\paragraph{Case 2 $k > 0$:}
For $k>0$, we rewrite the square root $\sqrt{-k} = i\sqrt{k}$ which can be inserted in the solution for $k<0$
\begin{align}
    \alpha_{\lambda}(\alpha_g) = \frac{\alpha_g}{2\alpha_1} \left[ b_0 - d_2 + i\sqrt{k} \tan \left( \frac{i\sqrt{k}}{2b_0} \log \left( \frac{\alpha_g}{\alpha_{g_0}} \right) + \tan^{-1} \left( \frac{k_0}{i\sqrt{k} \alpha_{g_0}} \right) \right) \right].
\end{align}
Using the identities \(\tan(iz) = i\tanh(z)\) and \(\tan^{-1}(iz) = i\tanh^{-1}(z)\), we rewrite the argument of the tangent. Let
\begin{align}
z = \frac{\sqrt{k}}{2b_0} \log \left( \frac{\alpha_g}{\alpha_{g_0}} \right) + \tanh^{-1} \left( \frac{k_0}{\sqrt{k} \alpha_{g_0}} \right).
\end{align}
Using the identity and substituting back in, we obtain
\begin{align}
\alpha_{\lambda}(\alpha_g) = \frac{\alpha_g}{2\alpha_1} \left[ b_0 - d_2 - \sqrt{k} \tanh(z) \right].
\end{align}
To express the hyperbolic tangent of a sum, the following relation is used
\begin{align}
\tanh(A+B) = \frac{\tanh(A) + \tanh(B)}{1 + \tanh(A)\tanh(B)}\;,
\end{align}
where
\begin{align}
A = \frac{\sqrt{k}}{2b_0} \log \left( \frac{\alpha_g}{\alpha_{g_0}} \right), \quad 
B = \tanh^{-1} \left( \frac{k_0}{\sqrt{k}\alpha_{g_0}} \right)\;.
\end{align}
Further, using the definition of $\tanh\left(x\right)=\frac{e^{2x}-1}{e^{2x}+1}$, the hyperbolic tangent of $A+B$ is thus
\begin{align}
    \tanh(A+B)=&\frac{ \frac{ \left( \frac{\alpha_g}{\alpha_{g_0}}\right)^{\frac{\sqrt{k}}{b_0}} - 1 }{ \left(\frac{\alpha_g}{\alpha_{g_0}} \right)^{\frac{\sqrt{k}}{b_0}} + 1 } + \frac{k_0}{\sqrt{k} \alpha_{g_0}} }{ 1 +\frac{ \left( \frac{\alpha_g}{\alpha_{g_0}} \right)^{\frac{\sqrt{k}}{b_0}} - 1 }{ \left(\frac{\alpha_g}{\alpha_{g_0}} \right)^{\frac{\sqrt{k}}{b_0}} + 1 }  \frac{k_0}{\sqrt{k} \, \alpha_{g_0}} }
    = \frac{
  \left( \left( \frac{\alpha_g}{\alpha_{g_0}}  \right)^{\frac{\sqrt{k}}{b_0}} - 1 \right) 
  + \frac{k_0}{\sqrt{k}\alpha_{g_0}} \left( \left( \frac{\alpha_g}{\alpha_{g_0}} \right)^{\frac{\sqrt{k}}{b_0}} + 1 \right)
}{
  \left( \left( \frac{\alpha_g}{\alpha_{g_0}}  \right)^{\frac{\sqrt{k}}{b_0}} + 1 \right) 
  + \frac{k_0}{\sqrt{k} \, \alpha_{g_0}} \left( \left( \frac{\alpha_g}{\alpha_{g_0}}  \right)^{\frac{\sqrt{k}}{b_0}} - 1 \right)
}\nonumber
\\&=  \frac{
   (k_0+\sqrt{k}\alpha_{g_0})+(k_0- \sqrt{k}\alpha_{g_0})\left(  \frac{\alpha_g}{\alpha_{g_0}}  \right)^{-\frac{\sqrt{k}}{b_0}}
}{
  (k_0+\sqrt{k}\alpha_{g_0})-(k_0- \sqrt{k}\alpha_{g_0} )\left(  \frac{\alpha_g}{\alpha_{g_0}}  \right)^{-\frac{\sqrt{k}}{b_0}} 
}\;.
\end{align}
Which gives the solution for $k>0$
\begin{equation}
    \boxed{\alpha_{\lambda} = \frac{\alpha_g}{2d_1} \left( b_0 - d_2 - \sqrt{k} \, \frac{ (k_0 + \sqrt{k}\alpha_{g_0}) + (k_0 - \sqrt{k}\alpha_{g_0}) \left( \frac{\alpha_g}{\alpha_{g_0}} \right)^{\sqrt{k}/b_0} }{ (k_0 + \sqrt{k}\alpha_{g_0}) - (k_0 - \sqrt{k}\alpha_{g_0}) \left( \frac{\alpha_g}{\alpha_{g_0}} \right)^{\sqrt{k}/b_0} } \right) , \quad k>0 }
    \label{eq:alphalamkg0}
\end{equation}
Let us now do the CAF analysis for this case. From Eq.\ \eqref{eq:alphalamkg0}, it can be seen that there are two possible fixed flows between the gauge and self-coupling. Namely, when $k_0+\sqrt{k}\alpha_g=0$ or when $k_0-\sqrt{k}\alpha_g=0$ which gives us the expression
\begin{equation}
    \alpha_\lambda =\frac{b_0-d_2\pm\sqrt{k}}{2d_1}\alpha_g\;.\label{eq:lamgfixed}
\end{equation}
If we only consider solutions for $\alpha_\lambda>0$ to be relevant solutions, we will see that these fixed-flow lines also create the boundaries of the asymptotically free region of the quartic coupling. First, if $\frac{b_0-d_2-\sqrt{k}}{2d_1}>0$, then it is asymptotically free along both fixed-flow lines. If $\frac{b_0-d_2+\sqrt{k}}{2d_1}>0$ and $\frac{b_0-d_2-\sqrt{k}}{2d_1}<0$, it will only be asymptotically free along one direction. Lastly, if $\frac{b_0-d_2+\sqrt{k}}{2d_1}<0$, it will not be asymptotically free along any of the fixed-flow directions. 

To ensure that no Landau poles are encountered within the fixed-flow lines, we consider the denominator of Eq.\ \eqref{eq:alphalamkg0} to find other possible asymptotically free flows. Take the denominator to be zero, then the gauge coupling has to uphold
\begin{equation}
    \left(\frac{\alpha_g}{\alpha_{g_0}}\right)^{-\frac{\sqrt{k}}{b_0}}=\frac{k_0+\sqrt{k}\alpha_{g_0}}{k_0-\sqrt{k}\alpha_{g_0}}\;.
\end{equation}
Thus, if there are any solutions, there will be Landau poles. First, the solution of $k_0+\sqrt{k}\alpha_{g_0}=0$ is not relevant, because this zero will be cancelled by a zero in the numerator, as shown by the fixed-flow analysis. Thus, for $\alpha_g=0$ there are no Landau poles. However, for $\alpha_g>0$ there still might be poles. Since the LHS will be positive, so must the RHS, this means that both the numerator and denominator must have the same sign. Furthermore, we have $\alpha_g\to0$, ensuring that the LHS is less than one. Essentially, for a pole in the UV, the RHS will have to be less than one, and if the RHS is larger than one, we have a pole in the IR. Thus, for a pole in the UV
\begin{equation}
    0<\frac{k_0+\sqrt{k}\alpha_{g_0}}{k_0-\sqrt{k}\alpha_{g_0}}<1\;,
\end{equation}
For this to be the case, both the denominator and the numerator would have to be negative, for the numerator to be less than the denominator. Thus, to ensure no poles in the UV, we required that $k_0+\sqrt{k}\alpha_{g_0}>0$. Considering possible poles in the IR, the condition for a pole will be
\begin{equation}
    1<\frac{k_0+\sqrt{k}\alpha_{g_0}}{k_0-\sqrt{k}\alpha_{g_0}}\;.
\end{equation}
For this to be the case, we need both numerator and denominator to be positive. Since the numerator needs to be positive for no poles in the UV. We can ensure no poles in the IR by constraining the denominator $k_0-\sqrt{k}\alpha_{g_0}<0$. The reason why we need to have no poles in the IR, even though we are only interested in asymptotic freedom, is that the coupling would blow up at minus infinity, creating a potential which is unbounded from below. To ensure a bounded potential when having only one quartic coupling, we need the coupling to be positive at all RG scales. 

Thus, combining this analysis with Eq.\ \eqref{eq:lamgfixed} the restriction on $\alpha_{\lambda_0}$ and $\alpha_{g_0}$ is given by
\begin{equation}
    \frac{b_0-d_2-\sqrt{k}}{2d_1}\le \frac{\alpha_{\lambda_0}}{\alpha_{g_0}}\le \frac{b_0-d_2+\sqrt{k}}{2d_1}\;.
\end{equation}
We also need to ensure a positive quartic coupling at all scales, resulting in the condition of a positive upper boundary
\begin{equation}
    {b_0-d_2+\sqrt{k}}>0\;.
\end{equation}
Thus, the quartic coupling can only be asymptotically free within and on the fixed-flow lines, and will have at least one positive asymptotically free solution if
\begin{equation}
    k>0\;,\quad\quad  \frac{b_0-d_2-\sqrt{k}}{2d_1}\le \frac{\alpha_{\lambda_0}}{\alpha_{g_0}}\le \frac{b_0-d_2+\sqrt{k}}{2d_1}, \quad\quad {b_0-d_2+\sqrt{k}}>0\;.
\end{equation}

\paragraph{Case 3 $k = 0$:}
To find the solution of the last case $\lim_{k \to 0^-}(\alpha_\lambda)$ we consider the solution $k<0$ (see Eq.\ \eqref{eq:alphalamkg0}), with the approximation $\sqrt{-k} =\epsilon\ll1$. The argument of $\tan^{-1}$ is then
\begin{align}
z &= -\tan^{-1}\left( \frac{k_0}{\epsilon\alpha_{g_0}} \right) + \frac{\epsilon}{2b_0} \log\frac{\alpha_g}{\alpha_{g_0}} .
\end{align}
Using that for large $z$, $\tan^{-1}(z) = \frac{\pi}{2} - \frac{1}{z} + O(z^{-3})$, we have
\begin{align}
z = -\frac{\pi}{2} +\epsilon\left( \frac{\alpha_{g_0}}{k_0} + \frac{1}{2b_0}\log\frac{\alpha_g}{\alpha_{g_0}}\right) + O(\epsilon^3) .
\end{align}
Let us call the second term for $\epsilon A$ where $A = \frac{\alpha_{g_0}}{k_0} + \frac{1}{2b_0}\log\frac{\alpha_g}{\alpha_{g_0}}$. Then for $\epsilon$ small
\begin{align}
\tan z = \tan\left(-\frac{\pi}{2} + \epsilon A\right) = -\cot\epsilon A = -\frac{1}{\epsilon A} + O(\mu) = -\frac{1}{\epsilon A} + O(\epsilon) .
\end{align}
Using that in front of tangent, there is a factor $\sqrt {-k}=\epsilon$
\begin{align}
\epsilon \tan z = \epsilon \left( -\frac{1}{\epsilon A} + O(\epsilon) \right) = -\frac{1}{A} + O(\epsilon^2) .
\end{align}
Inserting this back in the expression for $\alpha_\lambda$, we find
\begin{align}
\alpha_{\lambda} = \frac{\alpha_g}{2d_1} \left( b_0 - d_2 - \frac{1}{A} \right) = \frac{\alpha_g}{2d_1} \left( b_0 - d_2 - \frac{1}{ \frac{\alpha_{g_0}}{k_0} + \frac{1}{2b_0}\ln\frac{\alpha_g}{\alpha_{g_0}} } \right) .
\end{align}
From this we can impose the initial condition $\alpha_{\lambda} (\alpha_{g_0}) = \alpha_{\lambda0}$ to be
\begin{align}
\alpha_{\lambda0} = \frac{\alpha_{g_0}}{2d_1} \left( b_0 - d_2 - \frac{k_0}{\alpha_{g_0}} \right) .
\end{align}
This gives us an expression for $\frac{\alpha_{g_0}}{k_0}$ which we find to be
\begin{align}
    \frac{\alpha_{g_0}}{k_0} = \frac{1}{b_0 - d_2 - \frac{2d_1\alpha_{\lambda0}}{\alpha_{g_0}}} \;.\label{eq:appk0def}
\end{align}
Inserting this back into the expression for $\alpha_\lambda$ and simplifying, we obtain
\begin{align}
    \alpha_\lambda &= \frac{\alpha_g}{2d_1} \left[ b_0 - d_2 - \frac{1}{\frac{1}{\frac{-2d_1\alpha_{\lambda_0}}{\alpha_{g_0}} + b_0 - d_2}+ \frac{1}{2b_0} \ln \frac{\alpha_q}{\alpha_{g_0}}}  \right]\;,\nonumber
    \\&=\frac{\alpha_g}{2d_1} \left[ \frac{\frac{4b_0d_1\alpha_{\lambda_0}}{\alpha_{g_0}}+\left(\frac{-2d_1\alpha_{\lambda_0}}{\alpha_{g_0}} + b_0 - d_2\right)\ln \frac{\alpha_g}{\alpha_{g_0}}(b_0-d_2)}{{2b_0} + \left(\frac{-2d_1\alpha_{\lambda_0}}{\alpha_{g_0}} + b_0 - d_2\right) \ln \frac{\alpha_g}{\alpha_{g_0}}} \right]\label{eq:appk0}
\end{align}
Using the definition of $k_0$ in Eq.\ \eqref{eq:appk0def}, Eq.\ \eqref{eq:appk0} simplifies to
\begin{equation}
    \boxed{\alpha_{\lambda} = \frac{\alpha_g}{2d_1} \frac{ 4b_0 d_1 \alpha_{\lambda0} + k_0 (b_0 - d_2) \ln\frac{\alpha_g}{\alpha_{g_0}} }{ 2b_0 \alpha_{g_0} + k_0 \ln\frac{\alpha_g}{\alpha_{g_0}} } , \quad k=0 }
    \label{eq:alphalamkeq0}
\end{equation}
We have now obtained the last of the analytic solutions to Eq.\ \eqref{eq:betalamfracapp}, and we can move forward with the final CAF analysis. 

First, consider the denominator to be zero, this requirement yields
\begin{equation}
    \ln\frac{\alpha_g}{\alpha_{g_0}}=-\frac{2b_0\alpha_{g_0}}{k_0}\;.
\end{equation}
For $\alpha_g\ge0$, we again have two possible solutions, which will give Landau poles. For $\alpha_g<\alpha_{g_0}$, the RHS has to be negative, for a pole in the UV. To ensure no poles $k_0>0$. The case will be opposite for the IR, thus $k_0<0$ and the only possible solution, inserting this back in Eq.\ \eqref{eq:alphalamkeq0}, we obtain the restrictions for an asymptotically free quartic coupling in this case to be
\begin{equation}
    \alpha_{\lambda} = \frac{ \alpha_{\lambda0}  }{ \alpha_{g_0}  } \alpha_g\;,\quad\quad k_0=0\;, \quad\quad k=0\;.
\end{equation}
Thus for $k=0$, only one flow line would be asymptotically free. 

We have now considered the case of the gauge and Yukawa coupling off fixed flow. However, for the gauge and Yukawa coupling on fixed flow, the ratio of the beta function of the quartic and gauge coupling does not reduce
\begin{equation}
    \frac{\text{d} \alpha_\lambda}{\text{d}\alpha_g}=\frac{\alpha_\lambda}{b_0\alpha_g}\left(d_1\frac{\alpha_\lambda}{\alpha_g} + d_2+d_3\frac{\alpha_y}{\alpha_g}\right)+\frac{d_4}{b_0}+\frac{d_5\alpha_y^2}{b_0\alpha_g^2}\;.
\end{equation}
The CAF conditions in this case can be found by removing the Yukawa dependency using the fixed-flow relation of the Yukawa and gauge coupling in Eq.\ \eqref{eq:ygfixed2}. This lets us replace $\frac{\alpha_y}{\alpha_g}=\frac{b_0-c_1}{c_2}$ 
\begin{equation}
    \frac{\text{d} \alpha_\lambda}{\text{d}\alpha_g}=\frac{\alpha_\lambda}{b_0\alpha_g}\left(d_1\frac{\alpha_\lambda}{\alpha_g} + d_2+d_3\frac{b_0-c_1}{c_2}\right)+\frac{d_4}{b_0}+\frac{d_5(b_0-c_1)^2}{b_0c_2^2}\;.
\end{equation}
Now the equation can be rewritten as
\begin{equation}
    \frac{\text{d} \alpha_\lambda}{\text{d}\alpha_g}=\frac{\alpha_\lambda}{b_0\alpha_g}\left(d_1\frac{\alpha_\lambda}{\alpha_g} + d_2'\right)+\frac{d_4'}{b_0}\;,
\end{equation}
where 
\begin{align}
    d_2'=d_2+d_3\frac{b_0-c_1}{c_2}\;,\quad\quad d_4'&={d_4}+d_5\frac{(b_0-c_1)^2}{c_2^2}\;.
\end{align}
These new coefficients may change sign. Using the signs of coefficients $d_2,d_5\le 0$,  $d_3,d_4\ge0$, and $b_0-c_1>0$, we obtain
\begin{align}
     &d_2'>0 \quad \text{  if}\quad d_3\frac{b_0-c_1}{c_2}>|d_2|\;,
     \\&d_4'<0 \quad \text{  if}\quad |d_5|\frac{(b_0-c_1)^2}{c_2^2}>d_4\;.
\end{align}

After the CAF analysis, we have the following conditions when the Yukawa and gauge couplings are off fixed flow
\begin{equation}
    b_0<0,\quad\quad b_0-c_1>0,\quad\quad (b_0-d_2)^2-4d_1d_4\geq 0,\quad\quad b_0-d_2+\sqrt{(b_0-d_2)^2-4d_1d_4}>0\;, \label{eq:appCAF1con}
\end{equation}
for the coefficients of the beta functions, and the constraints on the gauge, Yukawa and self-coupling at an arbitrary energy scale 
\begin{align}
    &\text{For }k=0: \quad \alpha_{\lambda} = \frac{ \alpha_{\lambda0}  }{ \alpha_{g_0}  } \alpha_g\;,\quad\quad k_0=0\;.\label{eq:appke0con}\\&
    \text{For }k>0: \quad 
    \frac{b_0-d_2-\sqrt{k}}{2d_1}\le \frac{\alpha_{\lambda_0}}{\alpha_{g_0}}\le \frac{b_0-d_2+\sqrt{k}}{2d_1}\;. \label{eq:appkg0con}
\end{align}
Similarly, for on fixed-flow condition, the CAF conditions are given by
\begin{align}
    &b_0<0,\quad\quad b_0-c_1>0,\quad\quad \left(b_0-d_2-d_3\frac{b_0-c_1}{c_2}\right)^2-4d_1\left(d_4+d_5\left(\frac{b_0-c_1}{c_2}\right)^2\right)\geq 0,\\&b_0-d_2- d_3 \frac{b_0-c_1}{c_2}+\sqrt{\left(b_0-d_2 - d_3 \frac{b_0-c_1}{c_2}\right)^2-4d_1\left(d_4+d_5\left(\frac{b_0-c_1}{c_2}\right)^2\right)}>0\;.\label{eq:appCAF1conon}
\end{align}  

We have now obtained the CAF conditions in cases where the coupled ODE's can be solved analytically. Next, we move on to the more general case and how we can find CAF conditions utilizing the fixed-flow framework. 

\subsection{CAF Analysis Using the Fixed-Flow Framework} 
To utilise the fixed-flow approach, we redefine the couplings 
\begin{equation}
    \alpha_{g_i} = \frac{\tilde \alpha_{g_i}}{t}, \quad \quad  \alpha_{y_a} = \frac{\tilde \alpha_{y_a}}
    {t}, \quad \quad \alpha_{\lambda_m} = \frac{\tilde \alpha_{\lambda_m}}{t} \;.
\end{equation}
This gives us the following equations to solve
\begin{equation}
     \boldsymbol{x}_\infty=\begin{pmatrix}
        -\beta_{{\boldsymbol{g}}}(\tilde\alpha_{g\infty})\\-\beta_{{\boldsymbol{y}}}(\tilde\alpha_{g\infty},\tilde\alpha_{y\infty})\\-\beta_{\boldsymbol{\lambda}}(\tilde\alpha_{g\infty},\tilde\alpha_{y\infty},\tilde\alpha_{\lambda\infty})
    \end{pmatrix}\label{eq:solCAF2app}\;,
\end{equation}
effectively transforming the coupled ODE's to polynomial equations. 
\subsubsection{Two Quartic Couplings}\label{sec:appCafadj}
Using Eq.\ \eqref{eq:solCAF2app} we can write the polynomial equation for the quartic scalar coupling as
\begin{equation}
    \tilde\alpha_{\lambda _{m\infty}} + d^{\lambda}_{mnp}\tilde\alpha_{\lambda _{m\infty}}\tilde\alpha_{\lambda _{p\infty}}+ \tilde\alpha_{\lambda _{m\infty}}(d^{\lambda y}_m \tilde\alpha_{y _{\infty}} +d^{\lambda_g}_m\alpha_{y _{\infty}}) + d^{y}_m\alpha_{y\infty}^2+ d^{g}_m \alpha_{g\infty}^2 = 0    \;.\label{eq:CAF2quarticapp}
\end{equation}
We consider the case of $m=1,2$ corresponding to two quartic couplings. In this case, the algebraic equation \eqref{eq:CAF2quarticapp} can be written as two coupled quadratic equations
\begin{equation}
    a  {x}^2 + b  {z}^2 + c  {x}{z} + d  {x} + e = 0\;,
\end{equation}
\begin{equation}
    f  {z}^2 + g  {x}^2 + h  {x}{z} + i  {z} + j = 0\;,\label{eq:CAF2poly2app}
\end{equation}
where $x=\alpha_{\lambda_{1\infty}}$ and $z=\alpha_{\lambda_{2\infty}}$. The coefficients are given by
\begin{align}
    a=d^{\lambda}_{mmm},\quad b=d^{\lambda}_{mnn},\quad c=d^{\lambda}_{mmn},\quad d=(d^{\lambda y}_m \tilde\alpha_{y _{\infty}} +d^{\lambda_g}_m\tilde\alpha_{g _{\infty}}),\quad e=d^{y}_m\alpha_{y\infty}^2+ d^{g}_m \alpha_{g\infty}^2\;,\label{eq:coef2quarticapp}
\end{align}
where $m=1$ and $n=2$. For Eq.\ \eqref{eq:CAF2poly2app}, the coefficients $f$, $g$, $h$, $i$, and $j$ are defined analogously by exchanging the indices $m\Longleftrightarrow n$ of the coefficients in Eq.\ \eqref{eq:coef2quarticapp}.

The relevant case for our models has $b=0$
\begin{equation}
    a  {x}^2  + c  {x}{z} + d  {x} + e = 0 \;,   \label{eq:poly1}
\end{equation}
\begin{equation}
   f  {z}^2 + g  {x}^2 + h  {x}{z} + i  {z} + j = 0 \;. \label{eq:poly2}
\end{equation}
To determine whether there is any CAF region in this case, we must find the roots of the functions. This is done by inserting  ${z}$ from Eq.~\eqref{eq:poly1} into Eq.~\eqref{eq:poly2} and multiplying by $(d{x})^2$
\begin{equation}
    fT^2+gd^2 {x}^4 -hdT^2 {x}^2-idT  {x} +jd^2 {x}^2=0\;,
\end{equation}
where $T=ax^2+cx+e$. We have now obtained an uncoupled fourth-degree polynomial which we can express as
\begin{equation}
    \alpha  {x}^4  + \beta {x}^3 + \gamma  {x}^2 + \delta x+ \epsilon = 0  \;,  \label{eq:poly4th}
\end{equation}
where 
\begin{align}
&\alpha=fa^2+gd^2-had\;, \quad\beta= 2fac-hcd-iad\;, \quad\gamma=fc^2+2fae-hed-icd+jd^2\;, \\&\delta=e(2fc-id)\;, \quad\epsilon=fe^2\;.
\end{align}
By computing the sign of its determinant $\Delta$, given by \cite{Gellert1977}
\begin{align}
    \Delta = \ & 256\alpha^3 \epsilon^3 - 192\alpha^2 \beta\delta\epsilon^2 - 128\alpha^2 \gamma^2 \epsilon^2 + 144\alpha^2 \gamma\delta^2 \epsilon - 27\alpha^2 \delta^4 \\
& + 144\alpha\beta^2 \gamma\epsilon^2 - 6\alpha\beta^2 \delta^2 \epsilon - 80\alpha\beta\gamma^2 \delta\epsilon + 18\alpha\beta\gamma\delta^3 + 16\alpha\gamma^4 \epsilon \\
& - 4\alpha\gamma^3 \delta^2 - 27\beta^4 \epsilon^2 + 18\beta^3 \gamma\delta\epsilon - 4\beta^3 \delta^3 - 4\beta^2 \gamma^3 \epsilon + \beta^2 \gamma^2 \delta^2
\end{align}
one can find whether the solutions are real or complex. This depends on the following conditions \cite{Hansen:2017pwe}
\begin{itemize}
    \item For $\Delta < 0$, the polynomial will have two distinct real roots and two complex roots.
    \item For $\Delta > 0$, the polynomial has four real or complex roots.
    \begin{itemize}
        \item For $P < 0 \wedge  D < 0$: four real roots
        \item If $P > 0 \vee D > 0$: four complex roots.
    \end{itemize}
\end{itemize}
Where 
\begin{align}
    P &= 8\alpha\gamma - 3\beta^2\\
    D &= 64\alpha^3 \epsilon - 16\alpha^2 \gamma^2 + 16\alpha\beta^2 \gamma - 16\alpha^2 \beta\delta - 3\beta^4
\end{align}

Additionally, one should check whether the roots are real-positive or real-negative roots, which can be done using Descartes' rule of signs.

\paragraph{The determinant}of the quartic polynomial for the model described in sections \ref{sec:adj} and \ref{sec:adjNg} can be found using the beta functions of the theories and the definition of the coefficients of the polynomial. The boundary of the CAF region for both off and on fixed flow is found from the condition $b_0=0$ (Eqs. \eqref{eq:pmaxadj} and \eqref{eq:padjscalar}) and $\Delta=0$. The condition $\Delta=0$ with the Yukawa coupling off fixed flow is given by
\begin{align}
 \Delta=&-2176782336 N^4 P_0 - 725594112 A N^3 P_1 - 10077696 A^2 N^2 P_2 - 15116544 A^3 N P_3  \notag \\
&- 34992 A^4 P_4 - 139968 A^5 N P_5- 216 A^6 P_6 - 288 A^7 N P_7 - A^8 P_7  = 0\;,\label{eq:Delta0noY}
\end{align}
where $A = N (-21 + 4 N_g) + 4 (\mp3 N_g + p)$ and the polynomials are given by
\begin{align}
P_0 &= -217326564 - 56077596 N^2 + 23223267 N^4 - 2630538 N^6 + 166727 N^8 \notag\\&\;\;\;\;\; - 9548 N^{10} + 188 N^{12}\;, \\
P_1 &= -33434856 + 10958328 N^2 + 7518420 N^4 - 1773243 N^6 + 171254 N^8 \notag\\&\;\;\;\;\;- 9917 N^{10} + 350 N^{12}\;, \\
P_2 &= -64297800 + 83041848 N^2 + 15587964 N^4 - 10068291 N^6 \notag\\&\;\;\;\;\;+ 1385624 N^8 - 104921 N^{10} + 4472 N^{12} \;,\\
P_3 &= -571536 + 1910304 N^2 - 245412 N^4 - 262236 N^6 + 52675 N^8 - 5338 N^{10} + 247 N^{12} \\
P_4 &= -1714608 + 14362272 N^2 - 8214156 N^4 - 2139588 N^6 + 725257 N^8 \notag\\&\;\;\;\;\;- 98878 N^{10} + 4837 N^{12} \;,\\
P_5 &= 35964 - 46278 N^2 - 2949 N^4 + 3476 N^6 - 629 N^8 + 32 N^{10}\;, \\
P_6 &= 107892 - 365634 N^2 + 31473 N^4 + 27116 N^6 - 6143 N^8 + 320 N^{10}\;, \\
P_7 &= (-5 + N^2)^2 (-81 - 18 N^2 + 2 N^4)\;.
\end{align}
On fixed flow, we obtain
\begin{align} \Delta=&- B^4 D^2 N^2 (1 + N^2)^2 P_3^2 + B^2 D^3 N (7 + N^2) P_3^2 \nonumber\\ & - 64 C B^6 N^3 (1 + N^2)^3 P_3^3 + 72 C B^4 D N^2 (1 + N^2) (7 + N^2) P_3^3 + 432 C^2 B^4 N^2 (7 + N^2)^2 P_3^4 \nonumber\\& + B^2 D^3 N (1 + N^2)^2 P_4  - D^4 (7 + N^2) P_4 + 72 C B^4 D N^2 (1 + N^2)^3 P_3 P_4\nonumber \\ & - 80 C B^2 D^2 N (1 + N^2) (7 + N^2) P_3 P_4 + 96 C^2 B^4 N^2 (1 + N^2)^2 (7 + N^2) P_3^2 P_4 \label{eq:Delta0}\\ & - 576 C^2 B^2 D N (7 + N^2)^2 P_3^2 P_4 + 432 C^2 B^4 N^2 (1 + N^2)^4 P_4^2\nonumber\\& - 576 C^2 B^2 D N (1 + N^2)^2 (7 + N^2) P_4^2  + 128 C^2 D^2 (7 + N^2)^2 P_4^2  \nonumber\\&- 3072 C^3 B^2 N (1 + N^2) (7 + N^2)^2 P_3 P_4^2 - 4096 C^4 (7 + N^2)^3 P_4^3  = 0 \;,\nonumber
\end{align}
where the polynomials are given by
\begin{align} 
    P_1 &= -18 + N^2 (-3 + 4 N_g) + 4 N (\mp3 N_g + p) \;,\quad P_2 = -3 + N^2 + 2 N p \;,\\ P_3 &= 9 - 20 N^2 + N^4 \;,\quad  P_4 = 567 + 315 N^2 - 35 N^4 + N^6 \;, 
\end{align}
and the terms $B, C,D$ is given by
\begin{align}
    B &= (-3 + N^2) \big(N (15 + 4 N_g) + 4 (\mp3 N_g + p)\big) + \big(72 + N^2 (42 - 8 N_g) - 8 N (\mp3 N_g + p)\big) p\;, \\ C &= 2 P_1^2 p - 27 N P_2^2 \;,\\ D &= 23328 N P_2^2 - 12 N B^2 - 48 (3 - 2 N^2) C + (7 + N^2) \big(N B^2 + 8 (9 - N^2) C\big)\;.
\end{align}

\section{Supplementary Figures}
\subsection{RG Flows: BY model with an Anti-Fundamental Scalar}\label{app:flow}
\begin{figure}[b!]
    \centering
    \begin{subfigure}[b]{0.45\textwidth}
        \centering
        \includegraphics[width=\textwidth]{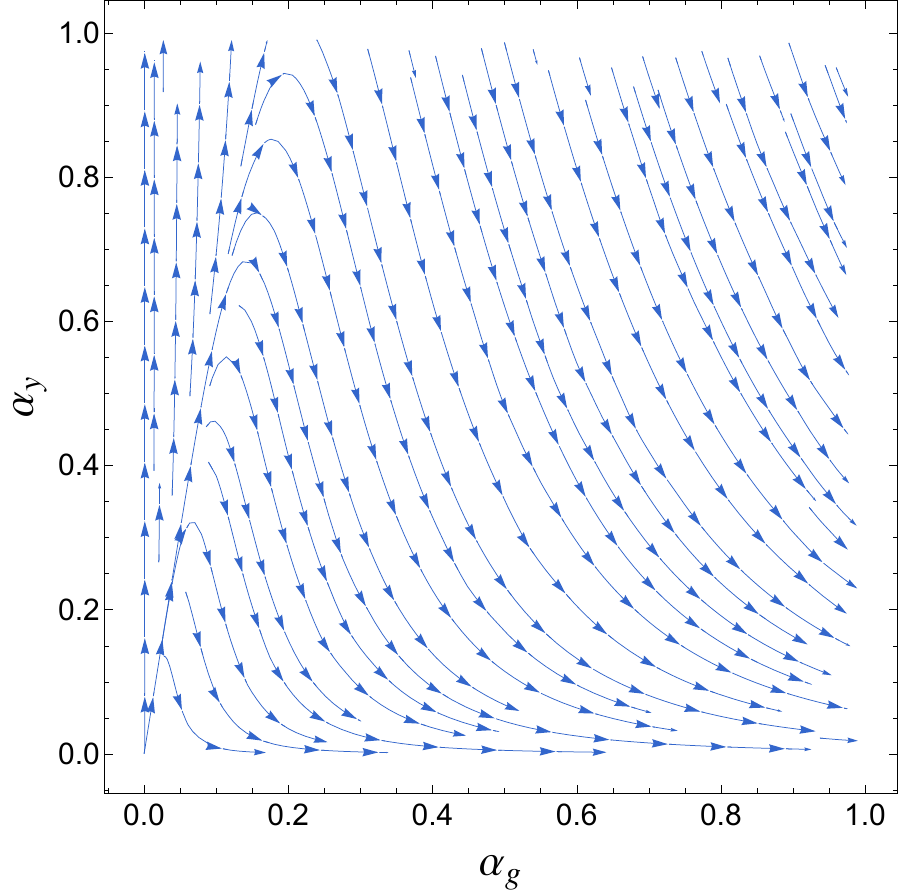}
        \caption{$(\alpha_g,\alpha_y)$} 
        \label{fig:flowgy_left}
    \end{subfigure}%
    \hfill
    \begin{subfigure}[b]{0.45\textwidth}
        \centering
        \includegraphics[width=\textwidth]{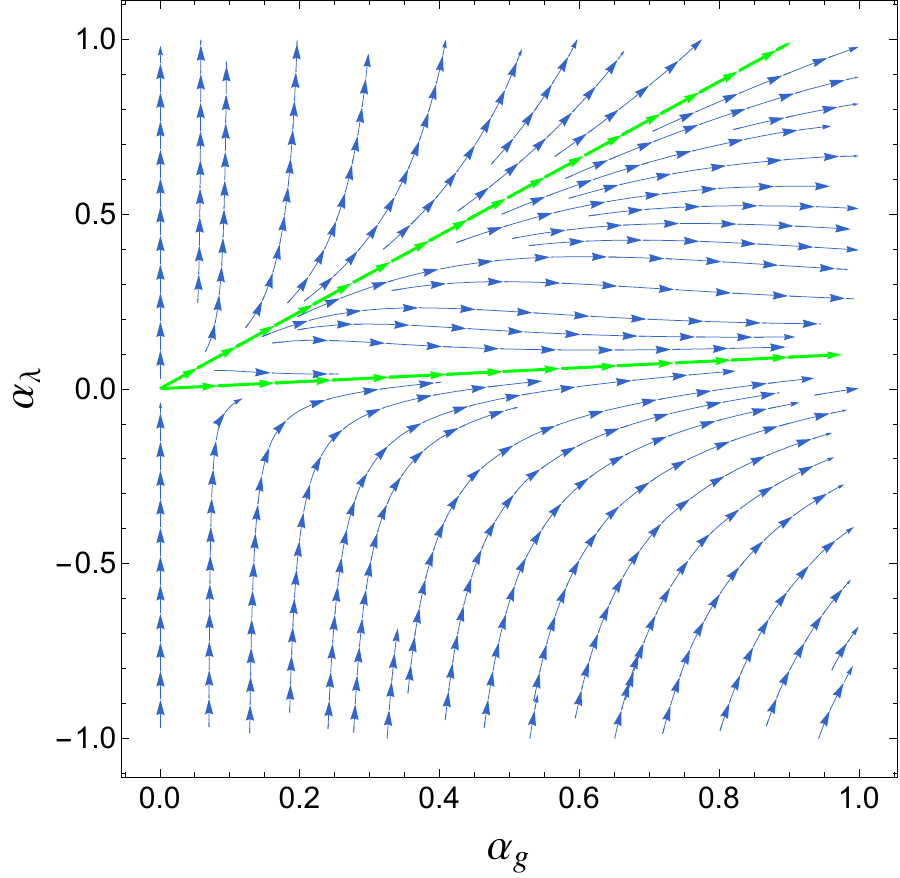}
        \caption{$(\alpha_g,\alpha_\lambda)$}
        \label{fig:flowgy_right}
    \end{subfigure}
    \caption{The figure displays flow diagrams for the BY model with $N=5$ and $p=30$ (a) for $(\alpha_g,\alpha_y)$ and (b) $(\alpha_g,\alpha_\lambda)$. The green flow lines represent fixed-flow trajectories of these couplings.}
\label{fig:flowgy}
\end{figure}
To better understand the CAF regions, the RG flow of the BY model (a similar analysis can be done for the GG model) with one anti-fundamental scalar can be analysed. Note that in all the flow diagrams, the arrows point towards the UV.

\begin{figure}[ht]
    \centering
    \begin{subfigure}[b]{0.49\textwidth}
        \centering
        \includegraphics[width=\textwidth]{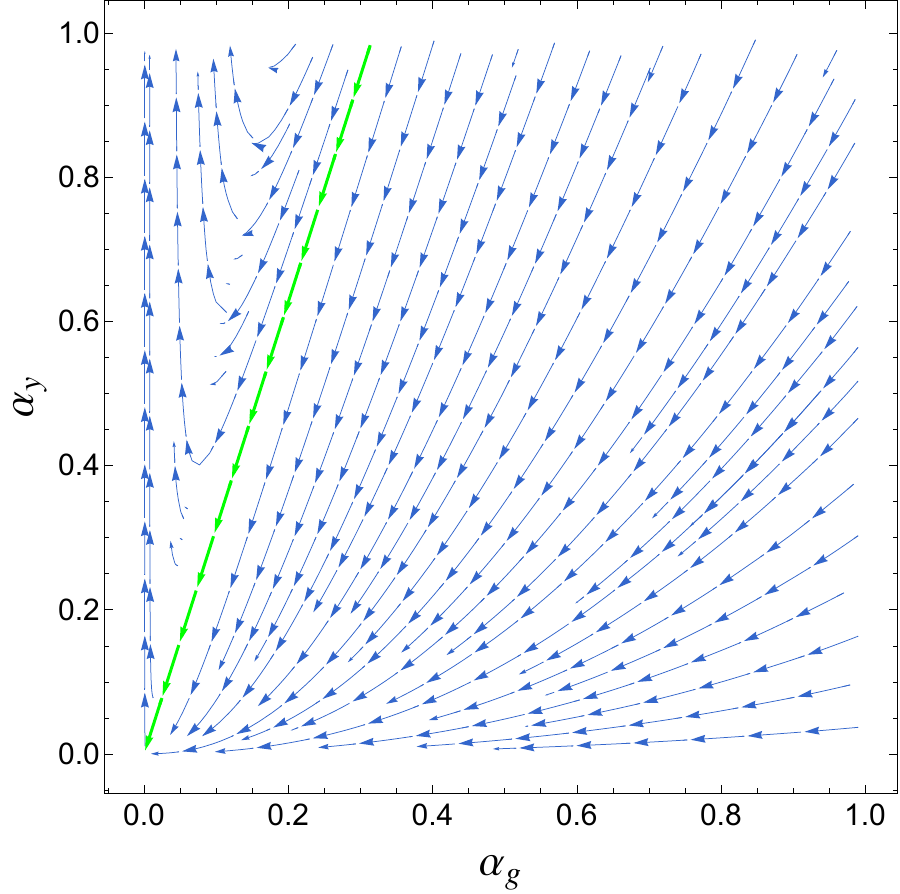}
        \caption{$(\alpha_g,\alpha_y)$}
        \label{fig:flowlam_left}
    \end{subfigure}%
    \hfill
    \begin{subfigure}[b]{0.49\textwidth}
        \centering
        \includegraphics[width=\textwidth]{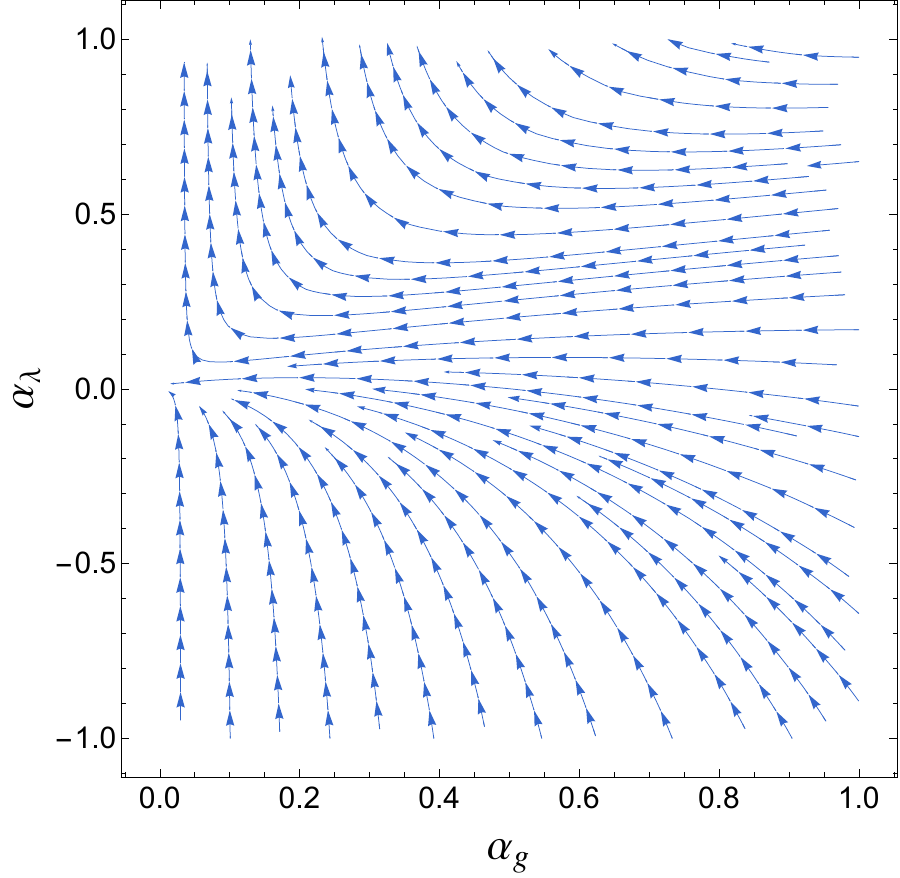}
        \caption{$(\alpha_g,\alpha_\lambda)$}
        \label{fig:flowlam_right}
    \end{subfigure}
    \caption{The figure displays flow diagrams for the BY model with $N=10$ and $p=12$(a) for $(\alpha_g,\alpha_y)$ and (b) $(\alpha_g,\alpha_\lambda)$. The green flow line represents the fixed-flow trajectory of these couplings.}
\label{fig:flowlam}
\end{figure}

First, consider Figure \ref{fig:flowgy_left}. The plot displays the gauge versus Yukawa coupling for $N=5$, $p=30$. This choice of parameters places the model outside the CAF region, where the gauge coupling loses asymptotic freedom. The flow plot also shows that there is no region within the plot where the gauge coupling is asymptotically free. Furthermore, one can see that for $\alpha_g=0$, the Yukawa coupling diverges, hitting a Landau pole in the UV. This is exactly as one would expect, since the gauge coupling is responsible for making the Yukawa coupling asymptotically free (see Section \ref{sec:CAF}). 

Now consider Figure \ref{fig:flowgy_right}, which has the same $N$ and $p$ but displays the gauge versus the quartic coupling. As already concluded from Figure \ref{fig:flowgy_left}, the gauge coupling cannot be asymptotically free. Thus, the quartic coupling cannot be asymptotically free\footnote{The assumption that $b_0<0$ is used in the CAF analysis regarding the quartic coupling, because it can not be asymptotically free without an asymptotically free gauge coupling. }. This is also evident from the flow plot in Figure \ref{fig:flowgy_right}. Here, none of the flows is asymptotically free in either of the couplings.
\begin{figure}[t!]
    \centering
    \begin{subfigure}[b]{0.5\textwidth}
        \centering
        \includegraphics[width=\textwidth]{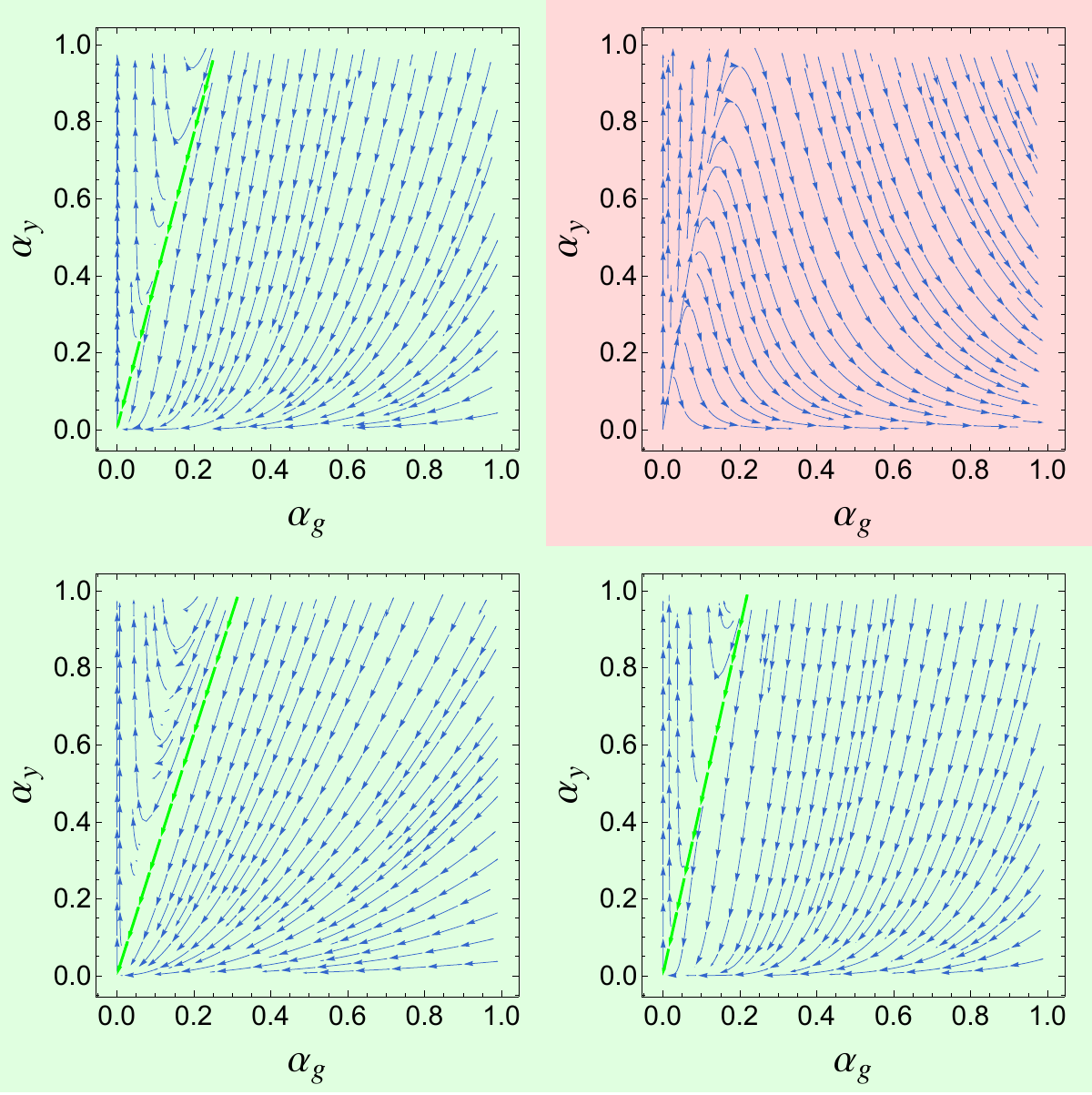}
        \caption{$(\alpha_g,\alpha_y)$}
        \label{fig:gridgy}
    \end{subfigure}%
    \hfill
    \begin{subfigure}[b]{0.5\textwidth}
        \centering
        \includegraphics[width=\textwidth]{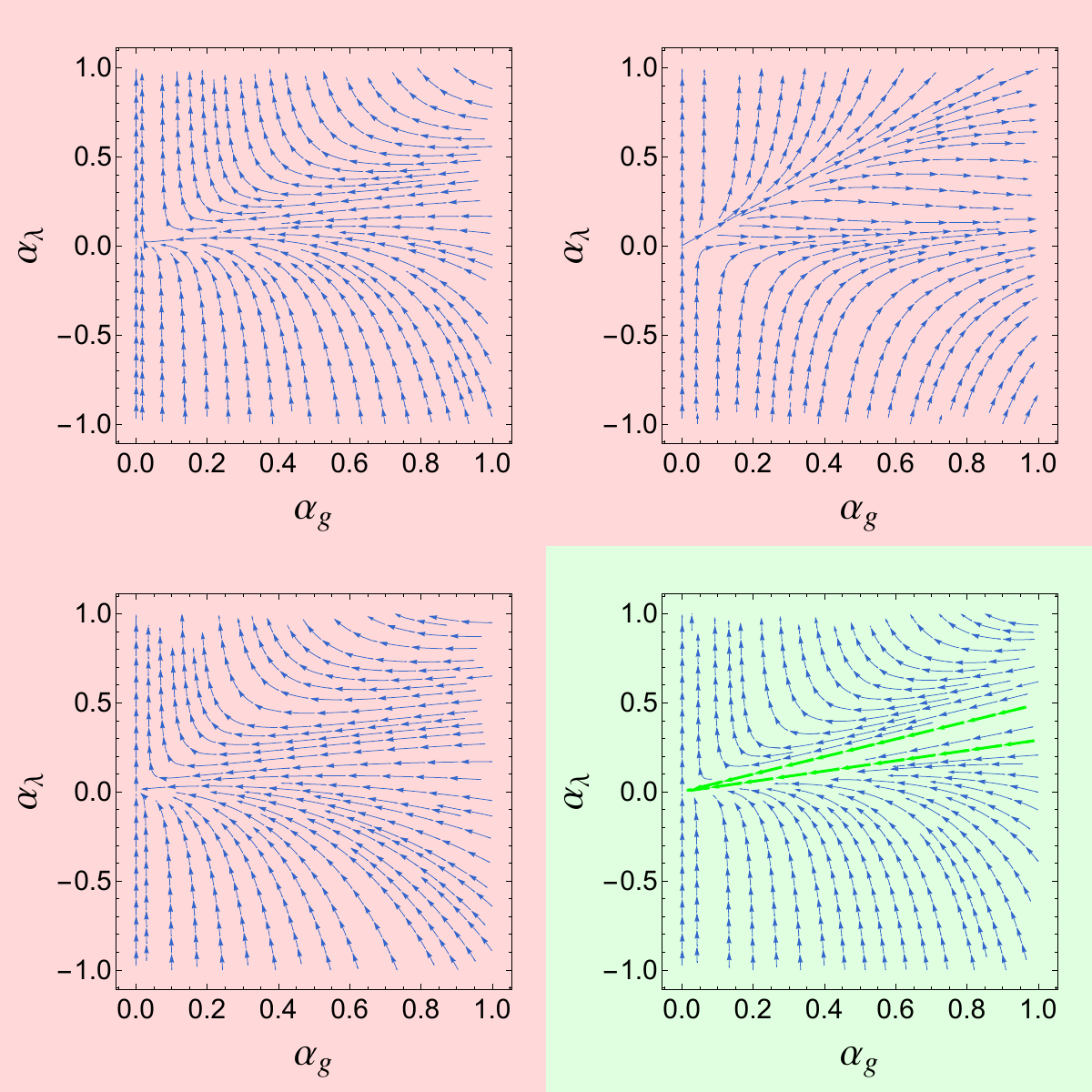}
        \caption{$(\alpha_g,\alpha_\lambda)$}
        \label{fig:gridlam}
    \end{subfigure}
    \caption{Flow diagrams of (a) $(\alpha_g,\alpha_y)$ and (b) $(\alpha_g,\alpha_\lambda)$ for $(N,p)=(5,12)$, $(5,10)$, $(10,12)$, and $(10,30)$ (top-left to bottom-right). Green (red) background regions indicate the presence (absence) of asymptotic freedom for both couplings. The green flow lines illustrate the fixed-flow trajectories of these couplings (a) $(\alpha_g,\alpha_y)$ and (b) of $(\alpha_g,\alpha_\lambda)$.}
\label{fig:flowgrid}
\end{figure}

In Figure \ref{fig:flowlam}, the case of $N=10,\;p=12$ is shown. These flow plots are below the light blue region on Figure \ref{fig:cafby}, where only the case of the Yukawa and the gauge coupling on fixed flow satisfies the CAF conditions. In Figure \ref{fig:flowlam_left}, a green dashed line can be seen, which shows the fixed flow of the gauge and Yukawa coupling. Everything underneath this line is asymptotically free in both couplings. Thus, the Yukawa and gauge coupling can be asymptotically free in this case. 

In Figure \ref{fig:flowlam_right}, the flow of the gauge and quartic coupling is plotted for $N=10$, $p=12$ as well. The flow diagram shows that for a positive quartic coupling, the flow always goes to infinity, and thus is not asymptotically free. Although we do not consider the region $\alpha_\lambda<0$, as it would lead to a potential unbounded from below, the RG flow in this region is directed towards $\alpha_\lambda=0$. However, in the CAF analysis, we required a positive value of $\alpha_\lambda$, and therefore there are no physical valid asymptotically free RG trajectories (see Section \ref{sec:CAF}).

\begin{figure}[t!]
    \centering
    \includegraphics[width=0.51\linewidth]{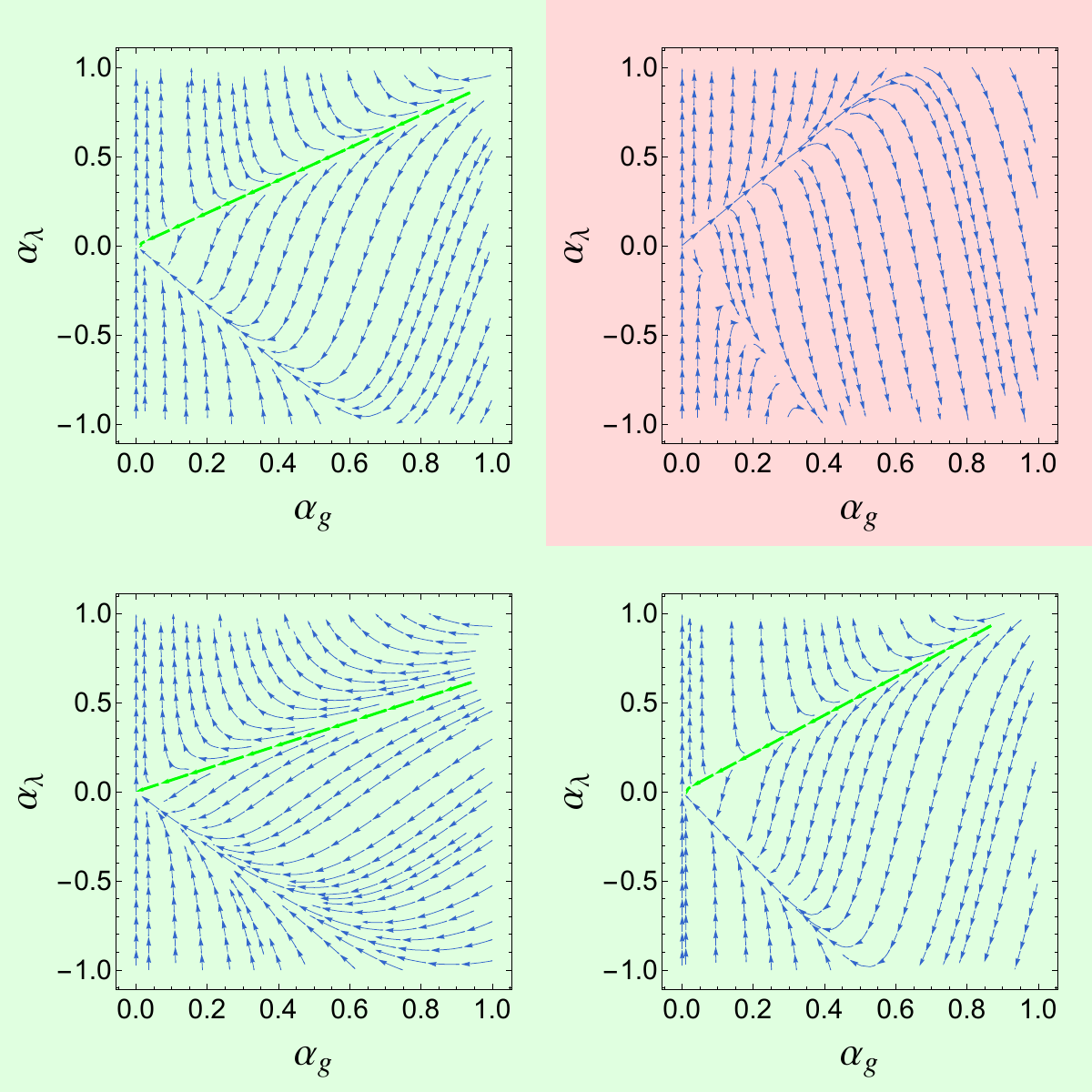}
    \caption{Flow diagrams of $(\alpha_g,\alpha_\lambda)$ for $(N,p)=(5,12),(5,10),(10,12), \,\text{and }(10,30)$ (top-left to bottom-right) for the gauge and Yukawa coupling on fixed flow. Green (red) background colour indicates the presence (absence) of asymptotic freedom for the scalar coupling, when the gauge and Yukawa coupling is on fixed flow. The green flow lines illustrate the fixed-flow trajectories of these couplings $(\alpha_g,\alpha_\lambda)$.}
    \label{fig:gridlamfixed}
\end{figure}
To compare more values of $N$ and $p$ using flow diagrams, we consider four scenarios in Figures \ref{fig:gridgy}, \ref{fig:gridlam} and \ref{fig:gridlamfixed}. Figure \ref{fig:gridgy} shows whether the gauge and Yukawa coupling are asymptotically free. Green (red) background colour indicates the presence (absence) of asymptotic freedom for both couplings. The same values of $N$, $p$ are shown in the second grid in Figure \ref{fig:gridlam}, here the coupling space is the gauge vs.\ the quartic coupling. In this case green (red) background colour indicates the presence (absence) of asymptotic freedom for the quartic coupling

Let us now consider the case Yukawa coupling on fixed flow. This changes nothing in the flow plots of the gauge vs.\ Yukawa coupling. The conditions of the quartic couplings do, however, change on fixed flow, as can be seen in Figure \ref{fig:gridlamfixed}. The three figures in the bottom left all exhibit similar behaviour. Below the line of the fixed flow (of the scalar and gauge coupling), the flow moves to a negative quartic coupling before flowing up to the origin, or the flow comes from negative infinity before reaching the origin. Even though this flow tends towards $\alpha_\lambda=0$, it does so from a negative quartic coupling, which is not physical. The RG flow above the fixed-flow line (green dashed line) goes towards a Landau pole. Thus, only exactly on the fixed-flow line is it possible to have a complete asymptotically free theory with a positive quartic coupling at all energy scales. These three flow plots also have asymptotically free gauge coupling, since the flow moves toward the origin. 

The plot in the top right of Figure \ref{fig:gridlamfixed} exhibits non-asymptotic freedom. This is evident from the RG flow, which never tends towards $\alpha_\lambda=0$. This is because the gauge coupling is not asymptotically free for these values of $N$ and $p$.

Now, combining the information from the three plots, we conclude that on fixed flow the theory is only CAF for one of the four parameter sets, namely when $N=10$ and $p=30$. Going back to Figure \ref{fig:cafby}, these values lie within the light blue CAF region. Additionally, on fixed flow three of the four parameter sets are CAF, effectively only the boundary coming from the gauge coupling influences whether the model is CAF in that case.   

Together, from the Figures \ref{fig:gridgy}, \ref{fig:gridlam} and \ref{fig:gridlamfixed}, we can understand the meaning of the CAF regions. Further, the conclusions drawn from these are consistent with Figure \ref{fig:cafby}.

\subsection{Chiral Models with an Adjoint Scalar} \label{sec:appGGplotsadj}
\begin{figure}[h!]
    \centering
    \begin{subfigure}[b]{0.4\textwidth}
        \centering
        \includegraphics[width=\textwidth]{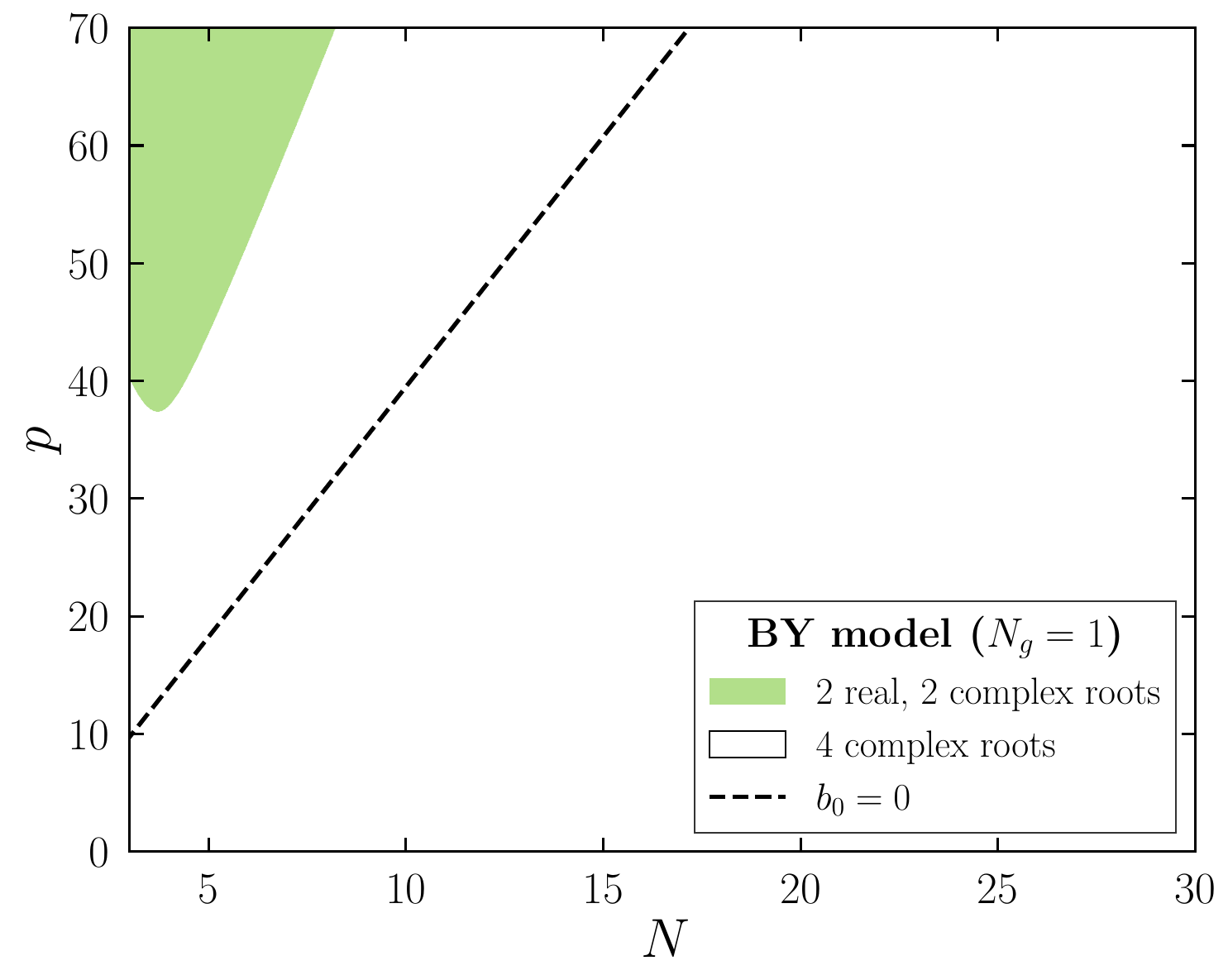}
        \caption{All couplings on fixed flow} 
        \label{fig:roots_fixedGG}
    \end{subfigure}
    \hspace{0.1cm}
    \begin{subfigure}[b]{0.4\textwidth}
        \centering
        \includegraphics[width=\textwidth]{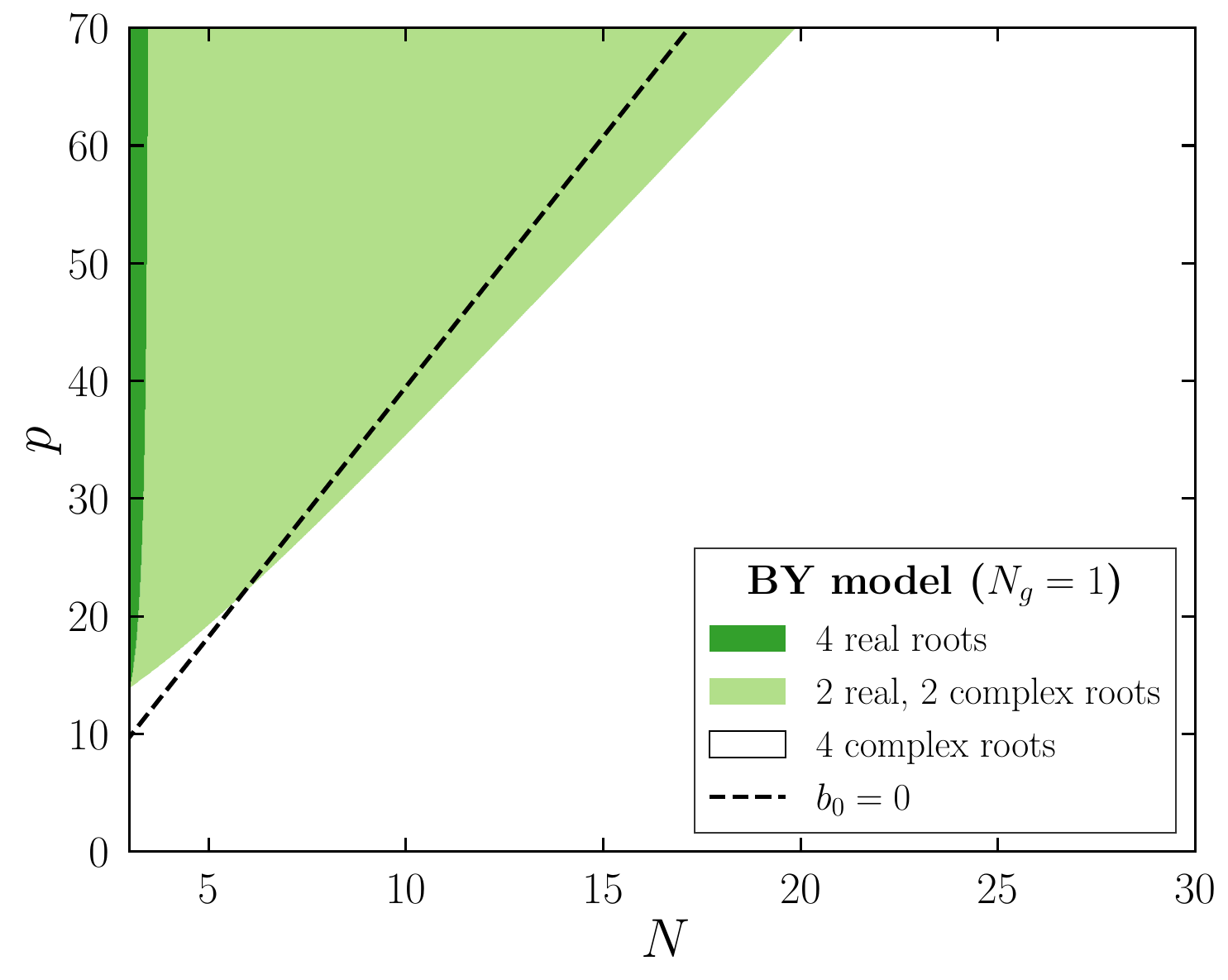}
        \caption{$\alpha_y$ off fixed flow} 
        \label{fig:roots_noyGG}
    \end{subfigure}
    
    \caption{ Root structure of the fourth-degree polynomial for the BY model, for (a) all couplings on fixed flow and (b) the Yukawa coupling off fixed flow. The shaded regions denote four complex roots (white), four real roots (dark green), and two real and two complex roots (light green). The dashed black line illustrates the $b_0=0$ condition.} 
    \label{fig:DeltaadjGG}
\end{figure}

In the model described in Section \ref{sec:adj}, we obtain Figure \ref{fig:DeltaadjGG} and \ref{fig:gg_combined_ng1}. Similarly to  the GG model, only the case of the Yukawa coupling off fixed flow yields a CAF region this can be seen by the root signature in Figure \ref{fig:DeltaadjGG}, where only \ref{fig:roots_noyGG} has real roots below the dashed line. 

    